\documentclass[10pt,nofootinbib,superscriptaddress,preprintnumbers,aps]{revtex4}
\usepackage{amsmath}
\usepackage{amssymb}
\usepackage{graphicx}
\usepackage{autobreak}

\makeatletter
\newcommand{\beq}{\begin{equation}}
\newcommand{\bX}{{\boldsymbol X}}
\newcommand{\bA}{{\boldsymbol A}}
\newcommand{\bG}{{\boldsymbol G}}
\newcommand{\bB}{{\boldsymbol B}}
\newcommand{\bV}{{\boldsymbol V}}
\newcommand{\bC}{{\boldsymbol C}}

\newcommand{\bk}{{\boldsymbol k}}

\newcommand{\bn}{{\boldsymbol n}}

\newcommand{\bP}{{\boldsymbol P}}
\newcommand{\bL}{{\boldsymbol L}}
\newcommand{\ble}{{\boldsymbol \ell}}
\newcommand{\blambda}{{\boldsymbol \lambda}}
\newcommand{\bJ}{{\boldsymbol J}}
\newcommand{\bq}{{\boldsymbol q}}
\newcommand{\bv}{{\boldsymbol v}}
\newcommand{\br}{{\boldsymbol r}}

\newcommand{\bx}{{\boldsymbol x}}

\newcommand{\bSigma}{{\boldsymbol \Sigma}}
\newcommand{\bS}{{\boldsymbol S}}
\newcommand{\ba}{{\boldsymbol a}}

\newcommand{\eeq}{\end{equation}}
\newcommand{\bea}{\begin{eqnarray}}

\newcommand{\eea}{\end{eqnarray}}
\usepackage{bm}
\usepackage{psfrag}
\usepackage{epsfig}
\usepackage{mathrsfs}
\usepackage{wasysym}
\usepackage{epstopdf}
\usepackage{comment}
\usepackage{pbox}
\usepackage{array}
\usepackage{xspace}
\usepackage[usenames,dvipsnames]{color}
\allowdisplaybreaks

\def\addUPitt{Pittsburgh Particle Physics, Astrophysics, and Cosmology Center, Department of Physics and Astronomy, University of Pittsburgh, Pittsburgh, PA 15260, USA}

\def\addDESY{Deutsches Elektronen-Synchrotron DESY, Notkestrasse 85, 22607 Hamburg, Germany}

\begin{document}
\preprint{DESY 21-039 }
\title{Gravitational radiation from inspiralling compact objects:\\ [.2cm] Spin-spin effects completed at the next-to-leading post-Newtonian order}
\author{Gihyuk Cho}
\affiliation{\addDESY}
\author{Brian Pardo}
\affiliation{\addUPitt}
\author{Rafael A.\ Porto}
\affiliation{\addDESY}

\begin{abstract}
Using the gravitational potential and source multipole moments bilinear in the spins, first computed to next-to-leading order (NLO) in the post-Newtonian (PN) expansion within the effective field theory (EFT) framework, we complete here the derivation of the dynamical invariants and flux-balance equations, including energy and angular momentum. We use these results to calculate spin-spin effects in the orbital frequency and accumulated phase to NLO for circular orbits. We~also derive the linear momentum and center-of-mass fluxes and associated kick-velocity, to the highest relevant PN order. We explicitly demonstrate the equivalence between the quadratic-in-spin source multipoles obtained using the EFT formalism and those rederived later with more traditional tools, leading to perfect agreement for spin-spin radiative observables to NLO among both approaches.\end{abstract}
\maketitle
\newpage
\tableofcontents
\newpage
\section{Introduction}

Binary systems composed of spinning compact objects are the primary source of gravitational waves (GWs) that are currently being detected by the LIGO/VIRGO collaboration \cite{LIGOV1,LIGOV2}, and will remain as key laboratories to explore gravity and the physics of black holes and neutron stars~\cite{buosathya,tune,music} (and perhaps even more exotic possibilities \cite{axiverse,qcd1,cardoso,salvo2,Ng:2020ruv,gcollider1,gcollider2}) with future GW observatories such as LISA \cite{lisa} and the Einstein Telescope \cite{et}. Yet, both the detection and accurate interpretation of the signal are dependent on precise theoretical predictions for the two-body problem in general relativity. This is notoriously important when spin effects are manifest, e.g.~\cite{salvo}, in particular due to the expectation that  binary black holes may be rapidly rotating, e.g.~\cite{Zackay:2019tzo}. While numerical simulations are required for the late stages of the dynamics, e.g.~\cite{Hinder:2018fsy}, the post-Newtonian (PN) expansion has provided the groundwork to tackle the weak-field/small-velocity inspiral regime \cite{blanchet,Schafer:2018kuf,walterLH,review}. So far, PN studies have been carried out notably in the conservative sector, both for nonspinning 
\cite{Blanchet:2003gy,nrgr3pn,Foffa:2012rn,nltail,nrgrG5,apparent,Damour:2014jta,Jaranowski:2015lha,Bernard:2015njp,Bernard:2016wrg,Marchand:2017pir,rec2_Foffa_2019,nrgr4pn2,5pn1,5pn2,Blumlein:2020pyo,blum,blum2,bini2,tail3} and spinning bodies \cite{Faye1,prl,Porto:2007px,nrgrss,nrgrs2,jan1,jan2,nrgrso,Levi:2015uxa,Levi:2016ofk,Levi:2020uwu,Levi:2020kvb,Antonelli:2020aeb, Antonelli:2020ybz}, using various tools \cite{blanchet,Schafer:2018kuf,walterLH,review}. In particular, the effective field theory (EFT) approach introduced in \cite{nrgr}, and later extended in \cite{nrgrs} to incorporate rotational degrees of freedom, has been instrumental to reach the present state-of-the-art.\footnote{The use of ideas from particle physics has also been reinvigorated lately due the repurposing of novel techniques from the theory of scattering amplitudes, e.g.~\cite{elvang,reviewdc}. In parallel with an extension of the EFT approach to the post-Minkowskian regime \cite{paper1,paper2,pmeft}, these novel approaches have extended the knowledge of the binary dynamics in the conservative sector, both for nonspinning \cite{cheung,donal,zvi1,zvi2,Cheung:2020gyp,soloncheung,3pmeft,tidaleft,Bern:2020uwk,Cheung:2020gbf,4pmzvi} and spinning objects~\cite{justin1,justin2,donalvines,Guevara:2018wpp,Guevara:2019fsj,Aoude:2020ygw,Chung:2020rrz,zvispin,luna,spinpm}. Radiation effects in the post-Minkowskian expansion have also been studied in e.g.~\cite{Parra2,Gabriele,janmogul,janmogul2,Mougiakakos:2021ckm}.} However, in the radiation sector, while the source multipoles needed to obtain the GW fluxes in an adiabatic expansion have been computed in some cases to fourth-order in the PN expansion for nonspinning bodies \cite{Blanchet:2004ek,Marchand:2020fpt}, the spin-dependent counterparts are known to next-to-next-to-leading order (N$^{2}$LO) at linear order in spin \cite{Faye2,nnloso,srad}, and {\it only} to NLO for bilinear in spins contributions \cite{srad,amps,Bohes2}.\footnote{The associated radiation-reaction spin-orbit and spin-spin effects in the dynamics have been computed at leading order  in \cite{natalia1,natalia2}.} This begs for more accurate computations of radiative observables in the case of spinning bodies.\vskip 4pt  All the necessary quadratic-in-spin source multipoles including finite-size effects were first obtained some time ago in \cite{srad,amps}, together with the spin-spin gravitational potentials \cite{prl,nrgrss,nrgrs2}, using the EFT formalism developed in \cite{nrgrs,andirad}. However, the derivation of the associated GW phase evolution was not carried out until recently. In \cite{pardo}, using the results obtained in \cite{nrgrso,srad} at linear order in the spin, the gravitational observables including the accumulated phase were computed to NLO and shown to agree with the values first derived in \cite{Faye1,Faye2} with more traditional tools. With the aim to move forward in precision towards higher PN orders, the purpose of this paper is to use the results in \cite{prl,nrgrss,nrgrs2,nrgrso,srad,amps} to complete the derivation of the spin-spin contributions to the dynamical invariants and flux-balance laws to NLO in generic configurations, including energy and angular momentum; and use the results to derive spin-spin effects in the orbital frequency and accumulated phase to 3PN for (quasi-)circular orbits. For completeness, we also compute the linear momentum and center-of-mass (CoM) radiated fluxes to the highest possible PN order using our  source multipoles, corresponding to leading spin-spin and NLO linear-in-spin corrections, with which we obtain the associated kick-velocity. We find agreement with a previous derivation of the radiated momentum~in~\cite{Racine} at linear order in the spin; however, we disagree when it comes to spin-spin contributions, even for the simplest case of binary black holes. We~trace the difference to missing terms from finite-size effects. On the other hand, we readily demonstrate the equivalence between the source multipole moments computed in \cite{srad,amps} within the EFT framework and those rederived later in \cite{Bohes2} using the approach in \cite{Faye1,Faye2}, yielding complete agreement for the linear and bilinear in spin radiative observables at NLO order among both formalisms. Our results here can be used to improve current waveform models for spinning bodies, notably for elliptic-like orbits \cite{Cho:2019brd,Chatziioannou:2017tdw,Csizmadia:2012wy}. The derivation of the N$^2$LO phase evolution is underway.\vskip 4pt

This paper is organized as follows. In sec.~\ref{sec:EFTsetup} we briefly overview the steps to compute the necessary ingredients to obtain the GW phase evolution in the EFT framework.
 In sec.~\ref{invariants} we start in the conservative sector and use the gravitational potential in \cite{prl,nrgrss,nrgrs2} to compute the conserved energy and angular momentum, as well as the position of the CoM, at quadratic order in the spins and to NLO in the PN expansion. In sec.~\ref{laws} we move to the radiation sector and compute the associated flux-balance equations using the source multipole moments bilinear in spin obtained in \cite{srad,amps}. We also calculate the radiated fluxes of linear momentum and position of the CoM, and derive the kick-velocity for (quasi-)circular orbits.  In sec.~\ref{phase} we use the results from previous sections to compute the bilinear-in-spin accumulated phase to 3PN order. We conclude in sec.~\ref{final} with a few remarks on future work. We~include several appendices with the coefficients of the lengthy expressions quoted in the main text. For the reader's convenience we add an ancillary file with the main results obtained with the aid of the \texttt{xAct} packages~\cite{martin2019xact}.
  \newpage
\section{EFT setup\label{sec:EFTsetup}}

Both the gravitational conservative dynamics and radiative multipole moments of spinning binary systems can be obtained using the EFT formalism introduced in \cite{nrgr,nrgrs,prl,nrgrss,nrgrs2,nrgrso,andirad,andirad2,srad,amps}.
We provide here a brief overview of the EFT framework and encourage the reader to consult the more comprehensive reviews in \cite{walterLH,Foffa:2013qca,review} for further details.

\subsection{Point-particle theory}

The main actor in the EFT approach is the point-particle action, 
\[
S_{\rm pp} = \sum_{A=\{1,2\}} \int {\cal R}_A d\sigma_A\,,
\]
with $\sigma_A$ an affine parameter, which for the case of spinning bodies is written in terms of a Routhian,
\begin{equation}
\label{rout}
\mathcal{R}_A= -\sum_{A}\biggl[m_{A}\sqrt{v_{A}^{2}}+\frac{1}{2}\omega_{\mu ab}S_{A}^{ab}v_{A}^{\mu}+\frac{1}{2m_{A}}R_{\nu cab}S_{A}^{ab}v_{A}^{\nu}S_{A}^{cd}v_{Ad}-  \frac{C^{(A)}_{ES^2}}{2m_A} \frac{E_{ab}}{\sqrt{v_A^2}} S_A^{ac}{S_{Ac}}^b  + \cdots\biggr],
\end{equation}
displayed here to quadratic order in the spins. The velocity is defined as $v^\mu \equiv \frac{dx^\mu}{d\sigma}$, and $w^{ab}_\mu$ are the Ricci rotation coefficients. The spin tensor, $S^{\mu\nu}$, has been projected onto a locally-flat frame, described by a tetrad field, $e_a^\mu$, such that $S^{ab} \equiv e^a_\mu e^b_\nu S^{\mu\nu}$. There are also two curvature-dependent terms. The first ensures the preservation of the covariant spin supplementary condition (SSC), to the order we work here, 
\beq
S_{A}^{ab} v_{Ab} = S_A^{ab} e_b^\mu v_{A\mu} =0 + {\cal O}(S^3)\, \label{ssc}.  
\eeq
The other one, depending on the electric component of the Weyl tensor, $E_{\mu\nu}$, encapsulates the spin-induced quadrupole moment of a rotating object. The $\smash{C^{(A)}_{ES^2}}$ parameters are so-called `Wilson coefficients,' which carry information about the short-distance physics of the compact body, e.g.~$C_{ES^2}=1$ for Kerr black holes \cite{nrgrss,nrgrs2}. The ellipsis contains an additional hierarchy of higher curvature corrections that include other finite-size multipolar corrections, beyond the quadrupole, as well as extra pieces required to enforce the SSC.\vskip 4pt

The equations of motion (EoM) for each particle can then be obtained via 
\beq
\frac{\delta S_{\rm pp}}{\delta x_A^{\mu}}\,\,, \quad\frac{dS_A^{ab}}{d\sigma_A}=\left\{ S_A^{ab},\mathcal{R}_A\right\}\, , \label{eom}
\eeq
with the spin algebra
\beq
\left\{ S_A^{ab},S_A^{cd}\right\} =\eta^{ac}S_A^{bd}+\eta^{bd}S_A^{ac}-\eta^{ad}S_A^{bc}-\eta^{bc}S_A^{ad}.
\eeq
After the dynamical equations are obtained, the SSC in \eqref{ssc} may be enforced, enabling us to rewrite the EoM in vectorial form. To the order we work in this paper, we have ($B\neq A$) \cite{nrgrss,nrgrs2,nrgrso}
\beq
S_{A}^{i0} = ({\bv}_A\times {\bS}_A)^i 
+ \frac{2Gm_B}{r}\bigl(({\bv}_A-{\bv}_B)\times{\bf S}_A\bigr)^i+\frac{G}{r^3} ({\bS}_B^i{\br}\cdot {\bS}_A-{\br}^i{\bS}_A\cdot{\bS}_B) +\cdots \,, \label{ssc2}
\eeq
where $\br\equiv \bx_A-\bx_B$, and the spin three-vector is defined as \beq \bS_A^i\equiv \frac{1}{2}\epsilon^{ijk}S_A^{jk}\,. \eeq

\subsection{Long-distance worldline action}
For the two-body problem in the inspiral regime the gravitational field is expanded around flat space in the weak-field approximation,
\begin{equation}
    g_{\mu\nu} = \eta_{\mu\nu}+h_{\mu\nu}\,,
\end{equation}
where $h_{\mu\nu}$ is further split into potential, $H_{\mu\nu}$, and radiation, $\bar h_{\mu\nu}$, modes scaling as $(k^0,\bk)_{\rm pot} \sim (1/r,v/r)$ and $(k^0,\bk)_{\rm rad} \sim (v/r,v/r)$, respectively \cite{nrgr}. 
The long-distance effective theory is obtained by `integrating out' the (off-shell) potential modes in a (classical) saddle-point approximation, and matching into a `long-distance' worldline theory, describing now the entire binary and depending only on the (on-shell) radiation field. The effective action takes the generic~form~\cite{andirad,andirad2}\footnote{Due the fact that both potential and radiation modes vary on the same time scales, the decoupling of short-distance effects occurs in space, but not in time. As as a result, the action is endowed with time-dependent `Wilson coefficients'. The same type of description can also be used to study absorption effects, e.g. \cite{dis1,dis2,dis3}.}
\bea
\label{Seff}
S_{\rm rad} &=& \int d \sigma \sqrt{(\eta_{\mu\nu} + \bar h _{\mu\nu})V^\mu V^\nu} \biggl[ -P^\mu(\sigma) V^\nu (\sigma) (1 + \bX^i \nabla_i)\bar h_{\mu\nu}(\sigma,\bX(\sigma)) - \frac{1}{2} \bar\omega^{\alpha\beta}_\mu (\sigma,\bX) J_{\alpha\beta}(\sigma) V^\mu (\sigma) \nonumber\\ &&+  
\sum_{\ell=2} \biggl( \frac{1}{\ell!} I^L_{\rm STF} (\sigma) \nabla_{L-2} \bar E_{i_{\ell-1}i_\ell}(\sigma,\bX(\sigma)) - \frac{2\ell}{(2\ell+1)!}J_{\rm STF} ^L(\sigma) \nabla_{L-2} \bar B_{i_{\ell-1}i_\ell}(\sigma,\bX(\sigma))\biggr)\biggr]\, ,
\eea 
around the CoM, $\bX(\sigma)$, of the binary, with $\sigma$ an affine parameter. Throughout this paper we use the notation $L \equiv {i_1\cdots i_L}$. 
 All the barred quantities are evaluated on~$\bar h_{\mu\nu}$. The first terms represent the total four-momentum, $P^\mu$, and angular momentum, $J^{\alpha\beta}(t)$, of the binary, with $V^\mu$ its CoM velocity. The higher-order multipole moments have electric-type, $I^{L}(t)$, and magnetic-type, $J^{L}(t)$, parity, and couple to derivatives of the electric and magnetic components of the Weyl tensor, respectively. We have also kept the term proportional to $\bX^i$ at linear order in derivatives, which will be useful later on to compute the position of the CoM. For most of the calculations, however, it is sufficient to have the binary at the origin, i.e. $\bX^i=0$. 

\subsection{Gravitational potential}
\begin{figure}[ht]
     \centering
         \includegraphics[width=8cm,height=3cm,keepaspectratio]{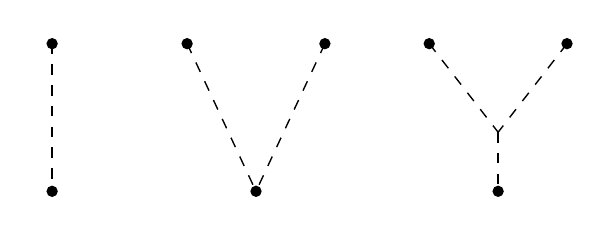}
         \caption{Topologies needed to match the gravitational potential to NLO (see text).}
         \label{fig1}
\end{figure}
The gravitational potential (along with the kinetic term) may be obtained by matching the $00$-component of the one-point function. Alternatively, it can be read off from the `vacuum binding energy,' where the outgoing radiation field is zero to zero.\footnote{In principle, radiation-reaction effects also contribute to the conservative binding energy through tail effects, e.g.~\cite{tail1,nltail}. However, to the order we work in this paper, we can safely ignore these type contributions.} Following the latter option, the relevant topologies are shown in Fig.~\ref{fig1} to NLO, where the dashed line represents the potential mode, whose propagator is expanded as $(k^0 \ll |\bk|)$
\beq
\frac{i}{(k^0)^2-\bk^2} = -\frac{i}{\bk^2}\biggl(1+ \frac{(k^0)^2}{\bk^2}+\cdots\biggr)\,,\label{expot}
\eeq
and subsequently truncated to the desired PN order. A similar power-counting also applies to the nonlinear couplings involving time derivatives of the $H_{\mu\nu}$ field, scaling as $\partial_0 H_{\mu\nu} \sim (v/r) H_{\mu\nu}$. The Feynman rules for the worldline vertices follow from the action/Routhian in \eqref{rout}, and include mass, spin, as well as finite-size and SSC-preserving contributions, which are likewise also PN expanded in powers of the velocity. Once the potential is known, the EoM follow from \eqref{eom}, which can afterwards be reduced to vectorial form using the SSC in \eqref{ssc2}. In principle, the gravitational potential also depends on time derivatives of the position and spin variables. As it is standard, these are either integrated by parts or reduced using lower-order EoM.\vskip 4pt

The procedure described above was carried out in \cite{nrgrs,prl,nrgrso,nrgrss,nrgrs2} to NLO in the PN expansion and to quadratic order in the spins. In Appendix~\ref{app:A} we provide expressions for the acceleration and spin EoM in the CoM frame to 3PN order. 


\subsection{Multipole moments}

\begin{figure}[ht]
     \centering
         \includegraphics[width=9cm,height=3.2cm,keepaspectratio]{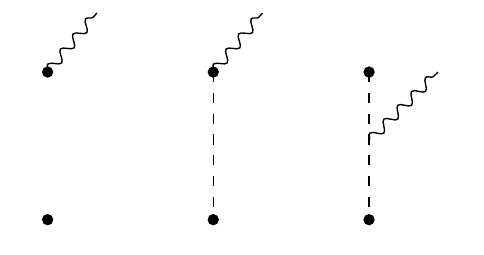}
         \caption{Topologies needed to match the one-point function to NLO (see text).}
         \label{fig2}
\end{figure}

It is convenient, in order to compute the multipole moments, to first obtain their generic dependence on moments of the (pseudo-)stress-energy tensor, which 
includes contributions from potential modes, and subsequently read off the latter via a matching computation with the one-point function (see \cite{andirad,andirad2} for details on this procedure). The necessary topologies to NLO are shown in Fig.~\ref{fig2} where, once again, the Feynman rules for the vertices follow from~\eqref{rout}. In addition to the potential modes, which are integrated out, the wavy line represents the radiation field that propagates to infinity.\footnote{Notice that the leading contribution is simply from the mass coupling to the radiation field in \eqref{rout}. This means that the NLO correction entails only `tree-level' potential exchanges, unlike the derivation of the gravitational potential in Fig.~\ref{fig1} which requires `one-loop' diagrams.} Importantly,  for the (momentum-conserving) interaction in the nonlinear cubic coupling, between an incoming potential mode with three-momentum $\bq$ and an outgoing radiation field with three-momentum $\bk$, in addition to the expansion of potentials in quasi-instantaneous interaction we alluded before in \eqref{expot} we must also expand the outgoing potential mode~as~\cite{nrgr}
\beq
\frac{i}{(\bk+\bq)^2} = \frac{i}{\bq^2}\biggl(1 - 2\frac{\bq\cdot\bk}{\bq^2} + \cdots\biggr)\,.
\eeq
At the level of the action, the above is related to a Taylor expansion in (spatial) derivatives of the radiation field around the CoM of the binary (placed at the origin) \cite{nrgr,andirad}
\beq
\bar h_{\mu\nu}(t,\bx) = \bar h_{\mu\nu}(t,\boldsymbol 0) + \partial_i \bar h_{\mu\nu}(t,\boldsymbol 0) \bx^i + \cdots \,.
\eeq
The procedure described here was carried out in \cite{srad,amps} to NLO in the PN expansion, and to linear and bilinear order in the spins. Together with the gravitational potential, the multipole moments are the last ingredient to compute all the GW fluxes. For completeness, we summarize  in Appendix \ref{app:A} the resulting multipoles in the CoM frame.

\section{Adiabatic invariants}\label{invariants}

Using the results for the gravitational potential it is straightforward to compute the conserved quantities of the system, which do not evolve in time when radiation fluxes are turned off. These are the objects, such as the binding energy and total angular momentum, that will be part of the balance equations in an adiabatic expansion, valid during the insipiral regime. It is somewhat convenient to express the values for these quantities in the CoM frame. This is achieved by first computing the binary's CoM, the same way we extract the multipoles, and then solving for the coordinates using the condition $\bX^i=0$. Using the notation $\br \equiv \bx_1-\bx_2$ and $\bv \equiv \dot \br$, we have solutions which can be PN expanded in the form
\begin{subequations}\label{eq:com}
\begin{align}
\label{xcom}
\bx_1^i&=\frac{m_2}{m}\,\br^i+\delta \br^i_\text{1PN}+\delta \br^i_\text{LO-SO}+\delta \br^i_\text{LO-SS}+\cdots\,,\\[1.5ex]
\bv_1^i&=\frac{m_2}{m}\,\bv^i+\delta \dot{\br}^i_\text{1PN}+\delta\dot{\br}^i_\text{LO-SO}+\delta\dot{\br}^i_\text{LO-SS}+\cdots\,,\end{align}
\end{subequations}
with $m\equiv m_1+m_2$, and similarly for the companion. To the order we work in this paper we find the known values (e.g.~\cite{kidder,buo1}),
\begin{subequations}
\begin{align}
\delta \br^i_\text{1PN}& =\frac{\nu \delta}{2}\,\br^i\Big(\,\bv^2-\frac{G\,m}{r}\,\Big)\,,\\[1ex]
\delta \br^i_\text{LO-SO} &=\frac\nu m(\bv\times\bSigma)^i\,,\\[1ex]
\delta \br^i_\text{LO-SS}&=0\,,
\end{align}
with $\nu\equiv \frac{m_1\,m_2}{m^2}$,  $\delta\equiv \frac{m_1-m_2}{m}$, and
\beq
\bSigma \equiv m \Big(\frac{\bS_2}{m_2}-\frac{\bS_1}{m_1}\Big)\,,\label{sigma}
\eeq
while for the velocities we have
\begin{align}
\delta \bv^i_\text{1PN}& = \frac{\delta\,\nu}{2}\bv^i\biggl(
\bv^2- \frac{G \,m }{r} 
\biggr) -  \frac{G \,m\,\delta\,\nu }{2\, r}\,\dot r \,\bn^i\,,\\[1ex]
\delta \bv^i_\text{LO-SO} &=-\frac{G\,\,\nu}{r^2}\,(\bn\times\bSigma)^i\,,
\end{align}
with $\bn \equiv \br/r$. The latter expressions follow by taking time derivatives of the position and using lower-order EoM.\vskip 4pt  For the remainder of the paper we will often quote the final results in terms of \eqref{sigma} and the total spin
\beq
\bS \equiv \bS_1+\bS_2\,. 
\eeq
\end{subequations}
For completeness, we also provide expressions for the 3PN correction at quadratic order in the spins in section \ref{secCoM}, which will be useful to express conserved quantities in the CoM frame at higher orders. The NLO spin-orbit correction was computed in \cite{pardo}.

\subsection{Binding energy}

From the gravitational potential we can derive the binding energy following the standard Euler-Lagrange procedure, yielding 
\begin{align}
E=E_\text{N}+E_\text{1PN}+E_\text{LO-SO}+\Big[\,E_\text{2PN}+E_\text{LO-SS}\,\Big]+E_\text{NLO-SO}+\Big[\,E_\text{3PN}+E_\text{NLO-SS}\Big]+\cdots\,
\end{align}
in a PN expansion. The computation is often  lengthy but straightforward. Using the potential in \cite{nrgrso}, the spin-orbit binding energy in the CoM at NLO was computed in \cite{pardo}, confirming the old result in \cite{buo1}. We compute here, using the results first obtained in \cite{prl,nrgrss,nrgrs2}, the contribution to the CoM binding energy to 3PN and quadratic order in the spins. We find 
\begin{align}
E_\text{LO-SS}=\frac{G \,\nu}{r^3}\,\frac{1}{4}\,e^0_4\,,
\end{align}
\begin{align}\label{3pnss}
E_\text{NLO-SS}=\frac{G \,\nu}{r^3}\,\Biggl[\,\frac{1}{8}\,e^0_6+\frac{G\,m}{r}\,\frac{1}{4}\,e^1_6\,\Biggr]\,,
\end{align}
where the $e^0_4,e^0_6, e^1_6$ coefficients are displayed in Appendix~\ref{app:B}. After rewriting the answer in terms of the conserved-norm spin variable,\footnote{We only quote the terms which are needed to match the multipoles. The reader should keep in mind, however, that higher-order corrections are necessary to achieve a precession form for the spin EoM; see, e.g.~\cite{nrgrso,nrgrss,nrgrs2}.}
\beq
\bS_{c} = \left(1-\frac{\bv^2}{2}\right) \bS  + \frac{1}{2} (\bv\cdot \bS) \bv + \cdots  \label{spinc} \,,
\eeq
the NLO result in \eqref{3pnss} is completely equivalent to the derivation in \cite{Bohes2}. This is expected, since the results in \cite{Bohes2} for the binding energy are obtained after confirming the equivalence with the value of the gravitational potential first computed in \cite{nrgrss,nrgrs2}).

\subsubsection{Nonprecessing (quasi-)circular orbits}

Many of the above expressions drastically simplify for the case of circular orbits, which we expect to provide a large fraction of the relevant sources of GWs once the binary enters the frequency band of present and future ground-based detectors. We proceed using the basis of vectors
$\{\,\bn,\,\blambda,\,\ble \,\}$,  defined by 
\begin{subequations}
\begin{align}
\ble &\equiv \frac{\br\times\bv}{|\br\times\bv|}\,,\\
\blambda&\equiv \ble\times\bn\,,
\end{align}
\end{subequations}
and $\bn$ the radial unit vector. Following the nonspinning case, we seek (quasi-)circular orbits obeying
\begin{subequations}
\begin{align} 
\label{dndt}
\frac{d\bn}{dt} &=\omega(t)\,\blambda\,, \\ 
\frac{d \blambda}{dt}&=-\omega(t) \,\bn\,,\\
\frac{d\ble}{dt}&=0\,.
\end{align}
\end{subequations}
The reader will immediately notice that these conditions cannot be fulfilled when spin effects are included, since only the total angular momentum is conserved.\footnote{When only spin-orbit corrections are included at leading order, orbits with $\dot r=0$ are still possible but not restricted to the plane.}  Hence, we must enforce additional constraints in order to find (quasi-)circular orbits. We use the `aligned-spin' simplifications
\begin{subequations}\label{eq:nonprespins}
\begin{align}
\bn\cdot\mathbf{S}_A=0\,,\\
\blambda\cdot\mathbf{S}_A=0\,,\\
\mathbf{S}_1\times\mathbf{S}_{2}=0\,,
\end{align}
\end{subequations}
to ensure the orbit is confined to the plane. In order to guarantee these are valid throughout the entire evolution of the binary, their time derivatives must also be consistent with the EoM to the desired PN order. The above conditions then imply that the spin vectors must be aligned with the orbital angular momentum, 
\beq 
\bS_A \equiv S_A \ble,
\eeq 
and remain constant in time ($\dot S_{1,2}=0$).\footnote{Notice that the covariant spin vector varies with time in generic orbits. However, since $\dot \bv^2=0$ for (quasi-)circular orbits when radiation-damping is omitted, the difference between $\bS$ and $\bS_c$ in \eqref{spinc} is simply an overall rescaling. Therefore, both vectors remain constant in time when initially aligned with the orbital angular momentum.} Moreover, because the radius is also constant ($\dot r=0$), there must be a direct relationship between the latter and the orbital frequency, which also obeys $\dot\omega=0$ as long as fluxes are ignored. Introducing the PN parameter $x \equiv (Gm\omega)^{2/3}$, we find
\begin{align}
\Big(\frac{G\,m}{r}\Big)_\text{N$+$1PN$+$LO-SO}=x+x^2\,\Big(\,1-\frac{\nu}{3}\,\Big)+\frac{x^{5/2}}{G\,m^2}\,\Big(\,\frac{5}{3}\,S+\delta\,\Sigma\,\Big)\,,
\end{align}
and for the spin-spin contributions we have
\begin{align}\label{eq:GMoverr}
\begin{autobreak}
\Big(\frac{G\,m}{r}\Big)_\text{SS}
=\frac{\,x^3}{4\, G^2 \,m^4}\,\bigg[\,(-4 - 2 \,\kappa_{+})\, \bS^2 + \,(-4 \,\delta + 2 \,\kappa_{-} - 2 \,\delta \,\kappa_{+}) \,(S \Sigma) + \,\big(\,\delta \,\kappa_{-} -  \,\kappa_{+} + \,(4 + 2 \,\kappa_{+}) \,\nu\big) \,\bSigma^2\,\bigg]
+\frac{\,x^4}{36\, G^2 \,m^4}\,\bigg[\,\big(74 - 42 \,\delta \,\kappa_{-} - 24 \,\kappa_{+} + \,(-48 - 24 \,\kappa_{+}) \,\nu\big)\, \bS^2
+ \,\big(78 \,\delta - 18 \,\kappa_{-} + 18 \,\delta \,\kappa_{+} + \,(-48 \,\delta + 192 \,\kappa_{-} - 24 \,\delta \,\kappa_{+}) \,\nu\big) (S \Sigma)
+ \,\big(36 - 9 \,\delta \,\kappa_{-} + 9 \,\kappa_{+} + \,(-90 + 54 \,\delta \,\kappa_{-} - 72 \,\kappa_{+}) \,\nu + \,(48 + 24 \,\kappa_{+}) \,\nu^2\big)\, \bSigma^2\bigg]\,,
\end{autobreak}
\end{align}
where we use the notation $(A B) \equiv \bA\cdot \bB$ (for $\bA\neq \bB$) adopted in, e.g.~\cite{Bohes2}. We also introduced the finite-size parameters $\kappa_\pm \equiv C_{ES^2}^{(1)} \pm   C_{ES^2}^{(2)} $. Replacing the expression \eqref{eq:GMoverr} into \eqref{3pnss}, we finally get the binding energy for (quasi-)circular orbits as a function of the orbital frequency,  
\begin{align}
\begin{autobreak}
(E)_\text{SS}
=\frac{\,x^3 \,\nu}{4 \,G^2 \,m^3}\,\Big(\,(4 + 2 \,\kappa_{+}) \,\bS^2 + \,(4 \,\delta - 2 \,\kappa_{-} + 2 \,\delta \,\kappa_{+}) \,(S\Sigma) + \,\big(- \,\delta \,\kappa_{-} + \,\kappa_{+} + \,(-4 - 2 \,\kappa_{+}) \,\nu\big)\Big) \,\bSigma^2
+\frac{\,x^4 \,\nu}{72 \,G^2 \,m^3}\,\Big(\,\big(-236 + 78 \,\delta \,\kappa_{-} + 132 \,\kappa_{+} + \,(12 + 6 \,\kappa_{+}) \,\nu\big) \,\bS^2
+ \,\big(-336 \,\delta - 54 \,\kappa_{-} + 54 \,\delta \,\kappa_{+} + \,(12 \,\delta - 318 \,\kappa_{-} + 6 \,\delta \,\kappa_{+}) \,\nu\big) \,(S\Sigma)
+ \,\big(-180 - 27 \,\delta \,\kappa_{-} + 27 \,\kappa_{+} + \,(396 - 81 \,\delta \,\kappa_{-} + 27 \,\kappa_{+}) \,\nu + \,(-12 - 6 \,\kappa_{+}) \,\nu^2\big) \,\bSigma^2\Big)\,.
\end{autobreak}
\end{align}
We can also transform the above expression as  function of the conserved-norm spin vector.  Using \eqref{spinc} we have,
\begin{align}
\big(E\big)_\text{SS}&=\frac{x^3\,\nu}{4\,G^2 \,m^3}\,\Bigg[\,(4 + 2 \,\kappa_{+}) \,\bS_\text{c}^2 + \,(4 \,\delta - 2 \,\kappa_{-} + 2 \,\delta \,\kappa_{+}) \,(S_\text{c} \,\Sigma_\text{c} )+ \,\big(- \,\delta \,\kappa_{-} + \,\kappa_{+} + \,(-4 - 2 \,\kappa_{+}) \,\nu\big) \,\bSigma_\text{c}^2\Bigg]\notag\\
&+\frac{x^4\,\nu}{72\,G^2 \,m^3}\,\Bigg[\,\big(-200 + 60 \,\delta \,\kappa_{-} + 150 \,\kappa_{+} + \,(-60 - 30 \,\kappa_{+}) \,\nu\big) \,\bS_\text{c}^2 \notag\\
&+ \,\big(-300 \,\delta - 90 \,\kappa_{-} + 90 \,\delta \,\kappa_{+} + \,(-60 \,\delta - 210 \,\kappa_{-} - 30 \,\delta \,\kappa_{+}) \,\nu\big) \,(S_\text{c} \,\Sigma_\text{c}) \notag\\
&+ \,\big(-180 - 45 \,\delta \,\kappa_{-} + 45 \,\kappa_{+} + \,(360 - 45 \,\delta \,\kappa_{-} - 45 \,\kappa_{+}) \,\nu + \,(60 + 30 \,\kappa_{+}) \,\nu^2\big) \,\bSigma_\text{c}^2\Bigg]\,,
\end{align}
which agrees with the result in \cite{Bohes2}. 

\subsection{Orbital angular momentum}\label{sec:orb}

The orbital angular momentum can be obtained either by matching the one-point function to the long-distance action in \eqref{Seff}, or by obtaining the Noether current using the gravitational potential. Following the latter, we arrive at an expression that can likewise be PN expanded as
\begin{align}
\bL^i=\bL^i_\text{N}+\bL^i_\text{1PN}+L^i_\text{LO-SO}+\Big[\,\bL^i_\text{2PN}+\bL^i_\text{LO-SS}\,\Big]+\bL^i_\text{NLO-SO}+\Big[\,\bL^i_\text{3PN}+\bL^i_\text{NLO-SS}\Big]+\cdots\,,
\end{align}
where
\begin{subequations}
\begin{align}
\begin{autobreak}
\bL^i_\text{N}
=\nu\,m\,r\,(\bn\times\bv)^{i}\,,
\end{autobreak}
\\[2ex]
\begin{autobreak}
\bL^i_\text{1PN}
=\nu\,m\,r\,(\bn\times\bv)^{i} \,\biggl(\frac{G \,m \,(3 + \,\nu)}{r} + \frac{\,(1 - 3 \,\nu ) \,\bv^2}{2}\biggr)\,,
\end{autobreak}
\\[2ex]
\begin{autobreak}
\bL^i_\text{LO-SO}
=\frac{G\,m\,\nu}{r}\Bigl(\,\bn^i \,\bigl(3 \,(nS) + \,\delta \,(n\Sigma)\bigr) - 3 \,\bS^i -  \,\delta \,\bSigma^i\,\Bigr)\,,
\end{autobreak}
\\[2ex]
\begin{autobreak}
\bL^i_\text{LO-SS}
=0\,.
\end{autobreak}
\end{align}
\end{subequations}
At 2.5PN and 3PN order, respectively, we have
\begin{eqnarray}
\bL^i_\text{NLO-SO}
&=&\frac{G\,m\,\nu}{2\,r}\,\biggl[\,\ell^0_5+\frac{G\,m}{r}\,\ell^1_5\,\biggr]\,,  \label{3pnLss1}
\\
\bL^i_\text{NLO-SS}
&=& \frac{G\,m\,\nu}{2\,r^2}\,\ell^0_6\,.  \label{3pnLss2}
\end{eqnarray}
The lengthy expression for the $\ell^{0}_5$, $\ell^{1}_5$ and $\ell^0_6$ coefficients can be found in Appendix~\ref{app:B}. Notice that in the orbital angular momentum, there is no spin-spin contribution at 2PN order. The 3PN expression for the spin-spin orbital angular momentum is presented here for the first time.  We have explicitly confirmed that the evolution of the total angular momentum obeys
\begin{align}
\frac{d}{dt}(\bL^i+\bS^i)=0\,,
\end{align}
using the EoM deriving from the gravitational potential to 3PN order \cite{nrgrs,prl,nrgrso,nrgrss,nrgrs2}, which provides a nontrivial check of our result.

\subsection{Center-of-mass position}\label{secCoM}

Finally, we also compute the correction to the CoM position. This is achieved, as we discussed in section \ref{invariants}, by matching the one-point function to the effective action in \eqref{Seff} and reading off the position of the CoM, $\bX^i$, by expanding in the soft-radiation mode at linear order $\bk$. 
We arrive at 
\begin{align}
   (\bX^i)_\mathrm{LO-SS}=0,\\
\begin{autobreak}
  m\, (\bX^i)_\mathrm{NLO-SS}=
    -\frac{G\nu}{8r^{2}}\biggl\{ \bS^i\bigl[-12\kappa_{-}(nS)+6(-2-\delta\kappa_{-}+\kappa_{+})(n\Sigma)\bigr]
    +\bSigma^i\bigl[6\bigl(2-\delta\kappa_{-}+\kappa_{+}\bigr)(nS)+6\bigl(\kappa_{-}(-1+2\nu)+\delta\kappa_{+}\bigr)(n\Sigma)\bigr]
    +\bn^i\bigl[\bigl(\kappa_{-}-4\delta\nu+\delta\kappa_{+}(-1-2\nu)\bigr)\bSigma^{2}+\bigl((4-16\nu)+2\delta\kappa_{-}+2\kappa_{+}(-1-4\nu)\bigr)(S\Sigma)
    +\bigl(4\delta+4\kappa_{-}+2\delta\kappa_{+}\bigr)\bS^{2}+\bigl(-12\delta-6\delta\kappa_{+}\bigr)(nS)^{2}
    +\bigl(12(-1+4\nu)+6\delta\kappa_{-}+6\kappa_{+}(-1+4\nu)\bigr)(nS)(n\Sigma)
    +\bigl(12\delta\nu+3\kappa_{-}(1-4\nu)+3(-1+2\nu)\delta\kappa_{+}\bigr)(n\Sigma)^{2}\bigr]\biggr\}\,,
    \end{autobreak}
\end{align}
from which we find the correction to \eqref{xcom} given by
\begin{align}
\begin{autobreak}
(\delta \br^i)_\text{NLO-SS}
=\bS^i \,\big(12 \,\kappa_{-} \,(nS) + \,(12 + 6 \,\delta \,\kappa_{-} - 6 \,\kappa_{+}) \,(n\Sigma)\big) 
+ \bSigma^i \,\Big[\,(-4 + 6 \,\delta \,\kappa_{-} - 6 \,\kappa_{+}) \,(nS) + \big(-6 \,\delta \,\kappa_{+} + \,\kappa_{-} \,(6 - 12 \,\nu)\big) \,(n\Sigma)\Big] 
+ \,\bn^i\, \Big[\,(12 \,\delta + 6 \,\delta \,\kappa_{+}) \,(nS)^2 + \,(12 \,\delta^2 - 6 \,\delta \,\kappa_{-} + 6 \,\delta^2 \,\kappa_{+}) \,(nS) \,(n\Sigma) 
+ \big(-3 \,\delta^2 \,\kappa_{-} + \,\delta \,\kappa_{+} \,(3 - 6 \,\nu) - 12 \,\delta \,\nu\big) \,(n\Sigma)^2 + \,(-4 \,\delta - 4 \,\kappa_{-} - 2 \,\delta \,\kappa_{+}) \,\bS^2
+ \big(-2 \,\delta \,\kappa_{-} + 4 \,(-3 + 4 \,\nu) + \,\kappa_{+} \,(2 + 8 \,\nu)\big) \,(S\Sigma) 
+ \big(-\kappa_{-} + 4 \,\delta \,\nu + \,\kappa_{+} \,(\,\delta + 2 \,\delta \,\nu)\big)\,\bSigma^2\Big]\,.
\end{autobreak}
\end{align}

\section{Flux-balance laws}\label{laws}

In this section we use the source multipole moments from \cite{srad} together with the EoM in \cite{nrgrs,prl,nrgrso,nrgrss,nrgrs2} to compute the flux-balance equations for the binding energy, linear momentum, and angular momentum to 3PN order. We do not include the contribution from the (spin-orbit) tail term, which corrects the radiative multipole moments~\cite{srad,amps}, whose effect can be found in, e.g.~\cite{tailluc}. The detailed expressions for generic orbits are lengthy and relegated to Appendix~\ref{app:C}. 
Explicit results are given in the main text for the case of nonprecessing (quasi-)circular orbits.

\subsection{Binding energy}
For the computation of the energy flux we use the well-known formula, e.g.,~\cite{andirad}
\begin{align}
\frac{dE}{dt}=-\frac{G}{5}\,\biggl(I^{(3)}_{ij}\,I^{(3)}_{ij}+\frac{16}{9}\,J^{(3)}_{ij}\,J^{(3)}_{ij}+\frac{5}{189}I^{(4)}_{ijk}I^{(4)}_{ijk}+\frac{5}{84}J^{(4)}_{ijk}\,J^{(4)}_{ijk}+\cdots\biggr)\,,
\end{align}
in terms of the source multipole moments. The explicit expression for the latter are given in Appendix~\ref{app:A} in the CoM frame. The time derivatives are order-reduced using the EoM, also found in Appendix~\ref{app:A}. The result can be expanded in the PN expansion,
\begin{align}
\begin{autobreak}
\frac{dE}{dt}
=\biggl(\frac{dE}{dt}\biggr)_\text{N}+\biggl(\frac{dE}{dt}\biggr)_\text{1PN}+\biggl(\frac{dE}{dt}\biggr)_\text{LO-SO}+\biggl[\,\biggl(\frac{dE}{dt}\biggr)_\text{2PN}+\biggl(\frac{dE}{dt}\biggr)_\text{LO-SS}\,\biggr]
+\biggl(\frac{dE}{dt}\biggr)_\text{NLO-SO}+\biggl[\,\biggl(\frac{dE}{dt}\biggr)_\text{3PN}+\biggl(\frac{dE}{dt}\biggr)_\text{NLO-SS}\biggr]+\cdots\,,
\end{autobreak}
\end{align}
and keeping only spin-spin contributions to the leading order, we find
\begin{align}\label{eq:energyflux2PN}
\biggl(\frac{dE}{dt}\biggr)_\text{LO-SS}
&=\frac{G^3\,m^2\,\nu^2}{105\,r^6}\,\bigg[\,(696 + 348 \,\kappa_{+}) \,(nS) \,(nv) \,(vS) \notag\\
&+ \,(348 \,\delta - 174 \,\kappa_{-} + 174 \,\delta \,\kappa_{+}) \,(nv) \,(n\Sigma) \,(vS) + \,(-144 - 72 \,\kappa_{+}) \,(vS)^2\notag\\
&+ \,\bS^2 \,\big(\,(312 + 156 \,\kappa_{+}) \,(nv)^2 + \,(-288 - 144 \,\kappa_{+}) \,\bv^2\big)\notag\\
&+ \,(nS)^2 \,\big(\,(-1632 - 816 \,\kappa_{+}) \,(nv)^2 + \,(1008 + 504 \,\kappa_{+}) \,\bv^2\big)\notag\\
&+ \,(S\Sigma) \,\big(\,(312 \,\delta - 156 \,\kappa_{-} + 156 \,\delta \,\kappa_{+}) \,(nv)^2 + \,(-288 \,\delta + 144 \,\kappa_{-} - 144 \,\delta \,\kappa_{+}) \,\bv^2\big)\notag\\
&+ \,(nS) \,(n\Sigma) \,\big(\,(-1632 \,\delta + 816 \,\kappa_{-} - 816 \,\delta \,\kappa_{+}) \,(nv)^2 
+ \,(1008 \,\delta - 504 \,\kappa_{-} + 504 \,\delta \,\kappa_{+}) \,\bv^2\big)\notag\\
& + \,(n\Sigma)^2 \,\Big[\,\big(-9 + 408 \,\delta \,\kappa_{-} + 1632 \,\nu
+ 408 \,\kappa_{+} \,(-1 + 2 \,\nu)\big) \,(nv)^2 \notag\\
&+ \,\big(-252 \,\delta \,\kappa_{-} + \,\kappa_{+} \,(252 - 504 \,\nu)- 1008 \,\nu\big) \,\bv^2\Big]\notag\\
& + \,(348 \,\delta - 174 \,\kappa_{-} + 174 \,\delta \,\kappa_{+}) \,(nS) \,(nv) \,(v\Sigma)+ \,\big(6 - 174 \,\delta \,\kappa_{-} + \,\kappa_{+} \,(174 - 348 \,\nu) - 696 \,\nu\big) \,(nv) \,(n\Sigma) \,(v\Sigma)\notag\\
&+ \,(-144 \,\delta + 72 \,\kappa_{-} - 72 \,\delta \,\kappa_{+}) \,(vS) \,(v\Sigma) + \,\big(-1 + 36 \,\delta \,\kappa_{-} 
+ 144 \,\nu + \,\kappa_{+} \,(-36 + 72 \,\nu)\big) \,(v\Sigma)^2\notag\\
& + \,\Big[\,\big(-9 - 78 \,\delta \,\kappa_{-} + \,\kappa_{+} \,(78 - 156 \,\nu)
- 312 \,\nu\big) \,(nv)^2 \notag\\
&+ \,\big(-3 + 72 \,\delta \,\kappa_{-} + 288 \,\nu + 72 \,\kappa_{+} \,(-1 + 2 \,\nu)\big) \,\bv^2\Big] \,\bSigma^2\,\bigg]\,.
\end{align}
At 3PN order, we have
\begin{align}\label{eq:energyflux3PN}
\bigg(\frac{dE}{dt}\bigg)_\text{NLO-SS}=\frac{G^3\,m^2\,\nu^2}{105\,r^6}\,\bigg[\,f_6^0+\frac{G\,m}{r}\,f_6^1+\frac{G^2\,m^2}{r^2}\,f_6^2\,\bigg]\,,
\end{align}
with the value of the $f_6^0, f_6^1$ and $f_6^2$ coefficients given in Appendix~\ref{app:C}.
After using the transformation in \eqref{spinc}, we have checked that the above expression is equivalent to the one obtained in \cite{Bohes2}.

\subsubsection{Nonprecessing (quasi-)circular orbits}

We consider here the case of aligned-spin (quasi-)circular orbits. Since radiation-reaction effects first enter at 2.5PN order beyond the leading effects, namely at 4PN and 4.5PN for spin-orbit and spin-spin contributions, respectively, we can safely ignore them here while working to 3PN order.  Hence, we can consistently replace $\dot{r}=0$ and $v^2=r^2\,\omega^2$ in \eqref{eq:energyflux2PN} and \eqref{eq:energyflux3PN} to obtain the emitted radiation. However, we must still face a radiation-reaction force which may induce the precession of the orbital plane. Even though the precession effects might be small, $d\ble/dt\leq\mathcal{O}(v^8)$, these may accumulate over time. Nonadiabatic methods, such as dynamical renormalization group, offer an alternative approach that remains valid over a longer time-scale by resumming secular effects \cite{chadDRG,zixinDRG}. In this work, however, we restrict our results to time scales {\it shorter} than those induced by the secular evolution of the orbital plane.  We find
\begin{align}
\begin{autobreak}
\bigg(\frac{dE}{dt}\bigg)_{\rm SS}
=-\frac{32\,x^7 \,\nu^2}{5 G^3 \,m^4}\,\Big[\,(4 + 2 \,\kappa_{+}) \,\bS^2 + \,(4 \,\delta - 2 \,\kappa_{-} + 2 \,\delta \,\kappa_{+}) \,(S\Sigma)
+ \,\big(\tfrac{1}{16} -  \,\delta \,\kappa_{-} + \,\kappa_{+} + \,(-4 - 2 \,\kappa_{+}) \,\nu\big) \,\bSigma^2\Big]
-\frac{32\,x^8 \,\nu^2}{\,5 G^3 \,m^4}\,\Big[\,\big(- \tfrac{6247}{504} + \tfrac{57}{16} \,\delta \,\kappa_{-} -  \tfrac{383}{112} \,\kappa_{+} + \,(- \tfrac{35}{2} -  \tfrac{35}{4} \,\kappa_{+}) \,\nu\big) \,\bS^2 
+ \,\big(- \tfrac{1865}{112} \,\delta + \tfrac{391}{56} \,\kappa_{-} -  \tfrac{391}{56} \,\delta \,\kappa_{+} + \,(- \tfrac{35}{2} \,\delta -  \tfrac{11}{2} \,\kappa_{-} -  \tfrac{35}{4} \,\delta \,\kappa_{+}) \,\nu\big) \,(S\Sigma) 
+ \,\big(- \tfrac{51}{16} + \tfrac{391}{112} \,\delta \,\kappa_{-} -  \tfrac{391}{112} \,\kappa_{+} + \,(\tfrac{6239}{336} + \tfrac{13}{16} \,\delta \,\kappa_{-} + \tfrac{691}{112} \,\kappa_{+}) \,\nu + \,(\tfrac{35}{2} + \tfrac{35}{4} \,\kappa_{+}) \,\nu^2\big) \,\bSigma^2\Big]\,,
\end{autobreak}
\end{align}
or in terms of the conserved-norm spin vectors,
\begin{align}
\bigg(\frac{dE}{dt}\bigg)_\text{SS}&=- \frac{32 \,x^7 \,\nu^2}{5 \,G^3 \,m^4} \,\Big[\,(4 + 2 \,\kappa_{+}) \,\bS_\text{c}^2 + \,(4 \,\delta - 2 \,\kappa_{-} + 2 \,\delta \,\kappa_{+}) \,(S_\text{c}\Sigma_\text{c}) + \,\big(\tfrac{1}{16} -  \,\delta \,\kappa_{-} + \,\kappa_{+} + \,(-4 - 2 \,\kappa_{+}) \,\nu\big) \,\bSigma_\text{c}^2\Big]\notag\\
&- \frac{32 \,x^8 \,\nu^2}{5 \,G^3 \,m^4} \,\Big[\,\big(- \tfrac{5239}{504} + \tfrac{41}{16} \,\delta \,\kappa_{-} -  \tfrac{271}{112} \,\kappa_{+} + \,(- \tfrac{43}{2} -  \tfrac{43}{4} \,\kappa_{+}) \,\nu\big) \,\bS_\text{c}^2\notag\\
& + \,\big(- \tfrac{817}{56} \,\delta + \tfrac{279}{56} \,\kappa_{-} -  \tfrac{279}{56} \,\delta \,\kappa_{+} + \,(- \tfrac{43}{2} \,\delta + \tfrac{1}{2} \,\kappa_{-} -  \tfrac{43}{4} \,\delta \,\kappa_{+}) \,\nu\big) \,(S_\text{c} \Sigma_\text{c} ) \notag\\
&+ \,\big(- \tfrac{25}{8} + \tfrac{279}{112} \,\delta \,\kappa_{-} -  \tfrac{279}{112} \,\kappa_{+} + \,(\tfrac{344}{21} + \tfrac{45}{16} \,\delta \,\kappa_{-} + \tfrac{243}{112} \,\kappa_{+}) \,\nu + \,(\tfrac{43}{2} + \tfrac{43}{4} \,\kappa_{+}) \,\nu^2\big) \,\bSigma_\text{c}^2\Big]\,.
\end{align}

\subsection{Angular momentum}
The source multipoles allow us also to compute the radiated angular momentum via 
\begin{align}
\frac{d\bJ^i}{dt}=-G\,\epsilon^{iab}\bigg(\frac{2}{5}\,I_{aj}^{(2)}\,I_{bj}^{(3)}+\frac{32}{45}\,J_{aj}^{(2)}\,J_{bj}^{(3)}+\frac{1}{63}\,I_{ajk}^{(3)}\,I_{bjk}^{(4)}+\frac{1}{28}\,J_{ajk}^{(3)}\,J_{bjk}^{(4)}+\cdots\bigg)\,,
\end{align}
which may be also expanded into PN contributions as
\begin{align}
\begin{autobreak}
\frac{d\bJ^{\,i}}{dt}
=\bigg(\frac{d\bJ^{\,i}}{dt}\bigg)_\text{N}+\bigg(\frac{d\bJ^{\,i}}{dt}\bigg)_\text{1PN}+\bigg(\frac{d\bJ^{\,i}}{dt}\bigg)_\text{LO-SO}+\bigg[\,\bigg(\frac{d\bJ^{\,i}}{dt}\bigg)_\text{2PN}+\bigg(\frac{d\bJ^{\,i}}{dt}\bigg)_\text{LO-SS}\,\bigg]+\bigg(\frac{d\bJ^{\,i}}{dt}\bigg)_\text{NLO-SO}
+\bigg[\,\bigg(\frac{d\bJ^{\,i}}{dt}\bigg)_\text{3PN}+\bigg(\frac{d\bJ^{\,i}}{dt}\bigg)_\text{NLO-SS}\bigg]+\cdots\,.
\end{autobreak}
\end{align}
For completeness, we include here both the spin-orbit and spin-spin terms to NLO in the PN expansion. The results can be written as
\begin{eqnarray}
\bigg(\frac{d\bJ^{\,i}}{dt}\bigg)_\text{LO-SO}&=&-\frac{G^2 \,m^2\,\nu^2}{15\,r^3}\,\bigg[\,g_3^{0\,i}+4\,\frac{G\,m}{r}\,g_3^{1\,i}+8\,\frac{G^2\,m^2}{r^2}\,g_3^{2\,i}\,\bigg]\,,\label{3pnJss1}\\
\bigg(\frac{d\bJ^{\,i}}{dt}\bigg)_\text{LO-SS}&=&-\frac{G^2 \,m\,\nu^2}{5\,r^4}\,\bigg[\,g_4^{0\,i}+\frac{G\,m}{r}\,g_4^{1\,i}\,\bigg]\,,\\
\bigg(\frac{d\bJ^{\,i}}{dt}\bigg)_\text{NLO-SO}&=&-\frac{G^2 \,m^2\,\nu^2}{105\,r^3}\,\bigg[\,g_5^{0\,i}+\frac{G\,m}{r}\,g_5^{1\,i}+\frac{G^2\,m^2}{r^2}\,g_5^{2\,i}+\frac{G^3\,m^3}{r^3}\,g_5^{3\,i}\,\bigg]\,,\\
\bigg(\frac{d\bJ^{\,i}}{dt}\bigg)_\text{NLO-SS}&=&-\frac{G^2 \,m\,\nu^2}{35\,r^4}\,\bigg[\,\frac{1}{2}\,g_6^{0\,i}+\frac{G\,m}{r}\,\frac{1}{3}\,g_6^{1\,i}+\frac{G^2\,m^2}{r^2}\,\frac{1}{3}\,g_6^{2\,i}\,\bigg]\,,\label{3pnJss2}
\end{eqnarray}
with the $g_3^{(0,1,2)i}, g_4^{(0,1)i}, g_5^{(0,1,2,3)i}$ and $g_6^{(0,1,2)i}$ coefficients displayed in Appendix~\ref{app:C}. The leading order spin-orbit and spin-spin expressions agree with corresponding results in \cite{kidder,Racine}.\vskip 4pt
As a nontrivial check, we have verified that the relationship 
\begin{align}
\frac{d\bJ^i}{dt}=\frac{1}{\omega}\,\frac{dE}{dt}\,\ble^{\,i}\,,
\end{align}
holds for spin effects to 3PN order for nonprecessing (quasi-)circular orbits. 

\subsection{Linear momentum \& Center-of-mass}\label{sec:mom}

Finally, we can compute the flux associated with linear momentum and CoM of the binary systems, which evolve due to the emission of GWs according to, e.g. \cite{Thorne},
\begin{align}
\frac{d\bP^i}{dt}=-G\,\bigg(\,\frac{2}{63}I^{(4)}_{ijk}I^{(3)}_{jk}+\frac{16}{45}\epsilon_{ijk}I^{(3)}_{jl}J^{(3)}_{kl}+\frac{1}{126}\epsilon_{ijk}I^{(4)}_{jlm}J^{(4)}_{klm}+\frac{4}{63}J^{(4)}_{ijk}J^{(3)}_{jk}+\cdots\,\bigg)\,,\label{dotp}
\end{align}
and (with $\bG^i \equiv m\bX^i$)
\begin{align}
\frac{d\bG^i}{dt}=\bP^i-G\,\bigg(\frac{1}{21}\big(I^{(3)}_{ijk}I^{(3)}_{jk}-I^{(4)}_{ijk}I^{(2)}_{jk}\big)+\frac{2}{21}\big(J^{(3)}_{ijk}J^{(3)}_{jk}-J^{(4)}_{ijk}J^{(2)}_{jk}\big)+\cdots\bigg)\,,
\end{align}
respectively. We can once again expand in various PN contributions, 
\begin{align}
\begin{autobreak}
\frac{d\bP^i}{dt}=\bigg(\frac{d\bP^i}{dt}\bigg)_\text{1PN}+\bigg(\frac{d\bP^i}{dt}\bigg)_\text{LO-SO}+\bigg(\frac{d\bP^i}{dt}\bigg)_\text{2PN}+\bigg(\frac{d\bP^i}{dt}\bigg)_\text{NLO-SO}+\bigg[\bigg(\frac{d\bP^i}{dt}\bigg)_\text{3PN}+\bigg(\frac{d\bP^i}{dt}\bigg)_\text{LO-SS}\bigg]+\cdots\,,
\end{autobreak}
\end{align}
and likewise,
\begin{align}
\begin{autobreak}
\frac{d\bG^i}{dt}=\bigg(\frac{d\bG^i}{dt}\bigg)_\text{1PN}+\bigg(\frac{d\bG^i}{dt}\bigg)_\text{2PN}+\bigg(\frac{d\bG^i}{dt}\bigg)_\text{LO-SO}+\bigg[\bigg(\frac{d\bG^i}{dt}\bigg)_\text{3PN}+\bigg(\frac{d\bG^i}{dt}\bigg)_\text{LO-SS}\bigg]+\cdots\,.
\end{autobreak}
\end{align}
Unfortunately, we do not have all the necessary source multipoles to complete the spin  corrections to NLO in the CoM, notably missing the NLO corrections to $J_{ijk}$, and therefore for the purpose of this paper we will keep spin-spin effects at leading order. We do, however, have all the information to compute the NLO spin-orbit contributions to the linear momentum. Inputing the source multipoles we find, in the CoM frame,\footnote{Technically speaking, the condition $\bX=0$ for the CoM frame becomes more subtle once we allow for non-inertial motion, due to GW emission. However, these effects can be ignored to 3PN order thanks to its non-secular behavior, i.e. $\bX=\mathcal{O}(x^{7/2})$, see e.g.~\cite{1983}.}
\begin{align}
\begin{autobreak}
\bigg(\frac{d\bP^i}{dt}\bigg)_\text{1PN}
=-\frac{8\delta\, G^3 m^4 \nu^2}{105r^4} \biggl[\bv^i \biggl(- \frac{8\,G\, m}{r} + 38 (nv)^2 - 50 \bv^2\biggr) 
+ \,\bn^i \,(nv)\biggl(\frac{12 G m }{r} -  45 (nv)^2 + 55  \,\bv^2\,\biggr)\,\biggr]\,,
\end{autobreak}
\\[2ex]
\begin{autobreak}
\bigg(\frac{d\bP^i}{dt}\bigg)_\text{LO-SO}
=-\frac{8 \,G^3 \,m^3 \,\nu^2}{15 r^5}\,\bigg[\,4 \,(nv) \,(\bv\times\bSigma)^{i} - 2 \,(\bn\times\bSigma)^{i} \,\bv^2 + \,\big(\bn\times\bv\big)^{i} \,(-3 \,(nv) \,(n\Sigma) - 2 \,(v\Sigma))\,\bigg]
\end{autobreak}
\\[2ex]
\begin{autobreak}
\bigg(\frac{d\bP^i}{dt}\bigg)_\text{NLO-SO}
=-\frac{4\,G^3 \,m^3 \,\nu^2}{945\,r^5}\,\bigg[\,h^{0\,i}_5+\frac{G\,m}{r}\,h^{1\,i}_5+\frac{G^2\,m^2}{r^2}\,h^{2\,i}_5\,\,\bigg]\,,
\end{autobreak}
\\[2ex]
\begin{autobreak}
\bigg(\frac{d\bP^i}{dt}\bigg)_\text{LO-SS}
=-\frac{2\,G^3 \,m^2 \,\nu^2}{105\,r^6} \,\bigg[\,h^{0\,i}_6+4\,\frac{G\,m}{r}\,h^{1\,i}_6\,\bigg]\,,
\end{autobreak}
\end{align}
whereas for the CoM position we have
\begin{align}
\begin{autobreak}
\bigg(\frac{d\bG^i}{dt}\bigg)_\text{1PN}
=-\,\bv^i \,\left(\frac{24 \,\delta \,G^3 \,m^4 \,\nu^2 \,(nv)}{35 r^3} + \frac{4 \,\delta \,G^2 \,m^3 \,\nu^2 \,(nv) \,(15 \,(nv)^2 - 29 \,\bv^2)}{105 r^2}\right) 
- \,\bn^i \,\Big(- \frac{16 \,\delta \,G^4 \,m^5 \,\nu^2}{35 r^4} + \frac{4 \,\delta \,G^3 \,m^4 \,\nu^2 \,(89 \,(nv)^2 - 101 \,\bv^2)}{105 r^3} 
-  \frac{2 \,\delta \,G^2 \,m^3 \,\nu^2 \,(225 \,(nv)^4 - 366 \,(nv)^2 \,\bv^2 + 113 \,(\bv^2)^2)}{105 r^2}\Big)\,,
\end{autobreak}
\\[2ex]
\begin{autobreak}
\bigg(\frac{d\bG^i}{dt}\bigg)_\text{LO-SO}
=-\frac{\,G^2 \,m^2 \,\nu^2}{105\,r^3}\,\bigg[\,k^{0\,i}_5+2\,\frac{G\,m}{r}\,k^{1\,i}_5+8\,\frac{G^2\,m^2}{r^2}\,k^{2\,i}_5\,\,\bigg]\,,
\end{autobreak}
\\[2ex]
\begin{autobreak}
\bigg(\frac{d\bG^i}{dt}\bigg)_\text{LO-SS}
=-\frac{G^2 \,m \,\nu^2}{105\,r^4} \,\bigg[\,k^{0\,i}_6+\,\frac{G\,m}{r}\,k^{1\,i}_6+\,\frac{G^2\,m^2}{r^2}\,k^{2\,i}_6\,\bigg]\,.
\end{autobreak}
\end{align}
The explicit expressions for the $h_5^{(0,1,2)i}, h_6^{(0,1)i}$ and $k_5^{(0,1,2)i}, k_6^{(0,1,2)i}$ coefficients are collected in Appendix~\ref{app:C}.\vskip 4pt

The calculation of the radiated linear momentum was also computed in \cite{Racine}, using the earlier results in \cite{Faye1,Faye2}. After transforming from the locally-flat and PN frames, through the relationship \cite{nrgrs2}
\beq
\bar{\bS}_A = \left(1-\frac{\bv^2_A}{2}-\frac{Gm_B}{r}\right) \bS_A  +  (\bv_A\cdot \bS_A) \bv_A + \cdots  \label{spinPN} \,,
\eeq
we find agreement for spin-orbit contributions to NLO in the PN expansion. However, we disagree on the spin-spin corrections for generic orbits. We have explicitly checked that the difference is due to the omission of the finite-size contributions to the current quadrupole and mass octupole moments. 
\subsubsection{Nonprecessing (quasi-)circular orbits}

The above expressions take on a much simpler form for the case of (quasi-)circular orbits. Using the conserved-norm spin vector, they become
\begin{align}
\frac{d\bP^i}{dt}&=-\frac{\blambda^i\,\nu^2}{315\,G}\Bigg\{\,(-1392 \,\delta x^{11/2} + \cdots)+\frac{336 x^6  \,(\Sigma_\text{c}\,\ell)}{\,G \,m^2} + x^7  \,\bigg(\frac{7520 \,\delta \,(S_\text{c}\,\ell )}{\,G \,m^2} + \frac{\,(1608 - 9888 \,\nu) \,(\Sigma_\text{c}\, \ell)}{\,G \,m^2}\bigg)\notag\\
& + x^{15/2}  \,\Bigg[\,\bigg(- \frac{6960 \,\delta}{\,G^2 \,m^4} -  \frac{240 \,\kappa_{-}}{\,G^2 \,m^4} -  \frac{3480 \,\delta \,\kappa_{+}}{\,G^2 \,m^4}\bigg) \,\bS_\text{c}^2 \notag\\
&+ \,\bigg(\frac{3240 \,\delta \,\kappa_{-}}{\,G^2 \,m^4} + \frac{120 \,\kappa_{+} \,(-27 + 116 \,\nu)}{\,G^2 \,m^4} + \frac{48 \,(-181 + 580 \,\nu)}{\,G^2 \,m^4}\bigg) \,(S_\text{c}\Sigma_\text{c}) \notag\\
&+ \,\Bigg( \frac{240 \,\delta \,(-3 + 29 \,\nu)}{\,G^2 \,m^4} + \frac{60 \,\delta \,\kappa_{+} \,(-27 + 58 \,\nu)}{\,G^2 \,m^4}- \frac{60 \,\kappa_{-} \,(-27 + 112 \,\nu)}{\,G^2 \,m^4}\Bigg) \,\bSigma_\text{c}^2\Bigg]\Bigg\}+\cdots,
\end{align}
for the linear momentum including leading spin-spin and up to NLO spin-orbit effects, whereas for the center-of-mass position we have
\begin{align}
\frac{d\bG^i}{dt}
&=-\frac{\bn^i\,\nu^2\,m}{640} \,\bigg\{(-4068 \,\delta x^4 + \cdots ) + x^{11/2} \,\bigg(\frac{21822 \,\delta \,(S_c\, \ell)}{\,G \,m^2} -  \frac{18 \,(-439 + 1441 \,\nu) \,(\Sigma_c\,\ell) }{\,G \,m^2}\bigg)\\
&+ x^6 \,\bigg[\,(- \frac{20340 \,\delta}{\,G^2 \,m^4} + \frac{72 \,\kappa_{-}}{\,G^2 \,m^4} -  \frac{10170 \,\delta \,\kappa_{+}}{\,G^2 \,m^4}) \,\bS_c^2 \\
&+ \,\bigg(\frac{10242 \,\delta \,\kappa_{-}}{\,G^2 \,m^4} -  \frac{180 \,(117 - 452 \,\nu)}{\,G^2 \,m^4} + \frac{18 \,\kappa_{+} \,(-569 + 2260 \,\nu)}{\,G^2 \,m^4}\bigg) \,(S_c\Sigma_c)\\
&+ \,\bigg(- \frac{180 \,\delta \,(4 - 113 \,\nu)}{\,G^2 \,m^4} + \frac{9 \,\delta \,\kappa_{+} \,(-569 + 1130 \,\nu)}{\,G^2 \,m^4} 
-  \frac{9 \,\kappa_{-} \,(-569 + 2268 \,\nu)}{\,G^2 \,m^4}\bigg) \,\bSigma_c^2\bigg]\bigg\} +\cdots\,.
\end{align}

\subsubsection{Kick-velocity}

An application of the above formula for the linear momentum flux is the derivation of the kick-velocity, $\bV^i_\text{kick}$, obtained via 
\begin{align}
\bV^i_\text{kick}=\frac{1}{m}\,\int^t_{-\infty} dt\,\bigg(\frac{d\bP^i}{dt}\bigg)\,,
\end{align}
which we compute taking advantage of the relation in \eqref{dndt}.
Performing the integration with the boundary condition $x=0$ at $t=-\infty$, we arrive at 
\beq
\begin{aligned}
\label{vkick}
\bigg(\bV^i_\text{kick}\bigg)_\text{Spin}
&=
  \bn^{i}\left\{-\frac{16 x^{9/2} \,\nu^2 \, \,(\Sigma_\text{c}\,\ell)}{15 \,G \,m^2}
+ \frac{8 x^{11/2} \,\nu^2 \, \,\bigg(-940 \,\delta \,(S_\text{c}\,\ell) + 3 \,(-67 + 412 \,\nu) \,(\Sigma_\text{c}\,\ell)\bigg)}{315 \,G \,m^2}\right. \\
&+ x^6  \,\left[\frac{8 \,\bigg(2 \,\kappa_{-} + 29 \,\delta \,(2 + \,\kappa_{+})\bigg) \,\nu^2 \,\bS_\text{c}^2}{21 \,G^2 \,m^4} -  \frac{8 \,\nu^2 \,\bigg(-362 + 135 \,\delta \,\kappa_{-} + 1160 \,\nu + 5 \,\kappa_{+} \,(-27 + 116 \,\nu)\bigg) \,(S_\text{c}\Sigma_\text{c})}{105 \,G^2 \,m^4}\right. \\
&+\left.\left. \frac{4 \,\nu^2 \,\Big[\,\delta \,\bigg(12 + \,\kappa_{+} \,(27 - 58 \,\nu) - 116 \,\nu\bigg) 
+ \,\kappa_{-} \,(-27 + 112 \,\nu)\Big] \,\bSigma_\text{c}^2}{21 \,G^2 \,m^4}\right]\right\}\,,
\end{aligned}
\eeq
where we have written the final expression in terms of the conserved-norm spin vectors. A similar result is given in \cite{Racine}, with full agreement in the spin-orbit sector. Yet, the disagreement for spin-spin contributions remains.  

\section{Accumulated phase} \label{phase}

Armed with the binding energy and radiated flux it is straightforward to compute the accumulated phase. Because, the norm of the covariant spin variables are not conserved during the radiation-reaction timescale, we provide here expressions in terms of the conserved-norm spin variable. We first obtain the evolution of the orbital frequency in time over a period, 
\begin{align}
\left(\frac{\dot{\omega}}{\omega^2}\right)_\text{SS}&=\frac{x^{9/2}}{\,G^2 \,m^4}  \,\Big[\,(192 + 96 \,\kappa_{+}) \,\nu \,\bS_\text{c}^2 + \,(192 \,\delta - 96 \,\kappa_{-} + 96 \,\delta \,\kappa_{+}) \,\nu \,(S_\text{c} \,\Sigma_\text{c})\notag\\
&+ \,\bigg(\,(\tfrac{6}{5} - 48 \,\delta \,\kappa_{-} + 48 \,\kappa_{+}) \,\nu + \,(-192 - 96 \,\kappa_{+}) \,\nu^2\bigg) \,\bSigma_\text{c}^2\Big]\notag\\
&+\frac{x^{11/2}}{\,G^2 \,m^4} \,\Big[\,\bigg(\,(\tfrac{102072}{35} + \tfrac{886}{5} \,\delta \,\kappa_{-} + \tfrac{10156}{35} \,\kappa_{+}) \,\nu + \,(- \tfrac{4128}{5} -  \tfrac{2064}{5} \,\kappa_{+}) \,\nu^2\bigg) \,\bS_\text{c}^2\notag\\
&+ \,\bigg(\,(\tfrac{14120}{7} \,\delta -  \tfrac{3282}{35} \,\kappa_{-} -  \tfrac{96}{5} \,\delta^2 \,\kappa_{-} + \tfrac{3954}{35} \,\delta \,\kappa_{+}) \,\nu + \,(- \tfrac{4128}{5} \,\delta -  \tfrac{1864}{5} \,\kappa_{-} -  \tfrac{2064}{5} \,\delta \,\kappa_{+}) \,\nu^2\bigg) \,(S_\text{c} \,\Sigma_\text{c})\notag \\
&+ \,\bigg(\,(\tfrac{789}{5} -  \tfrac{1977}{35} \,\delta \,\kappa_{-} + \tfrac{1977}{35} \,\kappa_{+}) \,\nu + \,(- \tfrac{47269}{35} + \tfrac{146}{5} \,\delta \,\kappa_{-} -  \tfrac{4976}{35} \,\kappa_{+}) \,\nu^2 + \,(\tfrac{4128}{5} + \tfrac{2064}{5} \,\kappa_{+}) \,\nu^3\bigg) \,\bSigma_\text{c}^2\Big]\,.
\end{align}
Hence, performing the standard integration\,,
\beq
\phi = \int dt \, \omega(t) = \int d\omega\, \frac{\omega(t)}{\dot \omega}
\eeq
we arrive at
\begin{align}
\left(\Delta\phi\right)_\text{SS}&=\frac{x^{-5/2}}{32\,\nu\,G^2\,m^4}\,\Bigg\{x^2 \,\Big[\,(-50 - 25 \,\kappa_{+}) \,\bS_\text{c}^2 + \,(-50 \,\delta + 25 \,\kappa_{-} - 25 \,\delta \,\kappa_{+}) \,(S_\text{c} \,\Sigma_\text{c}) \notag\\
&+ \,\bigg(- \tfrac{5}{16} + \tfrac{25}{2} \,\delta \,\kappa_{-} -  \tfrac{25}{2} \,\kappa_{+} + \,(50 + 25 \,\kappa_{+}) \,\nu\bigg) \,\bSigma_\text{c}^2\Big]\notag\\
&+ x^3 \,\Big[\,\bigg(- \tfrac{31075}{126} + \tfrac{2215}{48} \,\delta \,\kappa_{-} + \tfrac{15635}{84} \,\kappa_{+} + \,(60 + 30 \,\kappa_{+}) \,\nu\bigg) \,\bS_\text{c}^2\notag\\
&+ \,\bigg(- \tfrac{9775}{42} \,\delta -  \tfrac{47035}{336} \,\kappa_{-} + \tfrac{47035}{336} \,\delta \,\kappa_{+} + \,(60 \,\delta -  \tfrac{2575}{12} \,\kappa_{-} + 30 \,\delta \,\kappa_{+}) \,\nu\bigg) \,(S_\text{c} \,\Sigma_\text{c}) \notag\\
&+ \,\bigg(- \tfrac{410825}{2688} -  \tfrac{47035}{672} \,\delta \,\kappa_{-} + \tfrac{47035}{672} \,\kappa_{+} + \,(\tfrac{23535}{112} -  \tfrac{2935}{48} \,\delta \,\kappa_{-} -  \tfrac{4415}{56} \,\kappa_{+}) \,\nu + \,(-60 - 30 \,\kappa_{+}) \,\nu^2\bigg) \,\bSigma_\text{c}^2\Big]\Bigg\}\,,
\end{align}
which agrees with the result in \cite{Bohes2}.

\section{Conclusions}\label{final}

Building upon the quadratic-in-spin conservative and dissipative results obtained using the EFT approach in~\cite{prl,nrgrss,nrgrs2,srad,amps} to NLO in the PN expansion, in this paper we have completed the derivation of the EoM, adiabatic invariants and associated flux-balance equations for the energy and angular momentum of binary systems with spinning compact objects. We then used these results to compute the bilinear-in-spin evolution of the orbital frequency and accumulated phase for (quasi-)circular orbits to 3PN order, including finite-size effects, finding agreement with the value presented in \cite{Bohes2}. This is not surprising since, after all, the gravitational potential computed in \cite{Bohes2} was shown to be equivalent to the ones previously obtained in the EFT \cite{prl,nrgrss,nrgrs2} and ADM \cite{justin1,justin2} approaches; and moreover, as we demonstrated here, the source multipole moments in \cite{Bohes2} are in complete agreement with those derived before~in~\cite{srad}, after the latter are expressed in terms of conserved-norm spin variables.\vskip 4pt

Our results here include also the angular momentum flux for the first time, which is necessary to compute the phase evolution in elliptic-like orbits, allowing us to incorporate spin effects in the waveforms for generic configurations  \cite{Cho:2019brd,Chatziioannou:2017tdw}, thus extending the validity of the PN approximation towards higher frequencies. This will be crucial for proper data analysis with spinning binaries, since eccentricity can rapidly deteriorate the accuracy of waveforms relying on  (quasi-)circular approximations \cite{Csizmadia:2012wy}. Using the source multipoles and EoM we have also computed the radiated flux of linear momentum, with which we obtained the kick-velocity for (quasi-)circular orbits, including linear and bilinear spin effects to NLO and leading order, respectively. While perfect agreement is found with an earlier derivation in \cite{Racine} for spin-orbit effects, unfortunately we disagree in the spin-spin sector even for the case of black holes (with $\kappa_\pm=1$). We trace the discrepancy to finite-size contributions in the current quadrupole and mass octupole moments which were not included in the derivation in \cite{Racine}. As it turns out, only the quadratic-in-spin correction to $J_{ikl}$ at NLO is missing to complete the value of the kick-velocity at the same order. We are currently performing this calculation. Moreover, building on the recent rederivation of the spin-independent radiated fluxes at 2PN in \cite{radnrgr}, it is straightforward to continue pushing forward at the next order for spin effects in the EFT approach. The derivation of GW observables to N$^{2}$LO is underway.

\begin{center}
{\bf Acknowledgments}
\end{center}
The work of G.C. and R.A.P. is supported by the ERC Consolidator Grant ``Precision Gravity: From the LHC~to LISA,"  provided by the European Research Council (ERC) under the European Union's H2020 research and innovation programme, grant No.\,817791. G.C. and R.A.P. are also supported in part by the DFG under Germany's Excellence Strategy `Quantum Universe' (No.\,390833306). B.P. is supported by the National Science Foundation under Grant No. PHY-1820760. We~acknowledge extensive use of the \texttt{xAct} packages~\cite{martin2019xact}. 

\appendix

\section{Equations of motion \& Source multipoles}\label{app:A}
We summarize here the values for the spin-spin contributions to the acceleration and spin dynamics as well as the source multipole moments in the covariant SSC, first computed to NLO in \cite{prl,nrgrss,nrgrs2,nrgrso} and \cite{srad} within the EFT approach, respectively. As in the main text we use the short-hand notation $(A B) \equiv \bA\cdot \bB$ whenever $\bA \neq \bB$, as well as $(A B C) \equiv \bA\cdot (\bB \times \bC)$, throughout the appendices.

\subsection{Acceleration}

Spin effects in the acceleration enter at different PN orders,
\begin{subequations}\label{eq:acceleration}
\begin{align}
\ba=\ba_\text{N}+\ba_\text{1PN}+\ba_\text{LO-SO}+\Big[\,\ba_\text{2PN}+\ba_\text{LO-SS}
\,\Big]+\ba_\text{NLO-SO}+\Big[\,\ba_\text{3PN}+\ba_\text{NLO-SS}\,\Big]+\cdots\,,
\end{align}
where at Newtonian and 1PN order we have
\begin{align}
\ba_\text{N}^i=-\frac{G \,m\,\bn^i}{r^2}\,,
\end{align}
\begin{align}
\ba_\text{1PN}^i=- \frac{2 G \,m \,(-2 + \,\nu) \,(nv) \,\bv^i}{r^2} + \,\bn^i \,\Big[\,\frac{2 G^2 \,m^2 \,(2 + \,\nu)}{r^3} + \frac{G \,m \,\big(3 \,\nu \,(\,(nv)^2 - 2 \,\bv^2) - 2 \,\bv^2\big)}{2\, r^2}\,\Big]\,.
\end{align}
\end{subequations}
The leading order spin-spin acceleration is given by 
\begin{align}
\begin{autobreak}
\ba^i_\text{LO-SS}
=\frac{G}{m\,r^4}\,\frac{3}{4}\,\Bigg[\,\big(\,(-8 - 4 \,\kappa_{+}) \,(nS) + \,(-4 \,\delta + 2 \,\kappa_{-} - 2 \,\delta \,\kappa_{+}) \,(n\Sigma)\big) \,\bS^{i}
+ \,\Big[\,(-4 \,\delta + 2 \,\kappa_{-} - 2 \,\delta \,\kappa_{+}) \,(nS) + \,\big(2 \,\delta \,\kappa_{-} + 8 \,\nu + \,\kappa_{+} \,(-2 + 4 \,\nu)\big) \,(n\Sigma)\Big] \,\bSigma^{i} 
+ \,\bn^i \,\Big[\,(20 + 10 \,\kappa_{+}) \,(nS)^2 + \,(20 \,\delta - 10 \,\kappa_{-} + 10 \,\delta \,\kappa_{+}) \,(nS) \,(n\Sigma) + \,\big(-5 \,\delta \,\kappa_{-} + \,\kappa_{+} \,(5 - 10 \,\nu) - 20 \,\nu\big) \,(n\Sigma)^2 
+ \,(-4 - 2 \,\kappa_{+}) \,\bS^2 + \,(-4 \,\delta + 2 \,\kappa_{-} - 2 \,\delta \,\kappa_{+}) \,(S\Sigma) + \,\big(\,\delta \,\kappa_{-} + 4 \,\nu + \,\kappa_{+} \,(-1 + 2 \,\nu)\big) \,\bSigma^2\Big]\,\Bigg]\,,
\end{autobreak}
\end{align}
while we write the NLO correction as
\begin{align}
\label{eq:acc}
\ba^i_\text{NLO-SS}=\frac{G}{m\,r^4}\,\Bigg[\,\frac{1}{8}\,\bA^i_0+\frac{G\,m}{r}\,\frac{1}{2}\,\bA^i_1\,\Bigg]\,,
\end{align}
with 
\begin{subequations}
\begin{align}
\begin{autobreak}
\bA_0^i
=\bS^{i} \,\bigg\{\,(72 + 36 \,\delta \,\kappa_{-} - 36 \,\kappa_{+}) \,(nv) \,(vS) 
+ \,(nS) \Big[\big(-60 \,\delta \,\kappa_{-} + 120 \,\nu + 60 \,\kappa_{+} \,(1 + \,\nu)\big) \,(nv)^2 
+ \big(-24 \,\kappa_{+} \,(1 + 4 \,\nu) - 24 \,(1 + 8 \,\nu)\big) \,\bv^2\Big]
+ \,(n\Sigma) \Big[\big(60 \,\delta \,(-2 + \,\nu) + 30 \,\delta \,\kappa_{+} \,(2 + \,\nu)
+ \,\kappa_{-} \,(-60 + 90 \,\nu)\big) \,(nv)^2 + \big(-12 \,\delta \,\kappa_{+} \,(1 + 4 \,\nu)- 12 \,\delta \,(-1 + 8 \,\nu) + \,\kappa_{-} \,(12 + 48 \,\nu)\big) \,\bv^2\Big] 
+ \big(120 \,\delta - 36 \,\delta \,\kappa_{+} + \,\kappa_{-} \,(36 - 72 \,\nu)\big) \,(nv) \,(v\Sigma)\bigg\}
+ \bSigma^{i} \,\bigg\{\big(48 \,\delta - 36 \,\delta \,\kappa_{+} + \,\kappa_{-} \,(36 - 72 \,\nu)\big) \,(nv) \,(vS)
+ \,(nS) \Big[\big(60 \,\delta \,\nu + 30 \,\delta \,\kappa_{+} \,(2 + \,\nu) + \,\kappa_{-} \,(-60 + 90 \,\nu)\big) \,(nv)^2
+ \big(-12 \,\delta \,\kappa_{+} \,(1 + 4 \,\nu) - 12 \,\delta \,(1 + 8 \,\nu) + \,\kappa_{-} \,(12 + 48 \,\nu)\big) \,\bv^2\Big] 
+ \,(n\Sigma) \Big[\big(30 \,\delta \,\kappa_{-} \,(-2 + \,\nu) - 120 \,(-1 + \,\nu)^2 
- 30 \,\kappa_{+} \,(-2 + 5 \,\nu + 2 \,\nu^2)\big) \,(nv)^2 + \big(12 \,\delta \,\kappa_{-} \,(1 + 4 \,\nu)
+ 12 \,\kappa_{+} \,(-1 + 2 \,\nu) \,(1 + 4 \,\nu) + 24 \,(1 -  \,\nu + 8 \,\nu^2)\big) \,\bv^2\Big] 
+ \big(96 - 36 \,\delta \,\kappa_{-} \,(-1 + \,\nu) - 264 \,\nu + 36 \,\kappa_{+} \,(-1 + 3 \,\nu)\big) \,(nv) \,(v\Sigma)\bigg\}
+ \,\bv^{i} \Big\{\big(120 \,\kappa_{+} \,(-2 + \,\nu) + 240 \,\nu\big) \,(nS)^2 \,(nv) 
+ \big(-120 \,\kappa_{-} \,(-2 + \,\nu) + 120 \,\delta \,\kappa_{+} \,(-2 + \,\nu) + 240 \,\delta \,(1 + \,\nu)\big) \,(nS) \,(nv) \,(n\Sigma)
+ \big(-60 \,\delta \,\kappa_{-} \,(-2 + \,\nu) - 60 \,\kappa_{+} \,(-2 + \,\nu) \,(-1 + 2 \,\nu) - 240 \,(-1 + 2 \,\nu + \,\nu^2)\big) \,(nv) \,(n\Sigma)^2 
+ \big(-12 \,\delta \,\kappa_{-} + \,\kappa_{+} \,(60 - 24 \,\nu) - 48 \,(1 + \,\nu)\big) \,(nv) \,\bS^2
+ \big(-24 \,\delta \,\kappa_{+} \,(-3 + \,\nu) + 72 \,\kappa_{-} \,(-1 + \,\nu) - 48 \,\delta \,(3 + \,\nu)\big) \,(nv) \,(S\Sigma)
+ \big(72 + 12 \,\delta \,\kappa_{-} + \,\kappa_{+} \,(84 - 24 \,\nu) - 48 \,\nu\big) \,(nS) \,(vS)
+ \big(-12 \,\delta \,\kappa_{+} \,(-3 + \,\nu) - 24 \,\delta \,\nu - 12 \,\kappa_{-} \,(3 + \,\nu)\big) \,(n\Sigma) \,(vS)
+ \big(-12 \,\delta \,\kappa_{+} \,(-3 + \,\nu) - 24 \,\delta \,(-1 + \,\nu) - 12 \,\kappa_{-} \,(3 + \,\nu)\big) \,(nS) \,(v\Sigma)
+ \big(-36 \,\delta \,\kappa_{-} + 12 \,\kappa_{+} \,(3 - 6 \,\nu + 2 \,\nu^2) + 24 \,(-2 + \,\nu + 2 \,\nu^2)\big) \,(n\Sigma) \,(v\Sigma)
+ \big(12 \,\delta \,\kappa_{-} \,(-3 + 2 \,\nu) + 48 \,(-2 + 5 \,\nu + \,\nu^2) + 12 \,\kappa_{+} \,(3 - 8 \,\nu + 2 \,\nu^2)\big) \,(nv) \,\bSigma^2\Big\} 
+ \,\bn^i \,\bigg\{\,(-360 - 60 \,\delta \,\kappa_{-} + 60 \,\kappa_{+}) \,(nS) \,(nv) \,(vS)
+ \big(-240 \,\delta + 60 \,\delta \,\kappa_{+} + 60 \,\kappa_{-} \,(-1 + 2 \,\nu)\big) \,(nv) \,(n\Sigma) \,(vS) 
+ \big(12 \,\delta \,\kappa_{-} + 12 \,\kappa_{+} \,(-1 + \,\nu) + 24 \,(3 + \,\nu)\big) \,(vS)^2
+ \,(nS) \,(n\Sigma) \Big[\,(-420 \,\delta \,\nu + 210 \,\kappa_{-} \,\nu - 210 \,\delta \,\kappa_{+} \,\nu) \,(nv)^2
+ \big(-360 \,\kappa_{-} \,\nu + 240 \,\delta \,\kappa_{+} \,\nu + 60 \,\delta \,(1 + 8 \,\nu)\big) \,\bv^2\Big] 
+ \,(nS)^2 \,\Big[\,(-420 \,\nu - 210 \,\kappa_{+} \,\nu) \,(nv)^2 + \big(30 \,\delta \,\kappa_{-} + 60 \,(1 + 8 \,\nu) 
+ 30 \,\kappa_{+} \,(1 + 8 \,\nu)\big) \,\bv^2\Big]
+ \,(n\Sigma)^2 \Big[\big(105 \,\delta \,\kappa_{-} \,\nu + 420 \,\nu^2 + 105 \,\kappa_{+} \,\nu \,(-1 + 2 \,\nu)\big) \,(nv)^2 
+ \big(-150 \,\delta \,\kappa_{-} \,\nu - 30 \,\kappa_{+} \,\nu \,(-5 + 8 \,\nu) - 60 \,\nu \,(1 + 8 \,\nu)\big) \,\bv^2\Big]
+ \,\bS^2\, \Big[\big(30 \,\delta \,\kappa_{-} + 30 \,\kappa_{+} \,(-1 + \,\nu) + 60 \,(2 + \,\nu)\big) \,(nv)^2
+ \big(-12 \,\delta \,\kappa_{-} - 48 \,\kappa_{+} \,\nu - 12 \,(3 + 8 \,\nu)\big) \,\bv^2\Big] 
+ \,(S\Sigma) \Big[\big(\,\kappa_{-} \,(60 - 150 \,\nu) + 30 \,\delta \,\kappa_{+} \,(-2 + \,\nu) + 60 \,\delta \,(4 + \,\nu)\big) \,(nv)^2 
+ \big(12 \,\delta^3 \,\kappa_{+} - 12 \,\delta \,(5 + 8 \,\nu) + \,\kappa_{-} \,(-12 + 96 \,\nu)\big) \,\bv^2\Big] 
+ \big(-360 \,\delta + 60 \,\delta \,\kappa_{+} + 60 \,\kappa_{-} \,(-1 + 2 \,\nu)\big) \,(nS) \,(nv) \,(v\Sigma) 
+ \big(\,\kappa_{+} \,(60 - 180 \,\nu) + 60 \,\delta \,\kappa_{-} \,(-1 + \,\nu) + 120 \,(-2 + 7 \,\nu)\big) \,(nv) \,(n\Sigma) \,(v\Sigma)
+ \big(\,\kappa_{-} \,(24 - 60 \,\nu) + 12 \,\delta \,\kappa_{+} \,(-2 + \,\nu) + 24 \,\delta \,(5 + \,\nu)\big) \,(vS) \,(v\Sigma)
+ \big(-6 \,\delta \,\kappa_{-} \,(-2 + 3 \,\nu) - 24 \,(-2 + 7 \,\nu + \,\nu^2) - 6 \,\kappa_{+} \,(2 - 7 \,\nu + 2 \,\nu^2)\big) \,(v\Sigma)^2
+ \Big[\big(-15 \,\delta \,\kappa_{-} \,(-2 + 3 \,\nu) - 60 \,(-2 + 6 \,\nu + \,\nu^2) - 15 \,\kappa_{+} \,(2 - 7 \,\nu + 2 \,\nu^2)\big) \,(nv)^2 
+ \big(6 \,\delta \,\kappa_{-} \,(-1 + 6 \,\nu) + 6 \,\kappa_{+} \,(1 - 8 \,\nu + 8 \,\nu^2) + 12 \,(-2 + 7 \,\nu + 8 \,\nu^2)\big) \,\bv^2\Big] \,\bSigma^2\bigg\}\,,
\end{autobreak}
\\[2ex]
\begin{autobreak}
A_1^i
= S^{i}\,\Big[\big(82 - 12 \,\delta \,\kappa_{-} + 36 \,\nu + 18 \,\kappa_{+} \,(2 + \,\nu)\big) \,(nS)
+ \big(18 \,\delta \,(2 + \,\nu) + 3 \,\delta \,\kappa_{+} \,(8 + 3 \,\nu) + 3 \,\kappa_{-} \,(-8 + 5 \,\nu)\big) \,(n\Sigma)\Big]
+  \bSigma^i \,\Big[\big(6 \,\delta \,(7 + 3 \,\nu) + 3 \,\delta \,\kappa_{+} \,(8 + 3 \,\nu) + 3 \,\kappa_{-} \,(-8 + 5 \,\nu)\big) \,(nS) 
+ \big(3 \,\delta \,\kappa_{-} \,(-8 + \,\nu) - 2 \,\nu \,(37 + 18 \,\nu) - 3 \,\kappa_{+} \,(-8 + 17 \,\nu + 6 \,\nu^2)\big) \,(n\Sigma)\Big]
+ \,\bn^i\, \Big[\big(24 \,\delta \,\kappa_{-} - 48 \,\kappa_{+} \,(2 + \,\nu) - 6 \,(35 + 16 \,\nu)\big) \,(nS)^2 
+ \big(\,\kappa_{-} \,(120 - 48 \,\nu) - 24 \,\delta \,\kappa_{+} \,(5 + 2 \,\nu) - 6 \,\delta \,(33 + 16 \,\nu)\big) \,(nS) \,(n\Sigma)
+ \big(60 \,\delta \,\kappa_{-} + 6 \,\nu \,(31 + 16 \,\nu) + 12 \,\kappa_{+} \,(-5 + 10 \,\nu + 4 \,\nu^2)\big) \,(n\Sigma)^2
+ \big(-4 \,\delta \,\kappa_{-} + 10 \,\kappa_{+} \,(2 + \,\nu) + 4 \,(9 + 5 \,\nu)\big) \,\bS^2
+ \big(6 \,\kappa_{-} \,(-4 + \,\nu) + 4 \,\delta \,(9 + 5 \,\nu) + 2 \,\delta \,\kappa_{+} \,(12 + 5 \,\nu)\big) \,(S\Sigma)
+ \big(- \,\delta \,\kappa_{-} \,(12 + \,\nu) - 4 \,\nu \,(9 + 5 \,\nu) + \,\kappa_{+} \,(12 - 23 \,\nu - 10 \,\nu^2)\big) \,\bSigma^2\Big]\,.
\end{autobreak}
\end{align}
\end{subequations}
\subsection{Spin dynamics}\label{app:spindynamics}
In terms of the covariant spin vector, without yet transforming to a conserved-norm spin, 
the evolution equation is given by 
\begin{align}
\begin{autobreak}
\MoveEqLeft
\frac{d\bS^i_1}{dt}
= \Bigg[- \frac{G \,m_{2}^2 \,(nv) \,\bS_1^i}{\,m r^2} + \frac{2 G \,m_{1} \,m_{2} \,(nS_1) \,\bv^i}{\,m r^2} + \frac{2 G \,m_{2}^2 \,(nS_1) \,\bv^i}{\,m r^2} -  \frac{2 G \,m_{1} \,m_{2} \,\bn^i \,(vS_1)}{\,m r^2} -  \frac{G \,m_{2}^2 \,\bn^i \,(vS_1)}{\,m r^2}\Bigg]_\text{LO SO}
+ \Bigg[\,\frac{3 \,C_{{ES}^2}^{(1)} G \,m_{2} \,(\bn\times \bS_1)^{i} \,(nS_1)}{\,m_{1} r^3} + \frac{3 G \,(\bn\times \bS_1)^{i} \,(nS_2)}{r^3} + \frac{G \,(\bS_1\times \bS_2)^{i}}{r^3}\,\Bigg]_\text{LO SS}
+\Bigg[- \frac{\,\delta G^2 \,m \,m_{2} \,\nu \,\bn^i \,(nS_1) \,(nv)}{2 r^3} + \frac{G^2 \,m_{1} \,m_{2}^2 \,(nv) \,\bS_1^i}{\,m r^3} + \frac{\,\delta G^2 \,m \,m_{2} \,\nu \,(nv) \,\bS_1^i}{r^3} + \frac{3 G \,m_{1}^2 \,m_{2}^2 \,(nv)^3 \,\bS_1^i}{2 \,m^3 r^2} 
-  \frac{G^2 \,m_{1}^2 \,m_{2} \,(nS_1) \,\bv^i}{\,m r^3} -  \frac{3 G^2 \,m_{1} \,m_{2}^2 \,(nS_1) \,\bv^i}{2 \,m r^3} -  \frac{G^2 \,m_{2}^3 \,(nS_1) \,\bv^i}{2 \,m r^3} -  \frac{3 G \,m_{1}^3 \,m_{2} \,(nS_1) \,(nv)^2 \,\bv^i}{\,m^3 r^2} 
-  \frac{3 G \,m_{1}^2 \,m_{2}^2 \,(nS_1) \,(nv)^2 \,\bv^i}{\,m^3 r^2} -  \frac{G \,m_{1}^2 \,m_{2}^2 \,(nv) \,\bS_1^i \bv^2}{\,m^3 r^2} -  \frac{G \,m_{1} \,m_{2}^3 \,(nv) \,\bS_1^i \bv^2}{\,m^3 r^2} -  \frac{\,\delta G \,m_{2} \,\nu \,(nv) \,\bS_1^i \bv^2}{2 r^2} 
+ \frac{2 G \,m_{1}^3 \,m_{2} \,(nS_1) \,\bv^i \bv^2}{\,m^3 r^2} + \frac{4 G \,m_{1}^2 \,m_{2}^2 \,(nS_1) \,\bv^i \bv^2}{\,m^3 r^2} + \frac{2 G \,m_{1} \,m_{2}^3 \,(nS_1) \,\bv^i \bv^2}{\,m^3 r^2} + \frac{G^2 \,m_{1}^2 \,m_{2} \,\bn^i \,(vS_1)}{\,m r^3} 
+ \frac{G^2 \,m_{1} \,m_{2}^2 \,\bn^i \,(vS_1)}{2 \,m r^3} + \frac{G^2 \,m_{2}^3 \,\bn^i \,(vS_1)}{2 \,m r^3} -  \frac{\,\delta G^2 \,m \,m_{2} \,\nu \,\bn^i \,(vS_1)}{2 r^3} + \frac{3 G \,m_{1}^3 \,m_{2} \,\bn^i \,(nv)^2 \,(vS_1)}{\,m^3 r^2} 
+ \frac{3 G \,m_{1}^2 \,m_{2}^2 \,\bn^i \,(nv)^2 \,(vS_1)}{2 \,m^3 r^2} -  \frac{G \,m_{1}^2 \,m_{2}^2 \,(nv) \,\bv^i \,(vS_1)}{\,m^3 r^2} -  \frac{G \,m_{1} \,m_{2}^3 \,(nv) \,\bv^i \,(vS_1)}{\,m^3 r^2} -  \frac{2 G \,m_{1}^3 \,m_{2} \,\bn^i \bv^2 \,(vS_1)}{\,m^3 r^2} 
-  \frac{2 G \,m_{1}^2 \,m_{2}^2 \,\bn^i \bv^2 \,(vS_1)}{\,m^3 r^2} + \frac{\,\delta G \,m_{2} \,\nu \,\bn^i \bv^2 \,(vS_1)}{2 r^2} \Bigg]_\text{NLO SO}
+\Bigg[- \frac{G^2 \,m_{2} \,\bn^i \,(nS_1S_2)}{r^4} + \frac{G^2 \,m_{1} \,\nu \,\bn^i \,(nS_1S_2)}{r^4} + \frac{G^2 \,m_{2} \,\nu \,\bn^i \,(nS_1S_2)}{r^4} -  \frac{2 G^2 \,m_{2} \,(\bn\times \bS_1)^{i} \,(nS_1)}{r^4}
-  \frac{3 \,C_{{ES}^2}^{(1)} G^2 \,m_{2} \,(\bn\times \bS_1)^{i} \,(nS_1)}{r^4} -  \frac{12 \,C_{{ES}^2}^{(1)} G^2 \,m_{2}^2 \,(\bn\times \bS_1)^{i} \,(nS_1)}{\,m_{1} r^4} -  \frac{9 G^2 \,m \,(\bn\times \bS_1)^{i} \,(nS_2)}{r^4} 
+ \frac{G^2 \,m_{2} \,(\bn\times \bS_1)^{i} \,(nS_2)}{r^4} + \frac{15 \,C_{{ES}^2}^{(1)} G \,m_{2}^2 \,(\bn\times \bS_1)^{i} \,(nS_1) \,(nv)^2}{2 \,m^2 r^3} + \frac{15 G \,m_{1} \,m_{2} \,(\bn\times \bS_1)^{i} \,(nS_2) \,(nv)^2}{2 \,m^2 r^3} 
-  \frac{3 G^2 \,m \,(\bS_1\times \bS_2)^{i}}{r^4} + \frac{G^2 \,m_{2} \,(\bS_1\times \bS_2)^{i}}{r^4} -  \frac{3 G \,m_{1}^2 \,(nv)^2 \,(\bS_1\times \bS_2)^{i}}{\,m^2 r^3}-  \frac{3 G \,m_{1} \,m_{2} \,(nv)^2 \,(\bS_1\times \bS_2)^{i}}{2 \,m^2 r^3} 
+ \frac{3 G \,m_{2}^2 \,(nv) \,\bS_1^i \,(S_1nv)}{\,m \,m_{1} r^3} + \frac{3 G \,m_{1} \,m_{2} \,(nv) \,\bS_1^i \,(S_2nv)}{\,m^2 r^3} + \frac{3 G \,m_{2}^2 \,(nv) \,\bS_1^i \,(S_2nv)}{\,m^2 r^3} 
+ \frac{3 G \,m_{1} \,m_{2} \,(nS_1S_2) \,(nv) \,\bv^i}{\,m^2 r^3} + \frac{3 \,C_{{ES}^2}^{(1)} G \,m_{2}^3 \,(nS_1) \,(S_1nv) \,\bv^i}{\,m^2 \,m_{1} r^3} -  \frac{6 G \,m_{1} \,m_{2} \,(nS_1) \,(S_2nv) \,\bv^i}{\,m^2 r^3} -  \frac{3 G \,m_{2}^2 \,(nS_1) \,(S_2nv) \,\bv^i}{\,m^2 r^3} 
+ \frac{9 \,C_{{ES}^2}^{(1)} G \,m_{1} \,m_{2} \,(\bn\times \bS_1)^{i} \,(nS_1) \bv^2}{2 \,m^2 r^3} + \frac{21 \,C_{{ES}^2}^{(1)} G \,m_{2}^2 \,(\bn\times \bS_1)^{i} \,(nS_1) \bv^2}{2 \,m^2 r^3} + \frac{9 \,C_{{ES}^2}^{(1)} G \,m_{2}^3 \,(\bn\times \bS_1)^{i} \,(nS_1) \bv^2}{2 \,m^2 \,m_{1} r^3}
+ \frac{3 G \,m_{1}^2 \,(\bn\times \bS_1)^{i} \,(nS_2) \bv^2}{\,m^2 r^3} + \frac{9 G \,m_{1} \,m_{2} \,(\bn\times \bS_1)^{i} \,(nS_2) \bv^2}{2 \,m^2 r^3} + \frac{2 G \,m_{1}^2 \,(\bS_1\times \bS_2)^{i} \bv^2}{\,m^2 r^3} + \frac{5 G \,m_{1} \,m_{2} \,(\bS_1\times \bS_2)^{i} \bv^2}{2 \,m^2 r^3} 
-  \frac{6 \,C_{{ES}^2}^{(1)} G \,m_{2}^2 \,(nS_1) \,(nv) \,(\bv\times \bS_1)^{i}}{\,m^2 r^3} -  \frac{6 \,C_{{ES}^2}^{(1)} G \,m_{2}^3 \,(nS_1) \,(nv) \,(\bv\times \bS_1)^{i}}{\,m^2 \,m_{1} r^3} -  \frac{3 G \,m_{1}^2 \,(nS_2) \,(nv) \,(\bv\times \bS_1)^{i}}{\,m^2 r^3} 
-  \frac{3 G \,m_{1} \,m_{2} \,(nS_2) \,(nv) \,(\bv\times \bS_1)^{i}}{\,m^2 r^3} -  \frac{3 G \,m_{1} \,m_{2} \,\bv^i \,(vS_1S_2)}{\,m^2 r^3} -  \frac{G \,m_{2}^2 \,\bv^i \,(vS_1S_2)}{\,m^2 r^3} -  \frac{6 \,C_{{ES}^2}^{(1)} G \,m_{2}^2 \,(\bn\times \bS_1)^{i} \,(nv) \,(vS_1)}{\,m^2 r^3} 
-  \frac{3 G \,m_{2}^2 \,(\bn\times \bS_1)^{i} \,(nv) \,(vS_1)}{\,m \,m_{1} r^3} -  \frac{3 \,C_{{ES}^2}^{(1)} G \,m_{2}^3 \,(\bn\times \bS_1)^{i} \,(nv) \,(vS_1)}{\,m^2 \,m_{1} r^3} -  \frac{6 G \,m_{2}^2 \,\bn^i \,(S_1nv) \,(vS_1)}{\,m \,m_{1} r^3} 
+ \frac{3 \,C_{{ES}^2}^{(1)} G \,m_{2}^2 \,(\bv\times \bS_1)^{i} \,(vS_1)}{\,m^2 r^3} + \frac{3 G \,m_{2}^2 \,(\bv\times \bS_1)^{i} \,(vS_1)}{\,m \,m_{1} r^3} + \frac{3 \,C_{{ES}^2}^{(1)} G \,m_{2}^3 \,(\bv\times \bS_1)^{i} \,(vS_1)}{\,m^2 \,m_{1} r^3}
-  \frac{3 G \,m_{1}^2 \,(\bn\times \bS_1)^{i} \,(nv) \,(vS_2)}{\,m^2 r^3} -  \frac{6 G \,m_{1} \,m_{2} \,(\bn\times \bS_1)^{i} \,(nv) \,(vS_2)}{\,m^2 r^3} + \frac{2 G \,m_{1}^2 \,(\bv\times \bS_1)^{i} \,(vS_2)}{\,m^2 r^3} 
+ \frac{2 G \,m_{1} \,m_{2} \,(\bv\times \bS_1)^{i} \,(vS_2)}{\,m^2 r^3}\Bigg]_\text{NLO SS}+\cdots\,,
\end{autobreak}
\end{align}
and similarly for particle 2 after exchanging $1\leftrightarrow 2$.

\subsection{Source multipoles}
Using the results in \cite{srad} and translating to CoM coordinates, without yet performing the STF operation, we find the following expressions for the source multipoles needed to 3PN:
\begin{subequations}
\begin{align}
\begin{autobreak}
I^{ij}_\text{(2PN)}
=\frac{1}{2\,m}\Bigg\{\,(- \,\delta \,\kappa_{-} -  \,\kappa_{+}) \,\bS^i\, \bS^j + 4 \,\kappa_{-} \,\nu\, \bS^i\, \bSigma^j + \,(\,\delta \,\kappa_{-} \,\nu -  \,\kappa_{+} \,\nu)\, \bSigma^i\, \bSigma^j\Bigg\}\,,
\end{autobreak}
\\[2ex]
\begin{autobreak}
I^{ij}_\text{(3PN)}
=\frac{\nu}{84\,m}\,\Bigg\{ \,\bS^i\bn^j \,\Big[\,(nS) \,\big(\frac{84 \,\delta G \,\kappa_{-} \,m}{r} -  \frac{12 G \,m \,(11 + 23 \,\nu)}{r} -  \frac{6 G \,\kappa_{+} \,m \,(25 + 23 \,\nu)}{r}\big) 
+ \,(n\Sigma) \,\big(- \frac{46 \,\delta G \,m \,(-1 + 3 \,\nu)}{r} -  \frac{9 G \,\kappa_{-} \,m \,(-13 + 11 \,\nu)}{r} -  \frac{3 \,\delta G \,\kappa_{+} \,m \,(39 + 23 \,\nu)}{r}\big)\Big]  
+ \,\bn^i\bS^j \,\Big[\,(n\Sigma) \,\big(\frac{66 \,\delta G \,m \,(-1 + 3 \,\nu)}{r} + \frac{3 G \,\kappa_{-} \,m \,(-9 + 7 \,\nu)}{r} + \frac{9 \,\delta G \,\kappa_{+} \,m \,(3 + 11 \,\nu)}{r}\big)
+ \,(nS) \,\big(- \frac{60 \,\delta G \,\kappa_{-} \,m}{r} + \frac{132 G \,m \,(-1 + 3 \,\nu)}{r} + \frac{6 G \,\kappa_{+} \,m \,(-1 + 33 \,\nu)}{r}\big)\Big] 
+ \bS^i \bS^j \,\big(\frac{34 \,\delta G \,\kappa_{-} \,m}{r} + \frac{218 G \,\kappa_{+} \,m}{r} + \,(-13 \,\delta \,\kappa_{-} + 13 \,\kappa_{+}) \,\bv^2\big)
+  \,\bv^i\bS^j \,\Big[\,(-24 \,\delta \,\kappa_{-} + 24 \,\kappa_{+}) \,(vS) + \,\big(24 \,\delta \,\kappa_{+} + \,\kappa_{-} \,(-24 + 48 \,\nu)\big) \,(v\Sigma)\Big]
+ \, \bSigma^i\bn^j \,\Big[\,(nS) \,\big(- \frac{9 G \,\kappa_{-} \,m \,(-13 + 11 \,\nu)}{r} -  \frac{3 \,\delta G \,\kappa_{+} \,m \,(39 + 23 \,\nu)}{r} -  \frac{2 \,\delta G \,m \,(89 + 69 \,\nu)}{r}\big) 
+ \,(n\Sigma) \,\big(- \frac{3 \,\delta G \,\kappa_{-} \,m \,(-39 + 5 \,\nu)}{r} + \frac{4 G \,m \,\nu \,(61 + 69 \,\nu)}{r} + \frac{3 G \,\kappa_{+} \,m \,(-39 + 83 \,\nu + 46 \,\nu^2)}{r}\big)\Big] 
+ \bSigma^i \bS^j \,\Big[\frac{92 \,\delta G \,\kappa_{+} \,m}{r} -  \frac{4 G \,\kappa_{-} \,m \,(23 + 17 \,\nu)}{r} + \,\big(13 \,\delta \,\kappa_{+} + \,\kappa_{-} \,(-13 + 26 \,\nu)\big) \,\bv^2\Big] 
- \bSigma^i\bv^j \,28 \,\nu \,(v\Sigma) 
+ \,\bn^i \bSigma^j\,\Big[\,(nS) \,\big(\frac{66 \,\delta G \,m \,(-1 + 3 \,\nu)}{r} + \frac{3 G \,\kappa_{-} \,m \,(-9 + 7 \,\nu)}{r} + \frac{9 \,\delta G \,\kappa_{+} \,m \,(3 + 11 \,\nu)}{r}\big)
+ \,(n\Sigma) \,\big(- \frac{3 \,\delta G \,\kappa_{-} \,m \,(9 + 13 \,\nu)}{r} -  \frac{3 G \,\kappa_{+} \,m \,(-1 + 3 \,\nu) \,(9 + 22 \,\nu)}{r} -  \frac{4 G \,m \,\nu \,(-5 + 99 \,\nu)}{r}\big)\Big]
+ \bS^i\bSigma^j \,\Big[\frac{92 \,\delta G \,\kappa_{+} \,m}{r} -  \frac{4 G \,\kappa_{-} \,m \,(23 + 17 \,\nu)}{r} + \,\big(13 \,\delta \,\kappa_{+} + \,\kappa_{-} \,(-13 + 26 \,\nu)\big) \,\bv^2\Big] 
+ \,\bv^i\bSigma^j\,\Big[\,\big(24 \,\delta \,\kappa_{+} + \,\kappa_{-} \,(-24 + 48 \,\nu)\big) \,(vS) + \,\big(\,\kappa_{+} \,(24 - 72 \,\nu) + 24 \,\delta \,\kappa_{-} \,(-1 + \,\nu) - 140 \,\nu\big) \,(v\Sigma)\Big] 
+ \bSigma^i \bSigma^j \,\Big[\frac{112 G \,m \,\nu}{r} -  \frac{2 \,\delta G \,\kappa_{-} \,m \,(46 + 17 \,\nu)}{r} -  \frac{2 G \,\kappa_{+} \,m \,(-46 + 75 \,\nu)}{r}
+ \,\big(\,\kappa_{+} \,(13 - 39 \,\nu) + 13 \,\delta \,\kappa_{-} \,(-1 + \,\nu) + 140 \,\nu\big) \,\bv^2\Big] 
+ \,\bv^i \,\bv^j \,\Big[\,(22 \,\delta \,\kappa_{-} - 22 \,\kappa_{+}) \,\bS^2 + \,\big(-44 \,\delta \,\kappa_{+} + \,\kappa_{-} \,(44 - 88 \,\nu)\big) \,(S\Sigma) 
+ \,\big(-22 \,\delta \,\kappa_{-} \,(-1 + \,\nu) + 28 \,\nu + \,\kappa_{+} \,(-22 + 66 \,\nu)\big) \,\bSigma^2\Big]
+ \,\bn^i \,\bn^j \,\Big[\,(nS)^2 \,\big(\frac{42 \,\delta G \,\kappa_{-} \,m}{r} -  \frac{24 G \,m \,(-9 + 20 \,\nu)}{r} -  \frac{12 G \,\kappa_{+} \,m \,(-9 + 20 \,\nu)}{r}\big) 
+ \,(nS) \,(n\Sigma) \,\big(\frac{6 G \,\kappa_{-} \,m \,(-11 + 12 \,\nu)}{r} -  \frac{24 \,\delta G \,m \,(-9 + 20 \,\nu)}{r} -  \frac{6 \,\delta G \,\kappa_{+} \,m \,(-11 + 40 \,\nu)}{r}\big)
+ \,(n\Sigma)^2 \,\big(\frac{24 G \,m \,\nu \,(-9 + 20 \,\nu)}{r} + \frac{3 \,\delta G \,\kappa_{-} \,m \,(-11 + 26 \,\nu)}{r} + \frac{3 G \,\kappa_{+} \,m \,(11 - 48 \,\nu + 80 \,\nu^2)}{r}\big)
+ \,\big(- \frac{22 \,\delta G \,\kappa_{-} \,m}{r} + \frac{24 G \,m \,(-4 + 5 \,\nu)}{r} + \frac{4 G \,\kappa_{+} \,m \,(4 + 15 \,\nu)}{r}\big) \,\bS^2 
+ \,\big(\frac{24 \,\delta G \,m \,(-4 + 5 \,\nu)}{r} + \frac{2 G \,\kappa_{-} \,m \,(-19 + 14 \,\nu)}{r} + \frac{2 \,\delta G \,\kappa_{+} \,m \,(19 + 30 \,\nu)}{r}\big) \,(S\Sigma)
+ \,\big(- \frac{\,\delta G \,\kappa_{-} \,m \,(19 + 8 \,\nu)}{r} -  \frac{8 G \,m \,\nu \,(2 + 15 \,\nu)}{r} -  \frac{G \,\kappa_{+} \,m \,(-19 + 30 \,\nu + 60 \,\nu^2)}{r}\big) \,\bSigma^2\Big]
+ \delta^{ij} \,\big(\frac{112 G \,m \,\nu \,(n\Sigma)^2}{r} + 140 \,\nu \,(v\Sigma)^2 + \,(- \frac{112 G \,m \,\nu}{r} - 140 \,\nu \,\bv^2) \,\bSigma^2\big)\Bigg\}\,,
\end{autobreak}
\\[2ex]
\begin{autobreak}
I^{ijk}_\text{(2PN)}
=\frac{3\,\nu}{2\,m}\,\Bigg\{-2 \,\kappa_{-}  \, \bS^i\, \bS^j\,\br^{k}
+ \,(- \,\delta \,\kappa_{-}+ \,\kappa_{+} ) \, \bSigma^i\,\bS^j\,\br^{k} 
+ \,(- \,\delta \,\kappa_{-} + \,\kappa_{+} )\, \bS^i\, \bSigma^j\,\br^{k}
+ \,\big(\,\delta \,\kappa_{+} \, + \,\kappa_{-} \,(-1 + 2 \,\nu) \big) \, \bSigma^i\,\bSigma^j\,\br^{k}\Bigg\}\,,
\end{autobreak}
\\[2ex]
\begin{autobreak}
\label{Jij}
J^{ij}_\text{(2PN)}
=\frac{\nu}{2\,m}\Bigg\{2 \,\kappa_{-} (\bv\times\bS)^{i}\,\bS^j 
+ \,(\,\delta \,\kappa_{-} -  \,\kappa_{+})\,(\bv\times\bSigma)^{i} \,\bS^j
+ 3 \bS^i \,(\bv\times\bSigma)^{j} + \,(\,\delta \,\kappa_{-} -  \,\kappa_{+}) \,(\bv\times \bS)^{i} \,\bSigma^j 
+ \,(\bv\times\bSigma)^{i} \,\bSigma^j\,\big(- \,\delta \,\kappa_{+} + \,\kappa_{-} \,(1 - 2 \,\nu)\big) \Bigg\}\,.
\end{autobreak}
\end{align}
\end{subequations}

\section{Conservative sector}\label{app:B}
\subsection{Binding energy }

The coefficients of the binding energy in the CoM in \eqref{3pnss} are given by
\begin{subequations}
\begin{align}
\begin{autobreak}
e^0_4
=\,(12 + 6 \,\kappa_{+}) \,(nS)^2 + \,(12 \,\delta - 6 \,\kappa_{-} + 6 \,\delta \,\kappa_{+}) \,(nS) \,(n\Sigma) 
+ \big(-3 \,\delta \,\kappa_{-} + \,\kappa_{+} \,(3 - 6 \,\nu) - 12 \,\nu\big) \,(n\Sigma)^2
+ \,(-4 - 2 \,\kappa_{+}) \,\bS^2 + \,(-4 \,\delta + 2 \,\kappa_{-} - 2 \,\delta \,\kappa_{+}) \,(S\Sigma) 
+ \big(\,\delta \,\kappa_{-} + 4 \,\nu + \,\kappa_{+} \,(-1 + 2 \,\nu)\big) \,\bSigma^2
\end{autobreak}
\\[2ex]
\begin{autobreak}
e^0_6
=\,(-72 \,\delta - 24 \,\kappa_{-} + 24 \,\delta \,\kappa_{+}) \,(nv) \,(n\Sigma) \,(vS) 
+ 48 \,\kappa_{-} \,\nu \,(nv) \,(n\Sigma) \,(vS)
+ \big(8 \,\delta \,\kappa_{-} + 4 \,\kappa_{+} \,(-2 + \,\nu) + 8 \,(3 + \,\nu)\big) \,(vS)^2
+ \big(30 \,\kappa_{-} + 6 \,\delta \,\kappa_{+} \,(-5 + \,\nu)
+ 12 \,\delta \,(8 + \,\nu)\big) \,(nS) \,(n\Sigma) \,\bv^2
+ \,(nS)^2 \Big[\,(-60 \,\nu - 30 \,\kappa_{+} \,\nu) \,(nv)^2
+ \big(6 \,\delta \,\kappa_{-} + 6 \,\kappa_{+} \,(-4 + \,\nu) + 12 \,(4 + \,\nu)\big) \,\bv^2\Big]
+ \,\bS^2 \Big[\big(6 \,\delta \,\kappa_{-} + 6 \,\kappa_{+} \,(-1 + \,\nu) + 12 \,(2 + \,\nu)\big) \,(nv)^2 
+ \big(-4 \,\delta \,\kappa_{-} - 2 \,\kappa_{+} \,(-5 + \,\nu) - 4 \,(6 + \,\nu)\big) \,\bv^2\Big]
+ \,(S\Sigma) \Big[\big(\,\kappa_{-} \,(12 - 30 \,\nu) + 6 \,\delta \,\kappa_{+} \,(-2 + \,\nu)
+ 12 \,\delta \,(4 + \,\nu)\big) \,(nv)^2 + \big(-2 \,\delta \,\kappa_{+} \,(-7 + \,\nu) 
- 4 \,\delta \,(12 + \,\nu) + 2 \,\kappa_{-} \,(-7 + 9 \,\nu)\big) \,\bv^2\Big]
+ \,(n\Sigma)^2 \Big[\big(15 \,\delta \,\kappa_{-} \,\nu + 60 \,\nu^2 + 15 \,\kappa_{+} \,\nu \,(-1 + 2 \,\nu)\big) \,(nv)^2 
+ \big(-3 \,\delta \,\kappa_{-} \,(-5 + 3 \,\nu) + \,\kappa_{+} \,(-15 + 39 \,\nu - 6 \,\nu^2)
- 12 \,(-4 + 12 \,\nu + \,\nu^2)\big) \,\bv^2\Big]
+ \big(\,\kappa_{+} \,(24 - 72 \,\nu) + 24 \,\delta \,\kappa_{-} \,(-1 + \,\nu)
+ 72 \,(-1 + 3 \,\nu)\big) \,(nv) \,(n\Sigma) \,(v\Sigma)
+ \big(\,\kappa_{-} \,(16 - 36 \,\nu) + 4 \,\delta \,\kappa_{+} \,(-4 + \,\nu) 
+ 8 \,\delta \,(6 + \,\nu)\big) \,(vS) \,(v\Sigma)
+ \big(-2 \,\delta \,\kappa_{-} \,(-4 + 5 \,\nu) + \,\kappa_{+} \,(-8 + 26 \,\nu - 4 \,\nu^2)
- 8 \,(-3 + 9 \,\nu + \,\nu^2)\big) \,(v\Sigma)^2
+ \,(nS) \Big[\,(-72 - 24 \,\delta \,\kappa_{-} + 24 \,\kappa_{+}) \,(nv) \,(vS) 
+ \,(n\Sigma) \big(\,(-60 \,\delta \,\nu + 30 \,\kappa_{-} \,\nu
- 30 \,\delta \,\kappa_{+} \,\nu) \,(nv)^2 - 30 \,\kappa_{-} \,\nu \,\bv^2\big)
+ \big(-72 \,\delta + 24 \,\delta \,\kappa_{+} + \,\kappa_{-} \,(-24 + 48 \,\nu)\big) \,(nv) \,(v\Sigma)\Big]
+ \Big[\big(\,\delta \,\kappa_{-} \,(6 - 9 \,\nu) + \,\kappa_{+} \,(-6 + 21 \,\nu - 6 \,\nu^2) - 12 \,(-2 + 6 \,\nu + \,\nu^2)\big) \,(nv)^2
+ \big(\,\delta \,\kappa_{-} \,(-7 + 5 \,\nu) + 4 \,(-6 + 18 \,\nu + \,\nu^2)
+ \,\kappa_{+} \,(7 - 19 \,\nu + 2 \,\nu^2)\big) \,\bv^2\Big] \,\bSigma^2\,,
\end{autobreak}
\\[2ex]
\begin{autobreak}
e_6^1
=\,(-36 + 9 \,\delta \,\kappa_{-} - 15 \,\kappa_{+}) \,(nS)^2
+ \big(-32 \,\delta - 24 \,\delta \,\kappa_{+} + \,\kappa_{-} \,(24 - 36 \,\nu)\big) \,(nS) \,(n\Sigma)
+ \big(30 \,\nu - 3 \,\delta \,\kappa_{-} \,(-4 + 3 \,\nu) + \,\kappa_{+} \,(-12 + 33 \,\nu)\big) \,(n\Sigma)^2
+ \,(8 - 3 \,\delta \,\kappa_{-} + 5 \,\kappa_{+}) \,\bS^2 
+ \big(8 \,\delta + 8 \,\delta \,\kappa_{+} + \,\kappa_{-} \,(-8 + 12 \,\nu)\big) \,(S\Sigma)
+ \big(\,\kappa_{+} \,(4 - 11 \,\nu) - 10 \,\nu + \,\delta \,\kappa_{-} \,(-4 + 3 \,\nu)\big) \,\bSigma^2\,.
\end{autobreak}
\end{align}
\end{subequations}

\subsection{Angular momentum }

The coefficients of the angular momentum in the CoM in \eqref{3pnLss1} and \eqref{3pnLss2} are given by
\begin{subequations}
\begin{align}
\begin{autobreak}
\ell^0_5
=\,\bS^i \,\big(\,(4 + 5 \,\nu) \,(nv)^2 - 6 \,\bv^2\big) 
+ \,\bv^i \,\big(\,(-6 + 12 \,\nu) \,(nS) \,(nv) + 4 \,\delta \,\nu \,(nv) \,(n\Sigma) + 4 \,(vS) + 2 \,\delta \,\nu \,(v\Sigma)\big) 
+ \,\bn^i \,\big(\,(-2 - 8 \,\nu) \,(nv) \,(vS) + \,(nS) \,(-9 \,\nu \,(nv)^2 + 6 \,\bv^2) + \,(n\Sigma) \,(-6 \,\delta \,\nu \,(nv)^2 + 3 \,\delta \,\nu \,\bv^2) - 6 \,\delta \,\nu \,(nv) \,(v\Sigma)\big)\,,
\end{autobreak}
\\[2ex]
\begin{autobreak}
\ell^1_5
=\,\bn^i \,\big(\,(-1 - 4 \,\nu) \,(nS) - (\,\delta  + 3 \,\delta  \,\nu) \,(n\Sigma)\big) + \,\bS^i\,(1 + 4 \,\nu)  +\bSigma^i\, (\delta  + 3 \,\delta  \,\nu) 
\end{autobreak}
\\[2ex]
\begin{autobreak}
\ell^0_6
= \,\Big[\,\big(6 + \,\delta \,\kappa_{-} + \,\kappa_{+} \,(-1 + 2 \,\nu)\big) \,(nS)
+ \,\big(6 \,\delta + \,\kappa_{-} \,(1 - 3 \,\nu) + \,\delta \,\kappa_{+} \,(-1 + \,\nu) - 2 \,\nu\big) \,(n\Sigma)\Big] \,(\bv\times\bS)^{i} 
-6 \,\delta \,(Sn\Sigma) \,\bv^i+ \,\Big[\,\big(\,\kappa_{-} \,(1 - 3 \,\nu) + \,\delta \,\kappa_{+} \,(-1 + \,\nu) + 2 \,(3 \,\delta + \,\nu)\big) \,(nS)
+ \,\big(6 - 16 \,\nu + \,\kappa_{-} \,(\,\delta - 2 \,\delta \,\nu) + \,\kappa_{+} \,(-1 + 4 \,\nu - 2 \,\nu^2)\big) \,(n\Sigma)\Big] \,(\bv\times\bSigma)^{i} 
+ \,\bS^i \,\Big[\,\big(- \,\delta \,\kappa_{-} + \,\kappa_{+} \,(1 - 2 \,\nu) + 6 \,(2 + \,\nu)\big) \,(Snv) + \,\big(\,\kappa_{-} \,(1 - 3 \,\nu) + \,\delta \,\kappa_{+} \,(-1 + \,\nu) -  \,\nu - 3 \,\delta \,(2 + \,\nu)\big) \,(vn\Sigma)\Big]
+ \,\bn^i \,\bigg\{6 \,\delta \,(nv) \,(Sn\Sigma) + \,(n\Sigma) \,\Big[\,\big(3 \,\nu - 3 \,\delta \,(6 + \,\nu)\big) \,(Snv) - 6 \,(-2 + 7 \,\nu + \,\nu^2) \,(vn\Sigma)\Big]
+ \,(nS) \,\Big[-6 \,(3 + \,\nu) \,(Snv) + 3 \,\big(\,\nu + \,\delta \,(4 + \,\nu)\big) \,(vn\Sigma)\Big]\bigg\}
+ \,(\bn\times\bS)^{i} \,\bigg\{\,\big(-3 \,\delta \,\kappa_{-} + 6 \,(-3 + \,\nu) + \,\kappa_{+} \,(3 + 6 \,\nu)\big) \,(nS) \,(nv)
+ \,\Big[3 \,\kappa_{-} \,(-1 + \,\nu) + 3 \,\delta \,\kappa_{+} \,(1 + \,\nu) + 3 \,\big(\,\delta \,(-6 + \,\nu) + \,\nu\big)\Big] \,(nv) \,(n\Sigma)
+ \,\big(14 + 2 \,\delta \,\kappa_{-} + 6 \,\nu - 2 \,\kappa_{+} \,(1 + \,\nu)\big) \,(vS) 
+ \,\big(\,\kappa_{-} \,(2 - 3 \,\nu) -  \,\nu -  \,\delta \,\kappa_{+} \,(2 + \,\nu) + 3 \,\delta \,(4 + \,\nu)\big) \,(v\Sigma)\bigg\}
+ \,(\bn\times\bSigma)^{i} \,\Big[\,\big(3 \,\delta \,(-6 + \,\nu) + 3 \,\kappa_{-} \,(-1 + \,\nu) - 3 \,\nu + 3 \,\delta \,\kappa_{+} \,(1 + \,\nu)\big) \,(nS) \,(nv) 
+ \,\big(-3 \,\delta \,\kappa_{-} + \,\kappa_{+} \,(3 - 6 \,\nu - 6 \,\nu^2)- 6 \,(3 - 9 \,\nu + \,\nu^2)\big) \,(nv) \,(n\Sigma) 
+ \,\big(\,\kappa_{-} \,(2 - 3 \,\nu) + \,\nu -  \,\delta \,\kappa_{+} \,(2 + \,\nu) + \,\delta \,(14 + 3 \,\nu)\big) \,(vS)
 + \,\big(12 -  \,\delta \,\kappa_{-} \,(-2 + \,\nu) - 40 \,\nu - 6 \,\nu^2 + \,\kappa_{+} \,(-2 + 5 \,\nu + 2 \,\nu^2)\big) \,(v\Sigma)\Big]
+ \,\Big[\,\big(- \,\nu + 3 \,\delta \,(4 + \,\nu) + \,\kappa_{-} \,(-1 + 3 \,\nu) + \,\kappa_{+} \,(\,\delta -  \,\delta \,\nu)\big) \,(Snv)
+ \,\big(-6 + 26 \,\nu + 6 \,\nu^2 + \,\kappa_{-} \,(\,\delta - 2 \,\delta \,\nu) + \,\kappa_{+} \,(-1 + 4 \,\nu - 2 \,\nu^2)\big) \,(vn\Sigma)\Big] \,\bSigma^i
+ \,\big(\bn\times\bv\big)^{i} \,\Big[\,\big(24 - 3 \,\kappa_{+} \,(3 + \,\nu)\big) \,(nS)^2 + \,\big(48 \,\delta + 3 \,\kappa_{-} \,(3 + \,\nu) - 3 \,\delta \,\kappa_{+} \,(3 + \,\nu)\big) \,(nS) \,(n\Sigma) 
+ \,\big(24 - 72 \,\nu + \tfrac{3}{2} \,\delta \,\kappa_{-} \,(3 + \,\nu) + \tfrac{3}{2} \,\kappa_{+} \,(3 + \,\nu) \,(-1 + 2 \,\nu)\big) \,(n\Sigma)^2 + \,\big(3 \,\kappa_{+} \,(1 + \,\nu) - 2 \,(7 + 2 \,\nu)\big) \,\bS^2 
+ \,\big(-3 \,\kappa_{-} \,(1 + \,\nu) + 3 \,\delta \,\kappa_{+} \,(1 + \,\nu) - 2 \,\delta \,(13 + 2 \,\nu)\big) \,(S\Sigma)
+ \,\big(- \tfrac{3}{2} \,\delta \,\kappa_{-} \,(1 + \,\nu) -  \tfrac{3}{2} \,\kappa_{+} \,(1 + \,\nu) \,(-1 + 2 \,\nu) + 4 \,(-3 + 10 \,\nu + \,\nu^2)\big) \,\bSigma^2\Big]\,.
\end{autobreak}
\end{align}
\end{subequations}

\section{Radiation sector}\label{app:C} 

\subsection{Energy flux }
The coefficients of the radiated energy in \eqref{eq:energyflux3PN} are given by
\begin{subequations}
\begin{align}
\begin{autobreak}
f_6^0
=\,(vS)^2 \Big[\big(-8688 \,\delta \,\kappa_{-} + \,\kappa_{+} \,(11424 - 9720 \,\nu) - 6 \,(529 + 3240 \,\nu)\big) \,(nv)^2
+ \big(4462 + 2676 \,\delta \,\kappa_{-} + 3792 \,\nu + \,\kappa_{+} \,(-684 + 1896 \,\nu)\big) \,\bv^2\Big]
+ \,(nS) \,(vS) \Big[\big(25464 \,\delta \,\kappa_{-} + 3384 \,\kappa_{+} \,(-17 + 12 \,\nu) + 48 \,(-581 + 1692 \,\nu)\big) \,(nv)^3 
+ \big(4800 - 11832 \,\delta \,\kappa_{-} + \,\kappa_{+} \,(22200 - 8712 \,\nu) - 17424 \,\nu\big) \,(nv) \,\bv^2\Big] 
+ \,(n\Sigma) \,(vS) \Big[\big(\,\kappa_{-} \,(41496 - 71232 \,\nu) + 24 \,\delta \,\kappa_{+} \,(-1729 + 846 \,\nu) + 12 \,\delta \,(1273 + 3384 \,\nu)\big) \,(nv)^3
+ \big(-66 \,\delta \,(161 + 132 \,\nu) - 12 \,\delta \,\kappa_{+} \,(-1418 + 363 \,\nu) + \,\kappa_{-} \,(-17016 + 28020 \,\nu)\big) \,(nv) \,\bv^2\Big]
+ \,\bS^2 \Big[\big(-35436 - 2274 \,\delta \,\kappa_{-} + 28224 \,\nu + 6 \,\kappa_{+} \,(-2153 + 2352 \,\nu)\big) \,(nv)^4 
+ \big(49800 + 3684 \,\delta \,\kappa_{-} + \,\kappa_{+} \,(16452 - 15744 \,\nu) - 31488 \,\nu\big) \,(nv)^2 \,\bv^2 
+ \big(-16260 - 1374 \,\delta \,\kappa_{-} + 5136 \,\nu + 6 \,\kappa_{+} \,(-671 + 428 \,\nu)\big) \,(\bv^2)^2\Big]
+ \,(nS) \,(n\Sigma) \Big[\big(-60 \,\delta \,\kappa_{+} \,(-1725 + 1192 \,\nu) - 120 \,\delta \,(-591 + 1192 \,\nu) + 60 \,\kappa_{-} \,(-1725 + 2054 \,\nu)\big) \,(nv)^4
+ \big(-2448 \,\kappa_{-} \,(-32 + 25 \,\nu) + 288 \,\delta \,\kappa_{+} \,(-272 + 179 \,\nu) + 6 \,\delta \,(-10043 + 17184 \,\nu)\big) \,(nv)^2 \,\bv^2
+ \big(-12 \,\kappa_{-} \,(691 + 438 \,\nu) - 12 \,\delta \,\kappa_{+} \,(-691 + 548 \,\nu) - 24 \,\delta \,(-67 + 548 \,\nu)\big) \,(\bv^2)^2\Big]
+ \,(nS)^2 \Big[\big(81570 - 12930 \,\delta \,\kappa_{-} + \,\kappa_{+} \,(90570 - 71520 \,\nu) - 143040 \,\nu\big) \,(nv)^4
+ \big(2412 \,\delta \,\kappa_{-} + 12 \,(-4483 + 8592 \,\nu) + \,\kappa_{+} \,(-75924 + 51552 \,\nu)\big) \,(nv)^2 \,\bv^2
+ \big(2958 \,\delta \,\kappa_{-} + \,\kappa_{+} \,(11250 - 6576 \,\nu) - 6 \,(85 + 2192 \,\nu)\big) \,(\bv^2)^2\Big]
+ \,(S\Sigma) \Big[\big(\,\kappa_{-} \,(10644 - 5016 \,\nu) + 252 \,\delta \,(-201 + 112 \,\nu) + 12 \,\delta \,\kappa_{+} \,(-887 + 1176 \,\nu)\big) \,(nv)^4 
+ \big(336 \,\kappa_{-} \,(-38 + 3 \,\nu) - 96 \,\delta \,\kappa_{+} \,(-133 + 164 \,\nu) - 6 \,\delta \,(-11269 + 5248 \,\nu)\big) \,(nv)^2 \,\bv^2
+ \big(12 \,\delta \,\kappa_{+} \,(-221 + 214 \,\nu) + 2 \,\delta \,(-9377 + 2568 \,\nu) + \,\kappa_{-} \,(2652 + 2928 \,\nu)\big) \,(\bv^2)^2\Big]
+ \,(n\Sigma)^2 \Big[\big(90 \,\delta \,\kappa_{-} \,(-575 + 541 \,\nu) + 30 \,\kappa_{+} \,(1725 - 5073 \,\nu + 2384 \,\nu^2) + 60 \,(-307 - 1016 \,\nu + 2384 \,\nu^2)\big) \,(nv)^4 
+ \big(8238 + 67806 \,\nu - 103104 \,\nu^2 - 36 \,\delta \,\kappa_{-} \,(-1088 + 783 \,\nu) - 36 \,\kappa_{+} \,(1088 - 2959 \,\nu + 1432 \,\nu^2)\big) \,(nv)^2 \,\bv^2
+ \big(6 \,\delta \,\kappa_{-} \,(-691 + 55 \,\nu) + 6 \,(-515 - 638 \,\nu + 2192 \,\nu^2) + \,\kappa_{+} \,(4146 - 8622 \,\nu + 6576 \,\nu^2)\big) \,(\bv^2)^2\Big]
+ \,(vS) \Big[\big(-24 \,\delta \,\kappa_{+} \,(-838 + 405 \,\nu) - 12 \,\delta \,(1543 + 1620 \,\nu) + 24 \,\kappa_{-} \,(-838 + 1853 \,\nu)\big) \,(nv)^2
+ \big(-840 \,\kappa_{-} \,(-4 + 15 \,\nu) + 24 \,\delta \,\kappa_{+} \,(-140 + 79 \,\nu) + 2 \,\delta \,(3065 + 1896 \,\nu)\big) \,\bv^2\Big] \,(v\Sigma)
+ \,(nS) \Big[\big(\,\kappa_{-} \,(41496 - 71232 \,\nu) + 24 \,\delta \,\kappa_{+} \,(-1729 + 846 \,\nu) + 48 \,\delta \,(-233 + 846 \,\nu)\big) \,(nv)^3
+ \big(-12 \,\delta \,\kappa_{+} \,(-1418 + 363 \,\nu) - 6 \,\delta \,(-2023 + 1452 \,\nu) + \,\kappa_{-} \,(-17016 + 28020 \,\nu)\big) \,(nv) \,\bv^2\Big] \,(v\Sigma)
+ \,(n\Sigma) \Big[\big(-24 \,\delta \,\kappa_{-} \,(-1729 + 1907 \,\nu) - 192 \,(-174 + 193 \,\nu + 423 \,\nu^2) - 24 \,\kappa_{+} \,(1729 - 5365 \,\nu + 1692 \,\nu^2)\big) \,(nv)^3
+ \big(12 \,\delta \,\kappa_{-} \,(-1418 + 1349 \,\nu) + 12 \,\kappa_{+} \,(1418 - 4185 \,\nu + 726 \,\nu^2) + 12 \,(-652 + 275 \,\nu + 1452 \,\nu^2)\big) \,(nv) \,\bv^2\Big] \,(v\Sigma)
+ \Big[\big(12 \,\delta \,\kappa_{-} \,(-838 + 1129 \,\nu) + 48 \,(-296 + 668 \,\nu + 405 \,\nu^2) + 12 \,\kappa_{+} \,(838 - 2805 \,\nu + 810 \,\nu^2)\big) \,(nv)^2 
+ \big(2200 - 7646 \,\nu - 3792 \,\nu^2 - 24 \,\delta \,\kappa_{-} \,(-70 + 151 \,\nu) - 24 \,\kappa_{+} \,(70 - 291 \,\nu + 79 \,\nu^2)\big) \,\bv^2\Big] \,(v\Sigma)^2 
+ \Big[\big(-9714 + 64788 \,\nu - 28224 \,\nu^2 - 6 \,\delta \,\kappa_{-} \,(-887 + 797 \,\nu) - 6 \,\kappa_{+} \,(887 - 2571 \,\nu + 2352 \,\nu^2)\big) \,(nv)^4 
+ \big(12 \,\delta \,\kappa_{-} \,(-532 + 349 \,\nu) + 12 \,\kappa_{+} \,(532 - 1413 \,\nu + 1312 \,\nu^2) + 6 \,(2402 - 14085 \,\nu + 5248 \,\nu^2)\big) \,(nv)^2 \,\bv^2 
+ \big(-3754 + 21194 \,\nu - 5136 \,\nu^2 + 6 \,\delta \,\kappa_{-} \,(221 + 15 \,\nu) - 6 \,\kappa_{+} \,(221 - 427 \,\nu + 428 \,\nu^2)\big) \,(\bv^2)^2\Big] \,\bSigma^2\,,
\end{autobreak}
\\[2ex]
\begin{autobreak}
f^1_6
=\big(20472 \,\delta \,\kappa_{-} - 24 \,(5125 + 161 \,\nu) - 12 \,\kappa_{+} \,(5338 + 161 \,\nu)\big) \,(nS) \,(nv) \,(vS) 
+ \big(\,\kappa_{-} \,(42264 - 39978 \,\nu) - 6 \,\delta \,\kappa_{+} \,(7044 + 161 \,\nu)- 4 \,\delta \,(11608 + 483 \,\nu)\big) \,(nv) \,(n\Sigma) \,(vS) 
+ \big(27240 - 4664 \,\delta \,\kappa_{-} + 336 \,\nu + \,\kappa_{+} \,(14652 + 168 \,\nu)\big) \,(vS)^2
+ \,\bS^2 \Big[\big(2772 \,\delta \,\kappa_{-} + 8 \,(-6109 + 330 \,\nu) + \,\kappa_{+} \,(-23844 + 1320 \,\nu)\big) \,(nv)^2
+ \big(-2572 \,\delta \,\kappa_{-} + \,\kappa_{+} \,(21028 - 720 \,\nu) - 72 \,(-581 + 20 \,\nu)\big) \,\bv^2\Big]
+ \,(nS)^2 \Big[\big(300528 - 28788 \,\delta \,\kappa_{-} + \,\kappa_{+} \,(135588 - 2028 \,\nu) - 4056 \,\nu\big) \,(nv)^2
+ \big(-182752 + 12380 \,\delta \,\kappa_{-} + 3984 \,\nu + 24 \,\kappa_{+} \,(-3239 + 83 \,\nu)\big) \,\bv^2\Big]
+ \,(S\Sigma) \Big[\big(\,\kappa_{-} \,(26616 - 12408 \,\nu) + 264 \,\delta \,(-209 + 10 \,\nu) 
+ 24 \,\delta \,\kappa_{+} \,(-1109 + 55 \,\nu)\big) \,(nv)^2 
+ \big(-80 \,\delta \,\kappa_{+} \,(-295 + 9 \,\nu)- 8 \,\delta \,(-6109 + 180 \,\nu) + 16 \,\kappa_{-} \,(-1475 + 688 \,\nu)\big) \,\bv^2\Big] 
+ \,(nS) \,(n\Sigma) \Big[\big(-12 \,\delta \,\kappa_{+} \,(-13698 + 169 \,\nu) - 12 \,\delta \,(-23537 + 338 \,\nu) + 108 \,\kappa_{-} \,(-1522 + 1085 \,\nu)\big) \,(nv)^2
+ \big(\,\kappa_{-} \,(90116 - 51512 \,\nu) + 4 \,\delta \,\kappa_{+} \,(-22529 + 498 \,\nu) + 4 \,\delta \,(-45929 + 996 \,\nu)\big) \,\bv^2\Big]
+ \,(n\Sigma)^2 \Big[\big(6 \,\delta \,\kappa_{-} \,(-13698 + 4967 \,\nu) + 6 \,\kappa_{+} \,(13698 - 32363 \,\nu + 338 \,\nu^2) + 12 \,(467 - 22056 \,\nu + 338 \,\nu^2)\big) \,(nv)^2 
+ \big(-2 \,\delta \,\kappa_{-} \,(-22529 + 6688 \,\nu) - 2 \,\kappa_{+} \,(22529 - 51746 \,\nu + 996 \,\nu^2) - 4 \,(2425 - 46383 \,\nu + 996 \,\nu^2)\big) \,\bv^2\Big] 
+ \big(\,\kappa_{-} \,(42264 - 39978 \,\nu) - 6 \,\delta \,\kappa_{+} \,(7044 + 161 \,\nu) - 4 \,\delta \,(12686 + 483 \,\nu)\big) \,(nS) \,(nv) \,(v\Sigma) 
+ \big(9808 + 69804 \,\nu + 3864 \,\nu^2 - 6 \,\delta \,\kappa_{-} \,(-7044 + 3251 \,\nu) + 6 \,\kappa_{+} \,(-7044 + 17339 \,\nu + 322 \,\nu^2)\big) \,(nv) \,(n\Sigma) \,(v\Sigma)
+ \big(336 \,\delta \,(55 + \,\nu) + 4 \,\delta \,\kappa_{+} \,(4829 + 42 \,\nu) + 4 \,\kappa_{-} \,(-4829 + 4622 \,\nu)\big) \,(vS) \,(v\Sigma) 
+ \big(2 \,\delta \,\kappa_{-} \,(-4829 + 2290 \,\nu) + \,\kappa_{+} \,(9658 - 23896 \,\nu - 168 \,\nu^2) - 16 \,(299 + 613 \,\nu + 21 \,\nu^2)\big) \,(v\Sigma)^2
+ \Big[\big(-12 \,\delta \,\kappa_{-} \,(-1109 + 286 \,\nu)
- 12 \,\kappa_{+} \,(1109 - 2504 \,\nu + 110 \,\nu^2) - 8 \,(151 - 7717 \,\nu + 330 \,\nu^2)\big) \,(nv)^2
+ \big(4 \,\delta \,\kappa_{-} \,(-2950 + 733 \,\nu) + 8 \,(551 - 7036 \,\nu + 180 \,\nu^2) + 4 \,\kappa_{+} \,(2950 - 6633 \,\nu + 180 \,\nu^2)\big) \,\bv^2\Big] \,\bSigma^2\,,
\end{autobreak}
\\[2ex]
\begin{autobreak}
f^2_6
=\,(936 + 48 \,\delta \,\kappa_{-} + 432 \,\delta^2 \,\kappa_{+} - 3456 \,\nu) \,(nS)^2 
+ \big(-384 \,\delta^2 \,\kappa_{-} - 192 \,\delta \,\kappa_{+} \,(-2 + 9 \,\nu) - 16 \,\delta \,(-49 + 216 \,\nu)\big) \,(nS) \,(n\Sigma)
+ \big(-48 \,\delta^2 \,\kappa_{+} \,(-4 + 9 \,\nu) + 48 \,\delta \,\kappa_{-} \,(-4 + 17 \,\nu) - 16 \,\delta^2 \,(-1 + 54 \,\nu)\big) \,(n\Sigma)^2
+ \big(-16 \,\delta \,\kappa_{-} - 144 \,\delta^2 \,\kappa_{+} + 16 \,(-23 + 72 \,\nu)\big) \,\bS^2 
+ \big(128 \,\delta^2 \,\kappa_{-} + 64 \,\delta \,\kappa_{+} \,(-2 + 9 \,\nu) + 32 \,\delta \,(-7 + 36 \,\nu)\big) \,(S\Sigma) 
+ \big(16 \,\delta^2 \,\kappa_{+} \,(-4 + 9 \,\nu) + 24 \,\delta^2 \,(-1 + 12 \,\nu) - 16 \,\delta \,\kappa_{-} \,(-4 + 17 \,\nu)\big) \,\bSigma^2\,.
\end{autobreak}
\end{align}
\end{subequations}
\subsection{Angular momentum flux }

The coefficients for the radiated angular momentum in \eqref{3pnJss1}--\eqref{3pnJss2} are given by
\begin{subequations}
\begin{align}
\begin{autobreak}
g_3^{0\,i}
=\,\bS^i \,(-120 \,(nv)^4 + 264 \,(nv)^2 \,\bv^2 - 160 \,(\bv^2)^2)
+ \,\bv^i \,\big(\,(vS) \,(-348 \,(nv)^2 + 160 \,\bv^2) + \,(nS) \,(780 \,(nv)^3 - 444 \,(nv) \,\bv^2)
+ \,(n\Sigma) \,(420 \,\delta \,(nv)^3 - 228 \,\delta \,(nv) \,\bv^2) + \,(-204 \,\delta \,(nv)^2 + 88 \,\delta \,\bv^2) \,(v\Sigma)\big) 
+ \,\bn^i \,\big(\,(vS) \,(120 \,(nv)^3 + 84 \,(nv) \,\bv^2) + \,(nS) \,(-780 \,(nv)^2 \,\bv^2 + 444 \,(\bv^2)^2)
+ \,(n\Sigma) \,(-420 \,\delta \,(nv)^2 \,\bv^2 + 228 \,\delta \,(\bv^2)^2) + \,(120 \,\delta \,(nv)^3 + 12 \,\delta \,(nv) \,\bv^2) \,(v\Sigma)\big) 
+ \,(-120 \,\delta \,(nv)^4 + 192 \,\delta \,(nv)^2 \,\bv^2 - 88 \,\delta \,(\bv^2)^2) \,\bSigma^i\,,
\end{autobreak}
\\[2ex]
\begin{autobreak}
g_3^{1\,i}
=\,\bS^i \,(76 \,(nv)^2 - 72 \,\bv^2) 
+ \,\bv^i \,(-175 \,(nS) \,(nv) - 76 \,\delta \,(nv) \,(n\Sigma) + 113 \,(vS) + 44 \,\delta \,(v\Sigma))
+ \,\bn^i \,\big(-109 \,(nv) \,(vS) + \,(nS) \,(45 \,(nv)^2 + 122 \,\bv^2) + \,(n\Sigma) \,(27 \,\delta \,(nv)^2 + 47 \,\delta \,\bv^2) - 46 \,\delta \,(nv) \,(v\Sigma)\big)
+ \,(26 \,\delta \,(nv)^2 - 22 \,\delta \,\bv^2) \,\bSigma^i\,,
\end{autobreak}
\\[2ex]
\begin{autobreak}
g_3^{2\,i}
=\,\bn^i \,(2 \,(nS) -  \,\delta \,(n\Sigma)) - 2 \,\bS^i + \,\delta \,\bSigma^i\,,
\end{autobreak}
\\[2ex]
\begin{autobreak}
g_4^{0\,i}
=\,(\bv\times\bS)^{i} \,\Big[\,(-24 - 12 \,\kappa_{+}) \,(nv) \,(vS) + \,(nS) \big(\,(120 + 60 \,\kappa_{+}) \,(nv)^2 + \,(-72 - 36 \,\kappa_{+}) \,\bv^2\big)
+ \,(n\Sigma) \big(\,(60 \,\delta - 30 \,\kappa_{-} + 30 \,\delta \,\kappa_{+}) \,(nv)^2
+ \,(-36 \,\delta + 18 \,\kappa_{-} - 18 \,\delta \,\kappa_{+}) \,\bv^2\big)
+ \,(-12 \,\delta + 6 \,\kappa_{-} - 6 \,\delta \,\kappa_{+}) \,(nv) \,(v\Sigma)\Big] 
+ \,(\bv\times\bSigma)^{i} \,\bigg\{\,(-12 \,\delta + 6 \,\kappa_{-} - 6 \,\delta \,\kappa_{+}) \,(nv) \,(vS) + \,(nS) \big(\,(60 \,\delta - 30 \,\kappa_{-} + 30 \,\delta \,\kappa_{+}) \,(nv)^2 
+ \,(-36 \,\delta + 18 \,\kappa_{-} - 18 \,\delta \,\kappa_{+}) \,\bv^2\big)+ \,(n\Sigma) \,\Big[\big(-30 \,\delta \,\kappa_{-} + \,\kappa_{+} \,(30 - 60 \,\nu)- 120 \,\nu\big) \,(nv)^2
+ \big(18 \,\delta \,\kappa_{-} + 72 \,\nu + \,\kappa_{+} \,(-18 + 36 \,\nu)\big) \,\bv^2\Big]
+ \big(6 \,\delta \,\kappa_{-} + 24 \,\nu + \,\kappa_{+} \,(-6 + 12 \,\nu)\big) \,(nv) \,(v\Sigma)\bigg\}
+ \,\big(\bn\times\bv\big)^{i} \,\bigg\{\,(-360 - 180 \,\kappa_{+}) \,(nS) \,(nv) \,(vS) + \,(-180 \,\delta + 90 \,\kappa_{-} - 90 \,\delta \,\kappa_{+}) \,(nv) \,(n\Sigma) \,(vS)
+ \,(24 + 12 \,\kappa_{+}) \,(vS)^2 + \,(nS)^2 \big(\,(840 + 420 \,\kappa_{+}) \,(nv)^2 + \,(-240 - 120 \,\kappa_{+}) \,\bv^2\big) + \,\bS^2 \big(\,(-120 - 60 \,\kappa_{+}) \,(nv)^2
+ \,(48 + 24 \,\kappa_{+}) \,\bv^2\big)
+ \,(nS) \,(n\Sigma) \big(\,(840 \,\delta - 420 \,\kappa_{-} + 420 \,\delta \,\kappa_{+}) \,(nv)^2 + \,(-240 \,\delta + 120 \,\kappa_{-} - 120 \,\delta \,\kappa_{+}) \,\bv^2\big)
+ \,(S\Sigma) \big(\,(-120 \,\delta + 60 \,\kappa_{-} - 60 \,\delta \,\kappa_{+}) \,(nv)^2 + \,(48 \,\delta - 24 \,\kappa_{-} + 24 \,\delta \,\kappa_{+}) \,\bv^2\big)
+ \,(n\Sigma)^2 \,\Big[\big(-210 \,\delta \,\kappa_{-} + \,\kappa_{+} \,(210 - 420 \,\nu) - 840 \,\nu\big) \,(nv)^2
+ \big(60 \,\delta \,\kappa_{-} + 240 \,\nu + 60 \,\kappa_{+} \,(-1 + 2 \,\nu)\big) \,\bv^2\Big] + \,(-180 \,\delta + 90 \,\kappa_{-} - 90 \,\delta \,\kappa_{+}) \,(nS) \,(nv) \,(v\Sigma)
+ \big(90 \,\delta \,\kappa_{-} + 360 \,\nu + 90 \,\kappa_{+} \,(-1 + 2 \,\nu)\big) \,(nv) \,(n\Sigma) \,(v\Sigma)
+ \,(24 \,\delta - 12 \,\kappa_{-} + 12 \,\delta \,\kappa_{+}) \,(vS) \,(v\Sigma)
+ \big(-6 \,\delta \,\kappa_{-} + \,\kappa_{+} \,(6 - 12 \,\nu) - 24 \,\nu\big) \,(v\Sigma)^2 + \,\Big[\big(30 \,\delta \,\kappa_{-} + 120 \,\nu + \,\kappa_{+} \,(-30 + 60 \,\nu)\big) \,(nv)^2 
+ \big(-12 \,\delta \,\kappa_{-} + \,\kappa_{+} \,(12 - 24 \,\nu) - 48 \,\nu\big) \,\bv^2\Big] \,\bSigma^2\bigg\}\,,
\end{autobreak}
\\[2ex]
\begin{autobreak}
g_4^{1\,i}
=\big(\,(-48 - 24 \,\kappa_{+}) \,(nS) + \,(-24 \,\delta + 12 \,\kappa_{-} - 12 \,\delta \,\kappa_{+}) \,(n\Sigma)\big) \,(\bv\times\bS)^{i} 
+ \,\Big[\,(-24 \,\delta + 12 \,\kappa_{-} - 12 \,\delta \,\kappa_{+}) \,(nS) + \big(-2 + 12 \,\delta \,\kappa_{-} + 48 \,\nu + \,\kappa_{+} \,(-12 + 24 \,\nu)\big) \,(n\Sigma)\Big] \,(\bv\times\bSigma)^{i}
+ \,(\bn\times\bS)^{i} \big(\,(-24 - 12 \,\kappa_{+}) \,(nS) \,(nv) + \,(-12 \,\delta + 6 \,\kappa_{-} - 6 \,\delta \,\kappa_{+}) \,(nv) \,(n\Sigma)
+ \,(24 + 12 \,\kappa_{+}) \,(vS) + \,(12 \,\delta - 6 \,\kappa_{-} + 6 \,\delta \,\kappa_{+}) \,(v\Sigma)\big)
+ \,(\bn\times\bSigma)^{i} \,\Big[\,(-12 \,\delta + 6 \,\kappa_{-} - 6 \,\delta \,\kappa_{+}) \,(nS) \,(nv) + \big(6 \,\delta \,\kappa_{-} + 24 \,\nu + \,\kappa_{+} \,(-6 + 12 \,\nu)\big) \,(nv) \,(n\Sigma)
+ \,(12 \,\delta - 6 \,\kappa_{-} + 6 \,\delta \,\kappa_{+}) \,(vS) + \big(2 - 6 \,\delta \,\kappa_{-} + \,\kappa_{+} \,(6 - 12 \,\nu) - 24 \,\nu\big) \,(v\Sigma)\Big]
+ \,\big(\bn\times\bv\big)^{i} \,\Big[\,(-360 - 180 \,\kappa_{+}) \,(nS)^2 + \,(-360 \,\delta + 180 \,\kappa_{-} - 180 \,\delta \,\kappa_{+}) \,(nS) \,(n\Sigma)
+ \big(90 \,\delta \,\kappa_{-} + 360 \,\nu + 90 \,\kappa_{+} \,(-1 + 2 \,\nu)\big) \,(n\Sigma)^2 + \,(96 + 48 \,\kappa_{+}) \,\bS^2 + \,(96 \,\delta - 48 \,\kappa_{-} + 48 \,\delta \,\kappa_{+}) \,(S\Sigma)
+ \big(2 - 24 \,\delta \,\kappa_{-} + \,\kappa_{+} \,(24 - 48 \,\nu) - 96 \,\nu\big) \,\bSigma^2\Big]\,,
\end{autobreak}
\end{align}
\begin{align}
\begin{autobreak}
g_5^{0\,i}
=\bS^i\, \big(-105 \,(1 + 40 \,\nu) \,(nv)^6 + 15 \,(-43 + 716 \,\nu) \,(nv)^4 \,\bv^2 + \,(2589 - 10608 \,\nu) \,(nv)^2 \,(\bv^2)^2
+ \,(-1607 + 4004 \,\nu) \,(\bv^2)^3\big)+ 
\,\bv^i \,\Big[\,(vS) \big(\,(2145 - 7740 \,\nu) \,(nv)^4 + 30 \,(-65 + 334 \,\nu) \,(nv)^2 \,\bv^2 + \,(1607 - 4004 \,\nu) \,(\bv^2)^2\big)
+ \,(n\Sigma) \big(210 \,\delta \,(-4 + 35 \,\nu) \,(nv)^5 - 30 \,\delta \,(-49 + 388 \,\nu) \,(nv)^3 \,\bv^2 + 6 \,\delta \,(-217 + 787 \,\nu) \,(nv) \,(\bv^2)^2\big)
+ \,(nS) \big(1260 \,(-1 + 6 \,\nu) \,(nv)^5 - 30 \,(-53 + 496 \,\nu) \,(nv)^3 \,\bv^2 + \,(-2874 + 8376 \,\nu) \,(nv) \,(\bv^2)^2\big)
+ \big(-15 \,\delta \,(-79 + 455 \,\nu) \,(nv)^4 + 6 \,\delta \,(-147 + 1427 \,\nu) \,(nv)^2 \,\bv^2 + \,\delta \,(563 - 2477 \,\nu) \,(\bv^2)^2\big) \,(v\Sigma)\Big] 
+ \,\bn^i \,\Big[\,(vS) \big(105 \,(1 + 40 \,\nu) \,(nv)^5 - 1500 \,(1 + 2 \,\nu) \,(nv)^3 \,\bv^2 + \,(-639 + 588 \,\nu) \,(nv) \,(\bv^2)^2\big) 
+ \,(nS) \big(-1260 \,(-1 + 6 \,\nu) \,(nv)^4 \,\bv^2 + 30 \,(-53 + 496 \,\nu) \,(nv)^2 \,(\bv^2)^2 + \,(2874 - 8376 \,\nu) \,(\bv^2)^3\big)
+ \,(n\Sigma) \big(-210 \,\delta \,(-4 + 35 \,\nu) \,(nv)^4 \,\bv^2 + 30 \,\delta \,(-49 + 388 \,\nu) \,(nv)^2 \,(\bv^2)^2 - 6 \,\delta \,(-217 + 787 \,\nu) \,(\bv^2)^3\big) 
+ \big(525 \,\delta \,(5 + 13 \,\nu) \,(nv)^5 - 270 \,\delta \,(20 + 29 \,\nu) \,(nv)^3 \,\bv^2
+ 3 \,\delta \,(559 + 571 \,\nu) \,(nv) \,(\bv^2)^2\big) \,(v\Sigma)\Big]
+ \bSigma^i\,\big(-525 \,\delta \,(5 + 13 \,\nu) \,(nv)^6 + 15 \,\delta \,(281 + 977 \,\nu) \,(nv)^4 \,\bv^2 
- 15 \,\delta \,(53 + 685 \,\nu) \,(nv)^2 \,(\bv^2)^2 + \,\delta \,(-563 + 2477 \,\nu) \,(\bv^2)^3\big) \,,
\end{autobreak}
\\[2ex]
\begin{autobreak}
g_5^{1\,i}
=\bS^i\, \big(\,(-8577 + 7320 \,\nu) \,(nv)^4 + \,(7022 - 15068 \,\nu) \,(nv)^2 \,\bv^2 + 7 \,(-155 + 972 \,\nu) \,(\bv^2)^2\big)
+ \,\bv^i \,\Big[\,(vS) \big(\,(-588 + 10982 \,\nu) \,(nv)^2 + \,(3138 - 8238 \,\nu) \,\bv^2\big) 
+ \,(n\Sigma) \big(-9 \,\delta \,(1519 + 216 \,\nu) \,(nv)^3 + \,\delta \,(4789 + 1994 \,\nu) \,(nv) \,\bv^2\big) 
+ \,(nS) \big(-6 \,(2750 + 1699 \,\nu) \,(nv)^3 + \,(3542 + 6802 \,\nu) \,(nv) \,\bv^2\big) 
+ \big(\,\delta \,(4313 + 1486 \,\nu) \,(nv)^2 - 17 \,\delta \,(39 + 112 \,\nu) \,\bv^2\big) \,(v\Sigma)\Big]
+ \,\bn^i \,\Big[\,(vS) \big(\,(12570 - 10914 \,\nu) \,(nv)^3 + \,(-12836 + 9162 \,\nu) \,(nv) \,\bv^2\big) 
+ \,(nS) \big(-3 \,(1381 + 574 \,\nu) \,(nv)^4 + 6 \,(3919 + 2760 \,\nu) \,(nv)^2 \,\bv^2 + \,(-6057 - 11494 \,\nu) \,(\bv^2)^2\big) 
+ \,(n\Sigma) \big(3 \,\delta \,(-807 + \,\nu) \,(nv)^4 + 3 \,\delta \,(5909 + 1242 \,\nu) \,(nv)^2 \,\bv^2 -  \,\delta \,(6364 + 3679 \,\nu) \,(\bv^2)^2\big)
+ \big(-3 \,\delta \,(-759 + 296 \,\nu) \,(nv)^3 + 3 \,\delta \,(-1729 + 674 \,\nu) \,(nv) \,\bv^2\big) \,(v\Sigma)\Big] 
+  \bSigma^i\,\big(-3 \,\delta \,(-179 + 361 \,\nu) \,(nv)^4 - 10 \,\delta \,(336 + 85 \,\nu) \,(nv)^2 \,\bv^2 + \,\delta \,(2023 + 1117 \,\nu) \,(\bv^2)^2\big)\,,
\end{autobreak}
\\[2ex]
\begin{autobreak}
g_5^{2\,i}
=\bS^i\, \big(-2 \,(7711 + 272 \,\nu) \,(nv)^2 + \,(16218 + 2296 \,\nu) \,\bv^2\big) 
+ \,\bv^i\, \big(8 \,(6227 + 386 \,\nu) \,(nS) \,(nv) + 36 \,\delta \,(685 + 36 \,\nu) \,(nv) \,(n\Sigma)
- 2 \,(14701 + 728 \,\nu) \,(vS) - 10 \,\delta \,(1501 + 54 \,\nu) \,(v\Sigma)\big)
+ \,\bn^i \,\Big[\,(27954 + 536 \,\nu) \,(nv) \,(vS) + \,(n\Sigma) \big(-2 \,\delta \,(5157 + 847 \,\nu) \,(nv)^2 - 2 \,\delta \,(7151 + 163 \,\nu) \,\bv^2\big)
+ \,(nS) \big(-4 \,(5115 + 413 \,\nu) \,(nv)^2 - 12 \,(2392 + 189 \,\nu) \,\bv^2\big) - 2 \,\delta \,(-7561 + 96 \,\nu) \,(nv) \,(v\Sigma)\Big] 
+  \bSigma^i\,\big(2 \,\delta \,(-4040 + 541 \,\nu) \,(nv)^2 + 2 \,\delta \,(3962 + 187 \,\nu) \,\bv^2\big)\,,
\end{autobreak}
\\[2ex]
\begin{autobreak}
g_5^{3\,i}
=\,\bn^i \big(-8 \,(265 + 31 \,\nu) \,(nS) - 20 \,\delta \,(-12 + 7 \,\nu) \,(n\Sigma)\big) + 8 \,(265 + 31 \,\nu) \bS^i + 20 \,\delta \,(-12 + 7 \,\nu) \bSigma^i\,,
\end{autobreak}
\end{align}

\begin{align}
\begin{autobreak}
g_6^{0\,i}
=\,(\bv\times\bS)^{i} \,\bigg\{\,(vS) \Big[\big(-3020 \,\delta \,\kappa_{-} - 280 \,\kappa_{+} \,(1 + \,\nu) - 80 \,(-52 + 7 \,\nu)\big) \,(nv)^3 
+ \big(1916 \,\delta \,\kappa_{-} + \,\kappa_{+} \,(680 - 252 \,\nu) - 56 \,(68 + 9 \,\nu)\big) \,(nv) \,\bv^2\Big] 
+ \,(n\Sigma) \Big[\big(140 \,\delta \,\kappa_{+} \,(-23 + 2 \,\nu) + 280 \,\delta \,(-1 + 2 \,\nu) - 140 \,\kappa_{-} \,(-23 + 88 \,\nu)\big) \,(nv)^4 
+ \big(-10 \,\delta \,\kappa_{+} \,(-258 + 19 \,\nu) - 20 \,\delta \,(-169 + 19 \,\nu) + 30 \,\kappa_{-} \,(-86 + 349 \,\nu)\big) \,(nv)^2 \,\bv^2 
+ \big(\,\kappa_{-} \,(504 - 954 \,\nu) + 14 \,\delta \,\kappa_{+} \,(-36 + 19 \,\nu) + 4 \,\delta \,(-337 + 133 \,\nu)\big) \,(\bv^2)^2\Big] 
+ \,(nS) \Big[\big(6020 \,\delta \,\kappa_{-} + 1120 \,(-11 + \,\nu) + 140 \,\kappa_{+} \,(-3 + 4 \,\nu)\big) \,(nv)^4 + \big(16280 - 5140 \,\delta \,\kappa_{-} 
+ \,\kappa_{+} \,(20 - 380 \,\nu) - 760 \,\nu\big) \,(nv)^2 \,\bv^2 + \big(344 \,\delta \,\kappa_{-}+ 8 \,(-449 + 133 \,\nu) + \,\kappa_{+} \,(-664 + 532 \,\nu)\big) \,(\bv^2)^2\Big]
+ \Big[\big(-10 \,\delta \,\kappa_{+} \,(-137 + 14 \,\nu) - 20 \,\delta \,(57 + 14 \,\nu) + \,\kappa_{-} \,(-1370 + 6180 \,\nu)\big) \,(nv)^3 
+ \big(\,\kappa_{-} \,(618 - 3706 \,\nu) - 252 \,\delta \,\nu - 6 \,\delta \,\kappa_{+} \,(103 + 21 \,\nu)\big) \,(nv) \,\bv^2\Big] \,(v\Sigma)\bigg\} 
+ \,(\bv\times\bSigma)^{i} \,\bigg\{\,(vS) \Big[\big(-20 \,\delta \,(-223 + 14 \,\nu) - 10 \,\delta \,\kappa_{+} \,(-137 + 14 \,\nu)
+ \,\kappa_{-} \,(-1370 + 6180 \,\nu)\big) \,(nv)^3 + \big(\,\kappa_{-} \,(618 - 3706 \,\nu) - 84 \,\delta \,(46 + 3 \,\nu) - 6 \,\delta \,\kappa_{+} \,(103 + 21 \,\nu)\big) \,(nv) \,\bv^2\Big]
+ \,(nS) \Big[\big(560 \,\delta \,(-25 + \,\nu) + 140 \,\delta \,\kappa_{+} \,(-23 + 2 \,\nu) - 140 \,\kappa_{-} \,(-23 + 88 \,\nu)\big) \,(nv)^4 
+ \big(-20 \,\delta \,(-813 + 19 \,\nu) - 10 \,\delta \,\kappa_{+} \,(-258 + 19 \,\nu) + 30 \,\kappa_{-} \,(-86 + 349 \,\nu)\big) \,(nv)^2 \,\bv^2 
+ \big(\,\kappa_{-} \,(504 - 954 \,\nu) + 14 \,\delta \,\kappa_{+} \,(-36 + 19 \,\nu) + 4 \,\delta \,(-687 + 133 \,\nu)\big) \,(\bv^2)^2\Big]
+ \,(n\Sigma) \Big[\big(-140 \,\delta \,\kappa_{-} \,(-23 + 45 \,\nu) - 140 \,\kappa_{+} \,(23 - 91 \,\nu + 4 \,\nu^2) - 280 \,(7 - 58 \,\nu + 4 \,\nu^2)\big) \,(nv)^4
+ \big(10 \,\delta \,\kappa_{-} \,(-258 + 533 \,\nu) + 40 \,(84 - 568 \,\nu + 19 \,\nu^2) + 10 \,\kappa_{+} \,(258 - 1049 \,\nu + 38 \,\nu^2)\big) \,(nv)^2 \,\bv^2
+ \big(\,\delta \,\kappa_{-} \,(504 - 610 \,\nu) + \,\kappa_{+} \,(-504 + 1618 \,\nu - 532 \,\nu^2) - 8 \,(63 - 568 \,\nu + 133 \,\nu^2)\big) \,(\bv^2)^2\Big]
+ \Big[\big(10 \,\delta \,\kappa_{-} \,(-137 + 316 \,\nu) + 40 \,(-21 - 83 \,\nu + 14 \,\nu^2) + 10 \,\kappa_{+} \,(137 - 590 \,\nu + 28 \,\nu^2)\big) \,(nv)^3
+ \big(-2 \,\delta \,\kappa_{-} \,(-309 + 895 \,\nu) + 56 \,(-1 + 77 \,\nu + 9 \,\nu^2) + \,\kappa_{+} \,(-618 + 3026 \,\nu + 252 \,\nu^2)\big) \,(nv) \,\bv^2\Big] \,(v\Sigma)\bigg\}
+ \,(\bn\times\bv)^{i} \,\bigg\{\,(vS)^2 \Big[\big(2280 \,\delta \,\kappa_{-} + 40 \,(-3 + 142 \,\nu) + \,\kappa_{+} \,(-2100 + 2840 \,\nu)\big) \,(nv)^2
+ \big(-512 \,\delta \,\kappa_{-} + 8 \,(-99 + 19 \,\nu) + \,\kappa_{+} \,(-100 + 76 \,\nu)\big) \,\bv^2\Big]
+ \,(n\Sigma) \,(vS) \Big[\big(-70 \,\delta \,\kappa_{+} \,(-113 + 128 \,\nu) - 140 \,\delta \,(1 + 128 \,\nu)
+ 70 \,\kappa_{-} \,(-113 + 412 \,\nu)\big) \,(nv)^3 + \big(\,\kappa_{-} \,(1730 - 10090 \,\nu) + 10 \,\delta \,\kappa_{+} \,(-173 + 157 \,\nu) 
+ 20 \,\delta \,(36 + 157 \,\nu)\big) \,(nv) \,\bv^2\Big] + \,(nS) \,(vS) \Big[\big(7560 - 9940 \,\delta \,\kappa_{-} - 35840 \,\nu - 280 \,\kappa_{+} \,(-21 + 64 \,\nu)\big) \,(nv)^3 
+ \big(40 + 4260 \,\delta \,\kappa_{-} + 6280 \,\nu + \,\kappa_{+} \,(800 + 3140 \,\nu)\big) \,(nv) \,\bv^2\Big] 
+ \,\bS^2 \Big[\big(4060 + 1400 \,\delta \,\kappa_{-} - 8400 \,\nu - 70 \,\kappa_{+} \,(-23 + 60 \,\nu)\big) \,(nv)^4
+ \big(-1420 \,\delta \,\kappa_{-} + 40 \,(-90 + 181 \,\nu) + 20 \,\kappa_{+} \,(-69 + 181 \,\nu)\big) \,(nv)^2 \,\bv^2
+ \big(1068 + 248 \,\delta \,\kappa_{-} + \,\kappa_{+} \,(486 - 616 \,\nu) - 1232 \,\nu\big) \,(\bv^2)^2\Big]
+ \,(nS)^2 \Big[\big(8820 \,\delta \,\kappa_{-} + 1260 \,(-13 + 40 \,\nu) + 630 \,\kappa_{+} \,(-13 + 40 \,\nu)\big) \,(nv)^4 + \big(-4900 \,\delta \,\kappa_{-} 
- 420 \,\kappa_{+} \,(-7 + 29 \,\nu) - 280 \,(-49 + 87 \,\nu)\big) \,(nv)^2 \,\bv^2 + \big(-4660 - 140 \,\delta \,\kappa_{-} + 3600 \,\nu + 30 \,\kappa_{+} \,(-59 + 60 \,\nu)\big) \,(\bv^2)^2\Big]
+ \,(nS) \,(n\Sigma) \Big[\big(-1890 \,\kappa_{-} \,(-9 + 32 \,\nu) + 630 \,\delta \,\kappa_{+} \,(-27 + 40 \,\nu) + 1260 \,\delta \,(-13 + 40 \,\nu)\big) \,(nv)^4
+ \big(-840 \,\delta \,(-21 + 29 \,\nu) - 140 \,\delta \,\kappa_{+} \,(-56 + 87 \,\nu) + 140 \,\kappa_{-} \,(-56 + 227 \,\nu)\big) \,(nv)^2 \,\bv^2 
+ \big(\,\kappa_{-} \,(1630 - 1240 \,\nu) + 180 \,\delta \,(-29 + 20 \,\nu) + 10 \,\delta \,\kappa_{+} \,(-163 + 180 \,\nu)\big) \,(\bv^2)^2\Big] 
+ \,(S\Sigma) \Big[\big(-420 \,\delta \,(-19 + 20 \,\nu) - 210 \,\delta \,\kappa_{+} \,(-1 + 20 \,\nu) - 70 \,\kappa_{-} \,(3 + 20 \,\nu)\big) \,(nv)^4
+ \big(40 \,\delta \,(-174 + 181 \,\nu) + 20 \,\delta \,\kappa_{+} \,(2 + 181 \,\nu) + \,\kappa_{-} \,(-40 + 2060 \,\nu)\big) \,(nv)^2 \,\bv^2 
+ \big(\,\kappa_{-} \,(-238 - 376 \,\nu) - 14 \,\delta \,\kappa_{+} \,(-17 + 44 \,\nu) - 4 \,\delta \,(-351 + 308 \,\nu)\big) \,(\bv^2)^2\Big]
+ \,(n\Sigma)^2 \Big[\big(-1260 \,\nu \,(-13 + 40 \,\nu) - 315 \,\delta \,\kappa_{-} \,(-27 + 68 \,\nu) - 315 \,\kappa_{+} \,(27 - 122 \,\nu + 80 \,\nu^2)\big) \,(nv)^4
+ \big(70 \,\delta \,\kappa_{-} \,(-56 + 157 \,\nu) + 280 \,(14 - 77 \,\nu + 87 \,\nu^2) + 70 \,\kappa_{+} \,(56 - 269 \,\nu + 174 \,\nu^2)\big) \,(nv)^2 \,\bv^2
+ \big(\,\delta \,\kappa_{-} \,(815 - 760 \,\nu) - 20 \,(28 - 289 \,\nu + 180 \,\nu^2) - 5 \,\kappa_{+} \,(163 - 478 \,\nu + 360 \,\nu^2)\big) \,(\bv^2)^2\Big]
+ \,(vS) \Big[\big(\,\kappa_{-} \,(4380 - 11960 \,\nu) + 80 \,\delta \,(23 + 71 \,\nu) + 20 \,\delta \,\kappa_{+} \,(-219 + 142 \,\nu)\big) \,(nv)^2
+ \big(8 \,\delta \,(-134 + 19 \,\nu) + 4 \,\delta \,\kappa_{+} \,(103 + 19 \,\nu) + \,\kappa_{-} \,(-412 + 1972 \,\nu)\big) \,\bv^2\Big] \,(v\Sigma)
+ \,(nS) \Big[\big(-70 \,\delta \,\kappa_{+} \,(-113 + 128 \,\nu) - 140 \,\delta \,(-13 + 128 \,\nu) + 70 \,\kappa_{-} \,(-113 + 412 \,\nu)\big) \,(nv)^3
+ \big(\,\kappa_{-} \,(1730 - 10090 \,\nu) + 10 \,\delta \,\kappa_{+} \,(-173 + 157 \,\nu) + 20 \,\delta \,(8 + 157 \,\nu)\big) \,(nv) \,\bv^2\Big] \,(v\Sigma) 
+ \,(n\Sigma) \Big[\big(70 \,\delta \,\kappa_{-} \,(-113 + 270 \,\nu) + 280 \,(-21 + 15 \,\nu + 128 \,\nu^2) + 70 \,\kappa_{+} \,(113 - 496 \,\nu + 256 \,\nu^2)\big) \,(nv)^3
+ \big(-10 \,\delta \,\kappa_{-} \,(-173 + 583 \,\nu) - 40 \,(-21 + 43 \,\nu + 157 \,\nu^2) - 10 \,\kappa_{+} \,(173 - 929 \,\nu + 314 \,\nu^2)\big) \,(nv) \,\bv^2\Big] \,(v\Sigma)
+ \Big[\big(-10 \,\delta \,\kappa_{-} \,(-219 + 370 \,\nu) - 40 \,(-49 + 88 \,\nu + 142 \,\nu^2) - 10 \,\kappa_{+} \,(219 - 808 \,\nu + 284 \,\nu^2)\big) \,(nv)^2
+ \big(2 \,\delta \,\kappa_{-} \,(-103 + 237 \,\nu) + \,\kappa_{+} \,(206 - 886 \,\nu - 76 \,\nu^2) - 8 \,(35 - 162 \,\nu + 19 \,\nu^2)\big) \,\bv^2\Big] \,(v\Sigma)^2
+ \Big[\big(35 \,\delta \,\kappa_{-} \,(-3 + 20 \,\nu) + 140 \,(28 - 85 \,\nu + 60 \,\nu^2) + 35 \,\kappa_{+} \,(3 - 26 \,\nu + 120 \,\nu^2)\big) \,(nv)^4
+ \big(-10 \,\delta \,\kappa_{-} \,(2 + 39 \,\nu) + \,\kappa_{+} \,(20 + 350 \,\nu - 3620 \,\nu^2) - 40 \,(84 - 251 \,\nu + 181 \,\nu^2)\big) \,(nv)^2 \,\bv^2
+ \big(\,\delta \,\kappa_{-} \,(-119 + 60 \,\nu) + 4 \,(84 - 421 \,\nu + 308 \,\nu^2) + \,\kappa_{+} \,(119 - 298 \,\nu + 616 \,\nu^2)\big) \,(\bv^2)^2\Big] \,\bSigma^2\bigg\}
\end{autobreak}
\\[2ex]
\begin{autobreak}
g^{1\,i}_{6}
=\,(\bS\times \bSigma)^{i} \,(-792 \,\delta \,(nv)^3 + 904 \,\delta \,(nv) \,\bv^2)
+ \,(\bv\times\bS)^{i} \,\bigg\{\big(6648 - 1492 \,\delta \,\kappa_{-} + 48 \,\nu + \,\kappa_{+} \,(-92 + 24 \,\nu)\big) \,(nv) \,(vS)
+ \,(nS) \Big[\big(5772 \,\delta \,\kappa_{-} - 144 \,\kappa_{+} \,(54 + 11 \,\nu) - 36 \,(647 + 88 \,\nu)\big) \,(nv)^2 
+ \big(-2824 \,\delta \,\kappa_{-} + 4 \,(2530 + 279 \,\nu) + \,\kappa_{+} \,(3838 + 558 \,\nu)\big) \,\bv^2\Big]
+ \,(n\Sigma) \Big[\big(\,\kappa_{-} \,(6774 - 10752 \,\nu) - 12 \,\delta \,(877 + 132 \,\nu) - 6 \,\delta \,\kappa_{+} \,(1129 + 132 \,\nu)\big) \,(nv)^2
+ \big(\,\delta \,\kappa_{+} \,(3331 + 279 \,\nu) + \,\delta \,(3779 + 558 \,\nu) + \,\kappa_{-} \,(-3331 + 5369 \,\nu)\big) \,\bv^2\Big] 
+ \big(4 \,\delta \,\kappa_{+} \,(175 + 3 \,\nu) + 8 \,\delta \,(536 + 3 \,\nu) + \,\kappa_{-} \,(-700 + 2972 \,\nu)\big) \,(nv) \,(v\Sigma)\bigg\}
+ \,(\bv\times\bSigma)^{i} \,\bigg\{\big(4 \,\delta \,\kappa_{+} \,(175 + 3 \,\nu) + 2 \,\delta \,(1963 + 12 \,\nu) + \,\kappa_{-} \,(-700 + 2972 \,\nu)\big) \,(nv) \,(vS)
+ \,(nS) \Big[\big(\,\kappa_{-} \,(6774 - 10752 \,\nu) - 72 \,\delta \,(235 + 22 \,\nu) - 6 \,\delta \,\kappa_{+} \,(1129 + 132 \,\nu)\big) \,(nv)^2
+ \big(\,\delta \,\kappa_{+} \,(3331 + 279 \,\nu) + \,\delta \,(9419 + 558 \,\nu) + \,\kappa_{-} \,(-3331 + 5369 \,\nu)\big) \,\bv^2\Big] 
+ \,(n\Sigma) \Big[\big(-6 \,\delta \,\kappa_{-} \,(-1129 + 830 \,\nu) + 6 \,\kappa_{+} \,(-1129 + 3088 \,\nu + 264 \,\nu^2) + 6 \,(-447 + 5230 \,\nu
+ 528 \,\nu^2)\big) \,(nv)^2 + \big(\,\delta \,\kappa_{-} \,(-3331 + 2545 \,\nu) + \,\kappa_{+} \,(3331 - 9207 \,\nu - 558 \,\nu^2) - 4 \,(-381 + 3974 \,\nu 
+ 279 \,\nu^2)\big) \,\bv^2\Big] + \big(20 \,\delta \,\kappa_{-} \,(-35 + 74 \,\nu) - 4 \,\kappa_{+} \,(-175 + 720 \,\nu + 6 \,\nu^2)
- 6 \,(-205 + 1695 \,\nu + 8 \,\nu^2)\big) \,(nv) \,(v\Sigma)\bigg\} + \,(\bn\times\bS)^{i} \,\bigg\{\,(vS) \Big[\big(-8016 + 2238 \,\delta \,\kappa_{-} 
+ 1884 \,\nu + \,\kappa_{+} \,(-966 + 942 \,\nu)\big) \,(nv)^2 + \big(8012 - 862 \,\delta \,\kappa_{-} + 168 \,\nu + \,\kappa_{+} \,(448 + 84 \,\nu)\big) \,\bv^2\Big]
+ \,(n\Sigma) \Big[\big(6 \,\delta \,(399 + 103 \,\nu) + 3 \,\delta \,\kappa_{+} \,(1135 + 103 \,\nu) 
+ \,\kappa_{-} \,(-3405 + 8943 \,\nu)\big) \,(nv)^3 + \big(\,\kappa_{-} \,(2295 - 5601 \,\nu) 
- 15 \,\delta \,\kappa_{+} \,(153 + 65 \,\nu) - 6 \,\delta \,(782 + 325 \,\nu)\big) \,(nv) \,\bv^2\Big] + \,(nS) \Big[\big(13632 - 4626 \,\delta \,\kappa_{-} 
+ 1236 \,\nu + \,\kappa_{+} \,(2184 + 618 \,\nu)\big) \,(nv)^3 + \big(3288 \,\delta \,\kappa_{-} - 6 \,\kappa_{+} \,(217 + 325 \,\nu) - 12 \,(1371 + 325 \,\nu)\big) \,(nv) \,\bv^2\Big]
+ \Big[\big(\,\kappa_{-} \,(1602 - 4947 \,\nu) + 3 \,\delta \,\kappa_{+} \,(-534 + 157 \,\nu) + 6 \,\delta \,(-448 + 157 \,\nu)\big) \,(nv)^2 + \big(3 \,\delta \,(1133 + 28 \,\nu)
+ \,\delta \,\kappa_{+} \,(655 + 42 \,\nu) + \,\kappa_{-} \,(-655 + 1682 \,\nu)\big) \,\bv^2\Big] \,(v\Sigma)\bigg\} 
+ \,(\bn\times\bSigma)^{i} \,\bigg\{\,(vS) \Big[\big(\,\kappa_{-} \,(1602 - 4947 \,\nu) + 6 \,\delta \,(-940 + 157 \,\nu) + 3 \,\delta \,\kappa_{+} \,(-534 + 157 \,\nu)\big) \,(nv)^2
+ \big(\,\delta \,\kappa_{+} \,(655 + 42 \,\nu) + \,\delta \,(4769 + 84 \,\nu) + \,\kappa_{-} \,(-655 + 1682 \,\nu)\big) \,\bv^2\Big] 
+ \,(nS) \Big[\big(3 \,\delta \,\kappa_{+} \,(1135 + 103 \,\nu)
+ 6 \,\delta \,(2493 + 103 \,\nu) + \,\kappa_{-} \,(-3405 + 8943 \,\nu)\big) \,(nv)^3 + \big(\,\kappa_{-} \,(2295 - 5601 \,\nu) - 15 \,\delta \,\kappa_{+} \,(153 + 65 \,\nu) 
- 6 \,\delta \,(2468 + 325 \,\nu)\big) \,(nv) \,\bv^2\Big] + \,(n\Sigma) \Big[\big(3 \,\delta \,\kappa_{-} \,(-1135 + 1439 \,\nu) + \,\kappa_{+} \,(3405 - 11127 \,\nu - 618 \,\nu^2)
- 12 \,(-380 + 1779 \,\nu + 103 \,\nu^2)\big) \,(nv)^3 + \big(-9 \,\delta \,\kappa_{-} \,(-255 + 257 \,\nu) + 12 \,(-394 + 1839 \,\nu + 325 \,\nu^2)
+ 3 \,\kappa_{+} \,(-765 + 2301 \,\nu + 650 \,\nu^2)\big) \,(nv) \,\bv^2\Big] + \Big[\big(-9 \,\delta \,\kappa_{-} \,(-178 + 301 \,\nu) + \,\kappa_{+} \,(-1602 + 5913 \,\nu - 942 \,\nu^2)
- 6 \,(269 - 1702 \,\nu + 314 \,\nu^2)\big) \,(nv)^2 + \big(5 \,\delta \,\kappa_{-} \,(-131 + 164 \,\nu) + \,\kappa_{+} \,(655 - 2130 \,\nu - 84 \,\nu^2)
- 2 \,(-855 + 4373 \,\nu + 84 \,\nu^2)\big) \,\bv^2\Big] \,(v\Sigma)\bigg\} 
+ \,\big(\bn\times\bv\big)^{i} \,\bigg\{\big(-11784 \,\delta \,\kappa_{-} + 36 \,\kappa_{+} \,(767 + 243 \,\nu) 
+ 24 \,(3610 + 729 \,\nu)\big) \,(nS) \,(nv) \,(vS) + \big(42 \,\kappa_{-} \,(-469 + 457 \,\nu) + 12 \,\delta \,(3049 + 729 \,\nu)
+ 6 \,\delta \,\kappa_{+} \,(3283 + 729 \,\nu)\big) \,(nv) \,(n\Sigma) \,(vS)+ \big(1290 \,\delta \,\kappa_{-} - 6 \,\kappa_{+} \,(598 + 333 \,\nu) 
- 4 \,(2587 + 999 \,\nu)\big) \,(vS)^2 + \,(S\Sigma) \Big[\big(60 \,\delta \,\kappa_{+} \,(135 + 53 \,\nu)
+ 24 \,\delta \,(574 + 265 \,\nu) + \,\kappa_{-} \,(-8100 + 5316 \,\nu)\big) \,(nv)^2 + \big(\,\kappa_{-} \,(1476 - 2656 \,\nu) - 36 \,\delta \,\kappa_{+} \,(41 + 50 \,\nu)
- 50 \,\delta \,(115 + 72 \,\nu)\big) \,\bv^2\Big] + \,\bS^2 \Big[\big(-2124 \,\delta \,\kappa_{-} + 24 \,(543 + 265 \,\nu) + \,\kappa_{+} \,(5976 + 3180 \,\nu)\big) \,(nv)^2
+ \big(1114 \,\delta \,\kappa_{-} - 2 \,\kappa_{+} \,(181 + 900 \,\nu) - 4 \,(521 + 900 \,\nu)\big) \,\bv^2\Big] + \,(nS) \,(n\Sigma) \Big[\big(\,\kappa_{-} \,(78072 - 78042 \,\nu)
- 6 \,\delta \,\kappa_{+} \,(13012 + 2797 \,\nu) - 12 \,\delta \,(13956 + 2797 \,\nu)\big) \,(nv)^2 + \big(6 \,\delta \,\kappa_{+} \,(2653 + 1102 \,\nu) + 12 \,\delta \,(4023 + 1102 \,\nu)
+ 6 \,\kappa_{-} \,(-2653 + 3714 \,\nu)\big) \,\bv^2\Big] + \,(nS)^2 \Big[\big(23706 \,\delta \,\kappa_{-} - 6 \,\kappa_{+} \,(9061 + 2797 \,\nu) - 6 \,(29683 + 5594 \,\nu)\big) \,(nv)^2
+ \big(-7224 \,\delta \,\kappa_{-} + 6 \,(7003 + 2204 \,\nu) + \,\kappa_{+} \,(8694 + 6612 \,\nu)\big) \,\bv^2\Big] + \,(n\Sigma)^2 \Big[\big(-3 \,\delta \,\kappa_{-} \,(-13012 + 5105 \,\nu)
+ 6 \,(-1869 + 26316 \,\nu + 5594 \,\nu^2) + \,\kappa_{+} \,(-39036 + 93387 \,\nu + 16782 \,\nu^2)\big) \,(nv)^2 + \big(3 \,\delta \,\kappa_{-} \,(-2653 + 1306 \,\nu)
+ \,\kappa_{+} \,(7959 - 19836 \,\nu - 6612 \,\nu^2) - 6 \,(-1519 + 9138 \,\nu + 2204 \,\nu^2)\big) \,\bv^2\Big]
+ \big(36 \,\delta \,(923 + 243 \,\nu) + 42 \,\kappa_{-} \,(-469 + 457 \,\nu)+ 6 \,\delta \,\kappa_{+} \,(3283 + 729 \,\nu)\big) \,(nS) \,(nv) \,(v\Sigma)
+ \big(6 \,\delta \,\kappa_{-} \,(-3283 + 1235 \,\nu) - 24 \,(29 + 2235 \,\nu + 729 \,\nu^2)- 6 \,\kappa_{+} \,(-3283 + 7801 \,\nu + 1458 \,\nu^2)\big) \,(nv) \,(n\Sigma) \,(v\Sigma) 
+ \big(\,\kappa_{-} \,(4878 - 3162 \,\nu) - 18 \,\delta \,\kappa_{+} \,(271 + 111 \,\nu)- 2 \,\delta \,(2533 + 1998 \,\nu)\big) \,(vS) \,(v\Sigma) 
+ \big(1922 - 38 \,\nu + 3996 \,\nu^2 - 3 \,\delta \,\kappa_{-} \,(-813 + 97 \,\nu) + 3 \,\kappa_{+} \,(-813 + 1723 \,\nu + 666 \,\nu^2)\big) \,(v\Sigma)^2
+ \Big[\big(786 - 15180 \,\nu - 6360 \,\nu^2 + 6 \,\delta \,\kappa_{-} \,(-675 + 89 \,\nu) 
- 6 \,\kappa_{+} \,(-675 + 1439 \,\nu + 530 \,\nu^2)\big) \,(nv)^2
+ \big(\,\delta \,\kappa_{-} \,(738 - 214 \,\nu)+ 24 \,(-95 + 401 \,\nu + 150 \,\nu^2) + 2 \,\kappa_{+} \,(-369 + 845 \,\nu + 900 \,\nu^2)\big) \,\bv^2\Big] \,\bSigma^2\bigg\}\,,
\end{autobreak}
\\[2ex]
\begin{autobreak}
g^{2\,i}_6
=(\bv\times\bS)^{i}\,\Big[\big(-1268 \,\delta \,\kappa_{-} + 12 \,(941 + 30 \,\nu) + \,\kappa_{+} \,(6122 + 180 \,\nu)\big) \,(nS) 
+ \big(5 \,\delta \,\kappa_{+} \,(739 + 18 \,\nu) + 2 \,\delta \,(2543 + 90 \,\nu) + \,\kappa_{-} \,(-3695 + 2446 \,\nu)\big) \,(n\Sigma)\Big] 
+ (\bv\times\bSigma)^{i}\,\Big[\big(60 \,\delta \,(104 + 3 \,\nu) + 5 \,\delta \,\kappa_{+} \,(739 + 18 \,\nu) + \,\kappa_{-} \,(-3695 + 2446 \,\nu)\big) \,(nS) 
+ \big(\,\delta \,\kappa_{-} \,(-3695 + 1178 \,\nu) + \,\kappa_{+} \,(3695 - 8568 \,\nu - 180 \,\nu^2) - 4 \,(-54 + 2831 \,\nu + 90 \,\nu^2)\big) \,(n\Sigma)\Big]
+ \,(\bn\times\bS)^{i} \Big[\big(-2522 \,\delta \,\kappa_{-} + 8 \,\kappa_{+} \,(703 + 90 \,\nu) + 8 \,(1303 + 180 \,\nu)\big) \,(nS) \,(nv) + \big(18 \,\delta \,(159 + 40 \,\nu)
+ \,\delta \,\kappa_{+} \,(4073 + 360 \,\nu) + \,\kappa_{-} \,(-4073 + 4684 \,\nu)\big) \,(nv) \,(n\Sigma)
 + \big(1318 \,\delta \,\kappa_{-} + 8 \,(-926 + 27 \,\nu) + 4 \,\kappa_{+} \,(-901 + 27 \,\nu)\big) \,(vS)
+ \big(\,\kappa_{-} \,(2461 - 2690 \,\nu) + \,\delta \,\kappa_{+} \,(-2461 + 54 \,\nu) + 2 \,\delta \,(-1111 + 54 \,\nu)\big) \,(v\Sigma)\Big]
+ \,(\bn\times\bSigma)^{i} \Big[\big(24 \,\delta \,(187 + 30 \,\nu) + \,\delta \,\kappa_{+} \,(4073 + 360 \,\nu) + \,\kappa_{-} \,(-4073 + 4684 \,\nu)\big) \,(nS) \,(nv) 
+ \big(\,\delta \,\kappa_{-} \,(-4073 + 2162 \,\nu) + \,\kappa_{+} \,(4073 - 10308 \,\nu - 720 \,\nu^2) - 6 \,(195 + 779 \,\nu + 240 \,\nu^2)\big) \,(nv) \,(n\Sigma)
+ \big(\,\kappa_{-} \,(2461 - 2690 \,\nu) + \,\delta \,\kappa_{+} \,(-2461 + 54 \,\nu) + 2 \,\delta \,(-1661 + 54 \,\nu)\big) \,(vS)
+ \big(954 + \,\delta \,\kappa_{-} \,(2461 - 1372 \,\nu) + 4322 \,\nu - 216 \,\nu^2 + \,\kappa_{+} \,(-2461 + 6294 \,\nu - 108 \,\nu^2)\big) \,(v\Sigma)\Big] 
+ \,\big(\bn\times\bv\big)^{i} \Big[\big(98500 - 6486 \,\delta \,\kappa_{-} + 72 \,\nu + 6 \,\kappa_{+} \,(7469 + 6 \,\nu)\big) \,(nS)^2 
+ \big(36 \,\delta \,\kappa_{+} \,(1425 + \,\nu) + 2 \,\delta \,(48905 + 36 \,\nu) + \,\kappa_{-} \,(-51300 + 25908 \,\nu)\big) \,(nS) \,(n\Sigma) 
+ \big(2558 - 96542 \,\nu - 72 \,\nu^2 + 6 \,\delta \,\kappa_{-} \,(-4275 + 1078 \,\nu) - 6 \,\kappa_{+} \,(-4275 + 9628 \,\nu + 6 \,\nu^2)\big) \,(n\Sigma)^2
+ \big(1300 \,\delta \,\kappa_{-} + 4 \,\kappa_{+} \,(-2924 + 3 \,\nu) + 8 \,(-2891 + 3 \,\nu)\big) \,\bS^2 + \big(\,\kappa_{-} \,(12996 - 5212 \,\nu)
+ 12 \,\delta \,\kappa_{+} \,(-1083 + \,\nu) + 8 \,\delta \,(-3124 + 3 \,\nu)\big) \,(S\Sigma) + \big(-2 \,\delta \,\kappa_{-} \,(-3249 + 653 \,\nu) 
- 6 \,(201 - 4355 \,\nu + 4 \,\nu^2) - 2 \,\kappa_{+} \,(3249 - 7151 \,\nu + 6 \,\nu^2)\big) \,\bSigma^2\Big]\,.
\end{autobreak}
\end{align}
\end{subequations}

\subsection{Linear momentum and center-of-mass fluxes}\label{app:lmcomflux}

The coefficients for the expressions in sec.~\ref{sec:mom} are given by
\begin{subequations}
\begin{align}
\begin{autobreak}
h^{0\,i}_5
=\,(\bv\times\bS)^{i} \,(-669 \,\delta \,(nv)^3 + 251 \,\delta \,(nv) \,\bv^2) + \,(\bn\times\bS)^{i} \,(2940 \,\delta \,(nv)^4 - 5307 \,\delta \,(nv)^2 \,\bv^2 + 2659 \,\delta \,(\bv^2)^2) 
+ \,(\bv\times\bSigma)^{i} \,\big(-5331 \,(nv)^3 + 2513 \,(nv) \,\bv^2 + \,\nu \,(5952 \,(nv)^3 - 3464 \,(nv) \,\bv^2)\big) 
+ \,(\bn\times\bSigma)^{i} \,\big(9195 \,(nv)^4 - 9987 \,(nv)^2 \,\bv^2+ 2260 \,(\bv^2)^2 + \,\nu \,(-5442 \,(nv)^4 + 8610 \,(nv)^2 \,\bv^2 - 4594 \,(\bv^2)^2)\big) 
+ \,\big(\bn\times\bv\big)^{i} \,\big(-12 \,(nv)^3 \,(1541 \,\delta \,(nS) + 431 \,(n\Sigma))
+ 3 \,(nv) \,(5287 \,\delta \,(nS) + 1087 \,(n\Sigma)) \,\bv^2 
- 2 \,\bv^2 \,(2252 \,\delta \,(vS) + 1481 \,(v\Sigma))
 + \,(nv)^2 \,(6348 \,\delta \,(vS) + 6024 \,(v\Sigma))
+ \,\nu \,(32658 \,(nv)^3 \,(n\Sigma) - 27615 \,(nv) \,(n\Sigma) \,\bv^2 - 15717 \,(nv)^2 \,(v\Sigma) + 9247 \,\bv^2 \,(v\Sigma))\big) 
+ \,\bv^i \,\big(-21 \,(nv)^2 \,(463 \,\delta \,(Snv) + 274 \,(\Sigma nv))+ \,\bv^2 \,(9049 \,\delta \,(Snv) 
+ 4780 \,(\Sigma nv)) + \,\nu \,(22089 \,(nv)^2 \,(\Sigma nv) - 18265 \,\bv^2 \,(\Sigma nv))\big) 
+ \,\bn^i \,\big(-1056 \,(nv) \,\bv^2 \,(13 \,\delta \,(Snv) + 7 \,(\Sigma nv))+ \,(nv)^3 \,(14604 \,\delta \,(Snv) 
+ 8304 \,(\Sigma nv)) + \,\nu \,(-31146 \,(nv)^3 \,(\Sigma nv) + 27534 \,(nv) \,\bv^2 \,(\Sigma nv))\big)\,,
\end{autobreak}
\\[2ex]
\begin{autobreak}
h^{1\,i}_5
=496 \,\delta \,(nv) \,(\bv\times\bS)^{i} + \,(-6422 \,(nv) - 40 \,\nu \,(nv)) \,(\bv\times\bSigma)^{i}
+ \,(\bn\times\bSigma)^{i} \,\big(3557 \,(nv)^2 + \,\nu \,(4618 \,(nv)^2 - 3372 \,\bv^2) + 1389 \,\bv^2\big)
+ \,(\bn\times\bS)^{i} \,(-2092 \,\delta \,(nv)^2 + 1056 \,\delta \,\bv^2)
+ \,\big(\bn\times\bv\big)^{i} \,\big(\,(nv) \,(-1886 \,\delta \,(nS) + 4930 \,(n\Sigma)) + \,\nu \,(2807 \,(nv) \,(n\Sigma) - 759 \,(v\Sigma))
- 120 \,(\,\delta \,(vS) - 5 \,(v\Sigma))\big) + \,\bv^i \,(2568 \,\delta \,(Snv) + 669 \,(\Sigma nv) - 2391 \,\nu \,(\Sigma nv))
+ \,\bn^i \,\big(\,(nv) \,(-3460 \,\delta \,(Snv) - 1537 \,(\Sigma nv)) + 5563 \,\nu \,(nv) \,(\Sigma nv)\big)\,,
\end{autobreak}
\\[2ex]
\begin{autobreak}
h^{2\,i}_5
=-72 \,\delta \,(\bn\times\bS)^{i}\,,
\end{autobreak}
\\[2ex]
\begin{autobreak}
h^{0\,i}_6
=\,\bS^i \,\bigg\{\,(nv)^3 \,\Big[\,(1920 \,\delta + 4032 \,\kappa_{-} + 960 \,\delta \,\kappa_{+}) \,(nS)
+ \,\big(1536 \,\delta \,\kappa_{-} - 384 \,\kappa_{+} \,(4 + 5 \,\nu) - 60 \,(-45 + 64 \,\nu)\big) \,(n\Sigma)\Big]
+ \,(vS) \,(228 \,\delta \,\bv^2 + 1104 \,\kappa_{-} \,\bv^2 + 114 \,\delta \,\kappa_{+} \,\bv^2) 
+ \,(nv) \,\Big[\,(nS) \,(-1776 \,\delta \,\bv^2 - 3192 \,\kappa_{-} \,\bv^2 - 888 \,\delta \,\kappa_{+} \,\bv^2)
+ \,(n\Sigma) \,\big(-1152 \,\delta \,\kappa_{-} \,\bv^2 + 48 \,\kappa_{+} \,(24 + 37 \,\nu) \,\bv^2 + 12 \,(-103 + 296 \,\nu) \,\bv^2\big)\Big]
+ \,\big(495 \,\delta \,\kappa_{-} \,\bv^2 - 6 \,(-329 + 76 \,\nu) \,\bv^2 - 3 \,\kappa_{+} \,(165 + 76 \,\nu) \,\bv^2\big) \,(v\Sigma) 
+ \,(nv)^2 \,\Big[\,(-312 \,\delta - 1962 \,\kappa_{-} - 156 \,\delta \,\kappa_{+}) \,(vS)
+ \,\big(-903 \,\delta \,\kappa_{-} + 6 \,(-549 + 104 \,\nu) + 3 \,\kappa_{+} \,(301 + 104 \,\nu)\big) \,(v\Sigma)\Big]\bigg\}
+ \,\bigg\{\,(nv)^3 \,\Big[\,\big(1536 \,\delta \,\kappa_{-} - 384 \,\kappa_{+} \,(4 + 5 \,\nu) - 12 \,(251 + 320 \,\nu)\big) \,(nS)
+ \,\big(-192 \,\delta \,\kappa_{+} \,(8 + 5 \,\nu) - 192 \,\kappa_{-} \,(-8 + 11 \,\nu) - 60 \,\delta \,(19 + 32 \,\nu)\big) \,(n\Sigma)\Big]
+ \,(vS) \,\big(495 \,\delta \,\kappa_{-} \,\bv^2 - 8 \,(70 + 57 \,\nu) \,\bv^2 - 3 \,\kappa_{+} \,(165 + 76 \,\nu) \,\bv^2\big)
+ \,(nv) \,\Big[\,(n\Sigma) \,\big(24 \,\delta \,\kappa_{+} \,(48 + 37 \,\nu) \,\bv^2 + 24 \,\kappa_{-} \,(-48 + 59 \,\nu) \,\bv^2+ 12 \,\delta \,(131 + 148 \,\nu) \,\bv^2\big)
+ \,(nS) \,\big(-1152 \,\delta \,\kappa_{-} \,\bv^2 + 48 \,\kappa_{+} \,(24 + 37 \,\nu) \,\bv^2 + 12 \,(219 + 296 \,\nu) \,\bv^2\big)\Big]
+ \,\big(-3 \,\delta \,\kappa_{+} \,(165 + 38 \,\nu) \,\bv^2 - 2 \,\delta \,(-299 + 114 \,\nu) \,\bv^2 - 3 \,\kappa_{-} \,(-165 + 292 \,\nu) \,\bv^2\big) \,(v\Sigma)
+ \,(nv)^2 \,\Big[\,\big(-903 \,\delta \,\kappa_{-} + 156 \,(5 + 4 \,\nu) + 3 \,\kappa_{+} \,(301 + 104 \,\nu)\big) \,(vS)
+ \,\big(6 \,\delta \,(-185 + 52 \,\nu) + 3 \,\delta \,\kappa_{+} \,(301 + 52 \,\nu) + 3 \,\kappa_{-} \,(-301 + 550 \,\nu)\big) \,(v\Sigma)\Big]\bigg\} \,\bSigma^i
+ \,\bv^i \Bigg(\,(-888 \,\delta - 588 \,\kappa_{-} - 444 \,\delta \,\kappa_{+}) \,(vS)^2 + \,\bS^2 \,(-1248 \,\delta \,\bv^2 - 120 \,\kappa_{-} \,\bv^2 - 624 \,\delta \,\kappa_{+} \,\bv^2) 
+ \,(nS)^2 \,(4404 \,\delta \,\bv^2 - 156 \,\kappa_{-} \,\bv^2 + 2202 \,\delta \,\kappa_{+} \,\bv^2) 
+ \,(S\Sigma) \,\big(504 \,\delta \,\kappa_{-} \,\bv^2 + 24 \,\kappa_{+} \,(-21 + 104 \,\nu) \,\bv^2 + 4 \,(-449 + 1248 \,\nu) \,\bv^2\big) 
+ \,(n\Sigma)^2 \,\big(-3 \,\delta \,\kappa_{+} \,(-393 + 734 \,\nu) \,\bv^2 - 6 \,\delta \,(-71 + 734 \,\nu) \,\bv^2 + 3 \,\kappa_{-} \,(-393 + 1520 \,\nu) \,\bv^2\big)
+ \,(nS) \,(n\Sigma) \,\big(-2358 \,\delta \,\kappa_{-} \,\bv^2 - 6 \,\kappa_{+} \,(-393 + 1468 \,\nu) \,\bv^2 - 6 \,(-875 + 2936 \,\nu) \,\bv^2\big)
+ \,\big(-144 \,\delta \,\kappa_{-} + 48 \,\kappa_{+} \,(3 + 37 \,\nu) + 2 \,(-953 + 1776 \,\nu)\big) \,(vS) \,(v\Sigma)
+ \,\big(-12 \,\kappa_{-} \,(6 + 25 \,\nu) + 12 \,\delta \,\kappa_{+} \,(6 + 37 \,\nu) + 2 \,\delta \,(-257 + 444 \,\nu)\big) \,(v\Sigma)^2
+ \,(nv) \,\bigg\{\,\Big[\,(5268 \,\delta + 2190 \,\kappa_{-} + 2634 \,\delta \,\kappa_{+}) \,(nS) 
+ \,\big(-222 \,\delta \,\kappa_{-} - 6 \,\kappa_{+} \,(-37 + 878 \,\nu) - 6 \,(-127 + 1756 \,\nu)\big) \,(n\Sigma)\Big] \,(vS)
+ \,\Big[\,\big(-222 \,\delta \,\kappa_{-} - 24 \,(-93 + 439 \,\nu) - 6 \,\kappa_{+} \,(-37 + 878 \,\nu)\big) \,(nS) 
+ \,\big(-6 \,\delta \,\kappa_{+} \,(-37 + 439 \,\nu) + 6 \,\kappa_{-} \,(-37 + 513 \,\nu) - 6 \,\delta \,(183 + 878 \,\nu)\big) \,(n\Sigma)\Big] \,(v\Sigma)\bigg\}
+ \,\big(12 \,\delta \,\kappa_{+} \,(-21 + 52 \,\nu) \,\bv^2 - 12 \,\kappa_{-} \,(-21 + 94 \,\nu) \,\bv^2 + 8 \,\delta \,(-37 + 156 \,\nu) \,\bv^2\big)\,\bSigma^2 
+ \,(nv)^2 \,\Big[\,(-8988 \,\delta - 1632 \,\kappa_{-} - 4494 \,\delta \,\kappa_{+}) \,(nS)^2 + \,\big(2862 \,\delta \,\kappa_{-} 
+ 6 \,\kappa_{+} \,(-477 + 2996 \,\nu) + 6 \,(-989 + 5992 \,\nu)\big) \,(nS) \,(n\Sigma) 
+ \,\big(3 \,\delta \,\kappa_{+} \,(-477 + 1498 \,\nu) + 6 \,\delta \,(243 + 1498 \,\nu) - 3 \,\kappa_{-} \,(-477 + 2452 \,\nu)\big) \,(n\Sigma)^2
+ \,(1344 \,\delta + 468 \,\kappa_{-} + 672 \,\delta \,\kappa_{+}) \,\bS^2 + \,\big(-204 \,\delta \,\kappa_{-} - 12 \,\kappa_{+} \,(-17 + 224 \,\nu) - 12 \,(-211 + 448 \,\nu)\big) \,(S\Sigma)
+ \,\big(-24 \,\delta \,(-25 + 56 \,\nu) - 6 \,\delta \,\kappa_{+} \,(-17 + 112 \,\nu) + 6 \,\kappa_{-} \,(-17 + 146 \,\nu)\big) \,\bSigma^2\Big]\Bigg)
+ \,\bn^i\,\Bigg(\,(vS) \,\Big[\,(nS) \,(1992 \,\delta \,\bv^2 - 564 \,\kappa_{-} \,\bv^2 + 996 \,\delta \,\kappa_{+} \,\bv^2)
+ \,(n\Sigma) \,\big(-780 \,\delta \,\kappa_{-} \,\bv^2 - 24 \,(-89 + 166 \,\nu) \,\bv^2 - 12 \,\kappa_{+} \,(-65 + 166 \,\nu) \,\bv^2\big)\Big] 
+ \,\Big[\,(n\Sigma) \,\big(-12 \,\delta \,\kappa_{+} \,(-65 + 83 \,\nu) \,\bv^2 - 12 \,\delta \,(69 + 166 \,\nu) \,\bv^2 + 12 \,\kappa_{-} \,(-65 + 213 \,\nu) \,\bv^2\big)
+ \,(nS) \,\big(-780 \,\delta \,\kappa_{-} \,\bv^2 - 12 \,\kappa_{+} \,(-65 + 166 \,\nu) \,\bv^2 - 12 \,(151 + 332 \,\nu) \,\bv^2\big)\Big] \,(v\Sigma)
+ \,(nv)^2 \,\bigg\{\,\Big[\,(-8472 \,\delta - 2172 \,\kappa_{-} - 4236 \,\delta \,\kappa_{+}) \,(nS) + \,\big(1032 \,\delta \,\kappa_{-} + 24 \,\kappa_{+} \,(-43 + 353 \,\nu)
+ 6 \,(-1023 + 2824 \,\nu)\big) \,(n\Sigma)\Big] \,(vS) + \,\Big[\,\big(1032 \,\delta \,\kappa_{-} + 24 \,\kappa_{+} \,(-43 + 353 \,\nu) + 12 \,(-151 + 1412 \,\nu)\big) \,(nS)
+ \,\big(12 \,\delta \,\kappa_{+} \,(-86 + 353 \,\nu) - 12 \,\kappa_{-} \,(-86 + 525 \,\nu) + 6 \,\delta \,(31 + 1412 \,\nu)\big) \,(n\Sigma)\Big] \,(v\Sigma)\bigg\} 
+ \,(nv)^3 \,\Big[\,(12600 \,\delta - 1230 \,\kappa_{-} + 6300 \,\delta \,\kappa_{+}) \,(nS)^2 + \,\big(-7530 \,\delta \,\kappa_{-} - 90 \,(-143 + 560 \,\nu)
- 30 \,\kappa_{+} \,(-251 + 840 \,\nu)\big) \,(nS) \,(n\Sigma) + \,\big(-90 \,\delta \,(-3 + 140 \,\nu) - 15 \,\delta \,\kappa_{+} \,(-251 + 420 \,\nu) 
+ 15 \,\kappa_{-} \,(-251 + 922 \,\nu)\big) \,(n\Sigma)^2 + \,(-2016 \,\delta - 210 \,\kappa_{-} - 1008 \,\delta \,\kappa_{+}) \,\bS^2 
+ \,\big(798 \,\delta \,\kappa_{-} + 42 \,\kappa_{+} \,(-19 + 96 \,\nu) + 48 \,(-25 + 168 \,\nu)\big) \,(S\Sigma) 
+ \,\big(21 \,\delta \,\kappa_{+} \,(-19 + 48 \,\nu) + 36 \,\delta \,(11 + 56 \,\nu) - 21 \,\kappa_{-} \,(-19 + 86 \,\nu)\big) \,\bSigma^2\Big]
+ \,(nv) \,\Big[\,(1464 \,\delta + 1200 \,\kappa_{-} + 732 \,\delta \,\kappa_{+}) \,(vS)^2 + \,(nS)^2 \,(-7512 \,\delta \,\bv^2 + 2934 \,\kappa_{-} \,\bv^2 
- 3756 \,\delta \,\kappa_{+} \,\bv^2) + \,\bS^2 \,(1944 \,\delta \,\bv^2 - 126 \,\kappa_{-} \,\bv^2 + 972 \,\delta \,\kappa_{+} \,\bv^2) + \,(S\Sigma) \,\big(-1098 \,\delta \,\kappa_{-} \,\bv^2
- 72 \,(-7 + 108 \,\nu) \,\bv^2 - 18 \,\kappa_{+} \,(-61 + 216 \,\nu) \,\bv^2\big) 
+ \,(n\Sigma)^2 \,\big(3 \,\delta \,\kappa_{+} \,(-1115 + 1252 \,\nu) \,\bv^2
+ 6 \,\delta \,(-233 + 1252 \,\nu) \,\bv^2 - 3 \,\kappa_{-} \,(-1115 + 3482 \,\nu) \,\bv^2\big)
+ \,(nS) \,(n\Sigma) \,\big(6690 \,\delta \,\kappa_{-} \,\bv^2 + 6 \,\kappa_{+} \,(-1115 + 2504 \,\nu) \,\bv^2 + 6 \,(-1709 + 5008 \,\nu) \,\bv^2\big)
+ \,\big(468 \,\delta \,\kappa_{-} - 12 \,\kappa_{+} \,(39 + 244 \,\nu) - 6 \,(-751 + 976 \,\nu)\big) \,(vS) \,(v\Sigma)
 + \,\big(6 \,\kappa_{-} \,(39 + 44 \,\nu) - 6 \,\delta \,\kappa_{+} \,(39 + 122 \,\nu) 
- 6 \,\delta \,(-255 + 244 \,\nu)\big) \,(v\Sigma)^2 
+ \,\big(-36 \,\delta \,(19 + 54 \,\nu) \,\bv^2 - 9 \,\delta \,\kappa_{+} \,(-61 + 108 \,\nu) \,\bv^2
+ 9 \,\kappa_{-} \,(-61 + 230 \,\nu) \,\bv^2\big) \,\bSigma^2\Big]\Bigg)\,,
\end{autobreak}
\\[2ex]
\begin{autobreak}
h^{1\,i}_6
=\,\bS^i \,\bigg\{\,(nv) \,\Big[\,(60 \,\delta + 32 \,\kappa_{-} + 30 \,\delta \,\kappa_{+}) \,(nS) + \,\big(\,\delta \,\kappa_{-} + \,\kappa_{+} \,(-1 - 60 \,\nu) - 5 \,(1 + 24 \,\nu)\big) \,(n\Sigma)\Big] 
+ \,(-12 \,\delta - 44 \,\kappa_{-} - 6 \,\delta \,\kappa_{+}) \,(vS) + \,\big(59 - 19 \,\delta \,\kappa_{-} + 24 \,\nu + \,\kappa_{+} \,(19 + 12 \,\nu)\big) \,(v\Sigma)\bigg\}
+ \,\bigg\{\,(nv) \,\Big[\,\big(79 + \,\delta \,\kappa_{-} + \,\kappa_{+} \,(-1 - 60 \,\nu) - 120 \,\nu\big) \,(nS) + \,\big(\,\kappa_{-} \,(1 + 28 \,\nu) -  \,\delta \,\kappa_{+} \,(1 + 30 \,\nu) 
-  \,\delta \,(7 + 60 \,\nu)\big) \,(n\Sigma)\Big] + \,\big(-67 - 19 \,\delta \,\kappa_{-} + 24 \,\nu + \,\kappa_{+} \,(19 + 12 \,\nu)\big) \,(vS)
+ \,\big(-3 \,\delta^3 + \,\delta \,\kappa_{+} \,(19 + 6 \,\nu) + \,\kappa_{-} \,(-19 + 32 \,\nu)\big) \,(v\Sigma)\bigg\} \,\bSigma^i
+ \,\bv^i \,\Big[\,(192 \,\delta - 16 \,\kappa_{-} + 96 \,\delta \,\kappa_{+}) \,(nS)^2 + \,\big(-112 \,\delta \,\kappa_{-} - 16 \,\kappa_{+} \,(-7 + 24 \,\nu) - 4 \,(-49 + 192 \,\nu)\big) \,(nS) \,(n\Sigma) 
+ \,\big(-8 \,\delta \,\kappa_{+} \,(-7 + 12 \,\nu) + 8 \,\kappa_{-} \,(-7 + 26 \,\nu) - 3 \,\delta \,(1 + 64 \,\nu)\big) \,(n\Sigma)^2 + \,(-60 \,\delta + 20 \,\kappa_{-} - 30 \,\delta \,\kappa_{+}) \,\bS^2 
+ \,\big(50 \,\delta \,\kappa_{-} + 10 \,\kappa_{+} \,(-5 + 12 \,\nu) + 5 \,(-13 + 48 \,\nu)\big) \,(S\Sigma)
+ \,\big(5 \,\delta \,\kappa_{+} \,(-5 + 6 \,\nu) - 5 \,\kappa_{-} \,(-5 + 16 \,\nu) + 2 \,\delta \,(1 + 30 \,\nu)\big) \,\bSigma^2\Big]
+ \,\bn^i \,\bigg\{\,\Big[\,(24 \,\delta + 52 \,\kappa_{-} + 12 \,\delta \,\kappa_{+}) \,(nS)
 + \,\big(20 \,\delta \,\kappa_{-} - 4 \,\kappa_{+} \,(5 + 6 \,\nu) - 3 \,(-17 + 16 \,\nu)\big) \,(n\Sigma)\Big] \,(vS) 
+ \,\Big[\,\big(20 \,\delta \,\kappa_{-} - 4 \,\kappa_{+} \,(5 + 6 \,\nu) - 2 \,(13 + 24 \,\nu)\big) \,(nS) 
+ \,\big(-4 \,\delta \,\kappa_{+} \,(5 + 3 \,\nu) - 4 \,\kappa_{-} \,(-5 + 7 \,\nu) -  \,\delta \,(-1 + 24 \,\nu)\big) \,(n\Sigma)\Big] \,(v\Sigma)
+ \,(nv) \,\Big[\,(-282 \,\delta - 27 \,\kappa_{-} - 141 \,\delta \,\kappa_{+}) \,(nS)^2 
+ \,\big(114 \,\delta \,\kappa_{-} + 6 \,\kappa_{+} \,(-19 + 94 \,\nu) + 3 \,(-99 + 376 \,\nu)\big) \,(nS) \,(n\Sigma) 
+ \,\big(3 \,\delta \,\kappa_{+} \,(-19 + 47 \,\nu) + 6 \,\delta \,(1 + 47 \,\nu) - 3 \,\kappa_{-} \,(-19 + 85 \,\nu)\big) \,(n\Sigma)^2 
+ \,(66 \,\delta - 19 \,\kappa_{-} + 33 \,\delta \,\kappa_{+}) \,\bS^2 
+ \,\big(59 - 52 \,\delta \,\kappa_{-} - 264 \,\nu - 4 \,\kappa_{+} \,(-13 + 33 \,\nu)\big) \,(S\Sigma)
+ \,\big(-66 \,\delta \,\nu -  \,\delta \,\kappa_{+} \,(-26 + 33 \,\nu) + \,\kappa_{-} \,(-26 + 85 \,\nu)\big) \,\bSigma^2\Big]\bigg\}\,.
\end{autobreak}
\end{align}
\end{subequations}

\begin{subequations}
\begin{align}
\begin{autobreak}
k^{0\,i}_5
=\,(\bv\times\bS)^{i} \,(-2475 \,\delta \,(nv)^4 + 3726 \,\delta \,(nv)^2 \,\bv^2 - 1135 \,\delta \,(\bv^2)^2) 
+ \,(\bv\times\bSigma)^{i} \,\big(75 \,(-17 + 71 \,\nu) \,(nv)^4 - 6 \,(-313 + 1315 \,\nu) \,(nv)^2 \,\bv^2 + \,(-559 + 2325 \,\nu) \,(\bv^2)^2\big)
+ \,(\bn\times\bS)^{i} \,(3675 \,\delta \,(nv)^5 - 6210 \,\delta \,(nv)^3 \,\bv^2 + 2451 \,\delta \,(nv) \,(\bv^2)^2)
+ \,(\bn\times\bSigma)^{i} \,\big(-525 \,(-3 + 13 \,\nu) \,(nv)^5 + 30 \,(-87 + 379 \,\nu) \,(nv)^3 \,\bv^2 - 3 \,(-341 + 1483 \,\nu) \,(nv) \,(\bv^2)^2\big) 
+ \,\bv^i \,\Big[\,(nv)^3 \,\big(-1680 \,\delta \,(Snv) + 60 \,(-14 + 47 \,\nu) \,(\Sigma nv)\big) + \,(nv) \,\big(1716 \,\delta \,(Snv) \,\bv^2 - 12 \,(-73 + 253 \,\nu) \,\bv^2 \,(\Sigma nv)\big)\Big] 
+ \,\bn^i \,\Big[1878 \,\delta \,(Snv) \,(\bv^2)^2 - 6 \,(-167 + 645 \,\nu) \,(\bv^2)^2 \,(\Sigma nv) + \,(nv)^4 \,\big(5250 \,\delta \,(Snv) - 1050 \,(-3 + 11 \,\nu) \,(\Sigma nv)\big) 
+ \,(nv)^2 \,\big(-7320 \,\delta \,(Snv) \,\bv^2 + 60 \,(-70 + 261 \,\nu) \,\bv^2 \,(\Sigma nv)\big)\Big]
\end{autobreak}
\\[2ex]
\begin{autobreak}
k^{1\,i}_5
=\,(\bv\times\bS)^{i} \,(792 \,\delta \,(nv)^2 - 778 \,\delta \,\bv^2) + \,(\bv\times\bSigma)^{i} \,\big(-3 \,(-136 + 567 \,\nu) \,(nv)^2 + 3 \,(-134 + 567 \,\nu) \,\bv^2\big)
+ \,(\bn\times\bS)^{i} \,(-920 \,\delta \,(nv)^3 + 806 \,\delta \,(nv) \,\bv^2) + \,(\bn\times\bSigma)^{i} \,\big(\,(-404 + 1763 \,\nu) \,(nv)^3 -  \,(-354 + 1559 \,\nu) \,(nv) \,\bv^2\big)
+ \,\big(\bn\times\bv\big)^{i} \,\Big[\,(nv)^2 \,\big(-16 \,\delta \,(nS) - 2 \,(-7 + 13 \,\nu) \,(n\Sigma)\big) + 12 \,\delta \,(nS) \,\bv^2 + 2 \,(-7 + 15 \,\nu) \,(n\Sigma) \,\bv^2 
+ \,(nv) \,(4 \,\delta \,(vS) - 4 \,\nu \,(v\Sigma))\Big] + \,(nv) \,\bv^i \,\big(-48 \,\delta \,(Snv) + 4 \,(-21 + 116 \,\nu) \,(\Sigma nv)\big) 
+ \,\bn^i \,\Big[2025 \,\delta \,(Snv) \,\bv^2 - 5 \,(-175 + 662 \,\nu) \,\bv^2 \,(\Sigma nv) + \,(nv)^2 \,\big(-1955 \,\delta \,(Snv) + \,(-833 + 2978 \,\nu) \,(\Sigma nv)\big)\Big]\,,
\end{autobreak}
\\[2ex]
\begin{autobreak}
k^{2\,i}_5
=\,59 \,\delta \,(\bn\times\bS)^{i} \,(nv) + \,(21 - 95 \,\nu) \,(\bn\times\bSigma)^{i} \,(nv) + \,\big(\bn\times\bv\big)^{i} \,(\,\delta \,(nS) -  \,\nu \,(n\Sigma)) - 35 \,\delta \,(\bv\times\bS)^{i}
+ 5 \,(-3 + 13 \,\nu) \,(\bv\times\bSigma)^{i} + \,\bn^i \,\big(115 \,\delta \,(Snv) - 2 \,(-12 + 41 \,\nu) \,(\Sigma nv)\big)\,,
\end{autobreak}
\\[2ex]
\begin{autobreak}
k^{0\,i}_6
=\,\bS^i \,\bigg\{\,(nv)^4 \,\big(\,(-2100 \,\delta - 1050 \,\delta \,\kappa_{+}) \,(nS) + \,(-1050 \,\delta^2 + 525 \,\delta \,\kappa_{-} - 525 \,\delta^2 \,\kappa_{+}) \,(n\Sigma)\big)
+ \,(nS) \,(-852 \,\delta \,(\bv^2)^2 - 12 \,\kappa_{-} \,(\bv^2)^2 - 426 \,\delta \,\kappa_{+} \,(\bv^2)^2) + \,(n\Sigma) \,\big(-426 \,\delta^2 \,(\bv^2)^2 + 207 \,\delta \,\kappa_{-} \,(\bv^2)^2 
+ 3 \,\kappa_{+} \,(-69 + 284 \,\nu) \,(\bv^2)^2\big) + \,(nv)^2 \,\Big[\,(nS) \,(3000 \,\delta \,\bv^2 + 60 \,\kappa_{-} \,\bv^2 + 1500 \,\delta \,\kappa_{+} \,\bv^2)
+ \,(n\Sigma) \,\big(1500 \,\delta^2 \,\bv^2 - 720 \,\delta \,\kappa_{-} \,\bv^2 - 120 \,\kappa_{+} \,(-6 + 25 \,\nu) \,\bv^2\big)\Big] 
+ \,(nv)^3 \,\Big[\,(600 \,\delta - 300 \,\kappa_{-} + 300 \,\delta \,\kappa_{+}) \,(vS) + \,\big(300 \,\delta^2 - 300 \,\delta \,\kappa_{-} - 300 \,\kappa_{+} \,(-1 + 2 \,\nu)\big) \,(v\Sigma)\Big] 
+ \,(nv) \,\Big[\,(vS) \,(-576 \,\delta \,\bv^2 + 156 \,\kappa_{-} \,\bv^2 - 288 \,\delta \,\kappa_{+} \,\bv^2) + \,\big(-288 \,\delta^2 \,\bv^2 + 222 \,\delta \,\kappa_{-} \,\bv^2
+ 6 \,\kappa_{+} \,(-37 + 96 \,\nu) \,\bv^2\big) \,(v\Sigma)\Big]\bigg\} + \,\bigg\{\,(nv)^4 \,\Big[\,(-1050 \,\delta^2 + 525 \,\delta \,\kappa_{-} - 525 \,\delta^2 \,\kappa_{+}) \,(nS) 
+ \,\big(525 \,\delta^2 \,\kappa_{-} + 2100 \,\delta \,\nu + 525 \,\delta \,\kappa_{+} \,(-1 + 2 \,\nu)\big) \,(n\Sigma)\Big] + \,(n\Sigma) \,\big(852 \,\delta \,\nu \,(\bv^2)^2
+ 3 \,\delta \,\kappa_{+} \,(-69 + 142 \,\nu) \,(\bv^2)^2 - 3 \,\kappa_{-} \,(-69 + 280 \,\nu) \,(\bv^2)^2\big) + \,(nS) \,\big(-426 \,\delta^2 \,(\bv^2)^2 + 207 \,\delta \,\kappa_{-} \,(\bv^2)^2 
+ 3 \,\kappa_{+} \,(-69 + 284 \,\nu) \,(\bv^2)^2\big) + \,(nv)^2 \,\Big[\,(nS) \,\big(1500 \,\delta^2 \,\bv^2 - 720 \,\delta \,\kappa_{-} \,\bv^2 - 120 \,\kappa_{+} \,(-6 + 25 \,\nu) \,\bv^2\big)
+ \,(n\Sigma) \,\big(-3000 \,\delta \,\nu \,\bv^2 - 60 \,\delta \,\kappa_{+} \,(-12 + 25 \,\nu) \,\bv^2 + 60 \,\kappa_{-} \,(-12 + 49 \,\nu) \,\bv^2\big)\Big] 
+ \,(nv)^3 \,\Big[\,\big(300 \,\delta^2 - 300 \,\delta \,\kappa_{-} - 300 \,\kappa_{+} \,(-1 + 2 \,\nu)\big) \,(vS) + \,\big(-300 \,\delta \,\kappa_{+} \,(-1 + \,\nu) 
- 600 \,\delta \,\nu + 300 \,\kappa_{-} \,(-1 + 3 \,\nu)\big) \,(v\Sigma)\Big] + \,(nv) \,\Big[\,(vS) \,\big(-288 \,\delta^2 \,\bv^2 + 222 \,\delta \,\kappa_{-} \,\bv^2 
+ 6 \,\kappa_{+} \,(-37 + 96 \,\nu) \,\bv^2\big) + \,\big(576 \,\delta \,\nu \,\bv^2 + 6 \,\delta \,\kappa_{+} \,(-37 + 48 \,\nu) \,\bv^2 - 6 \,\kappa_{-} \,(-37 + 122 \,\nu) \,\bv^2\big) \,(v\Sigma)\Big]\bigg\} \,\bSigma^i
+ \,\bv^i \Biggl(\,(vS) \,\Big[\,(nS) \,(-984 \,\delta \,\bv^2 - 24 \,\kappa_{-} \,\bv^2 - 492 \,\delta \,\kappa_{+} \,\bv^2) + \,(n\Sigma) \,\big(-492 \,\delta^2 \,\bv^2
+ 234 \,\delta \,\kappa_{-} \,\bv^2 + 6 \,\kappa_{+} \,(-39 + 164 \,\nu) \,\bv^2\big)\Big] + \,\Big[\,(n\Sigma) \,\big(984 \,\delta \,\nu \,\bv^2
+ 6 \,\delta \,\kappa_{+} \,(-39 + 82 \,\nu) \,\bv^2 - 6 \,\kappa_{-} \,(-39 + 160 \,\nu) \,\bv^2\big) + \,(nS) \,\big(-492 \,\delta^2 \,\bv^2 + 234 \,\delta \,\kappa_{-} \,\bv^2
+ 6 \,\kappa_{+} \,(-39 + 164 \,\nu) \,\bv^2\big)\Big] \,(v\Sigma) + \,(nv)^2 \,\bigg\{\,\Big[\,(2880 \,\delta + 120 \,\kappa_{-} + 1440 \,\delta \,\kappa_{+}) \,(nS)
+ \,\big(1440 \,\delta^2 - 660 \,\delta \,\kappa_{-} - 60 \,\kappa_{+} \,(-11 + 48 \,\nu)\big) \,(n\Sigma)\Big] \,(vS)
+ \,\Big[\,\big(1440 \,\delta^2 - 660 \,\delta \,\kappa_{-} - 60 \,\kappa_{+} \,(-11 + 48 \,\nu)\big) \,(nS)
+ \,\big(-2880 \,\delta \,\nu - 60 \,\delta \,\kappa_{+} \,(-11 + 24 \,\nu) + 60 \,\kappa_{-} \,(-11 + 46 \,\nu)\big) \,(n\Sigma)\Big] \,(v\Sigma)\bigg\}
+ \,(nv)^3 \,\Big[\,(-5460 \,\delta - 2730 \,\delta \,\kappa_{+}) \,(nS)^2 + \,(-5460 \,\delta^2 + 2730 \,\delta \,\kappa_{-} - 2730 \,\delta^2 \,\kappa_{+}) \,(nS) \,(n\Sigma) 
+ \,\big(1365 \,\delta^2 \,\kappa_{-} + 5460 \,\delta \,\nu + 1365 \,\delta \,\kappa_{+} \,(-1 + 2 \,\nu)\big) \,(n\Sigma)^2 + \,(660 \,\delta + 60 \,\kappa_{-} + 330 \,\delta \,\kappa_{+}) \,\bS^2
+ \,\big(660 \,\delta^2 - 270 \,\delta \,\kappa_{-} - 30 \,\kappa_{+} \,(-9 + 44 \,\nu)\big) \,(S\Sigma) + \,\big(-660 \,\delta \,\nu 
- 15 \,\delta \,\kappa_{+} \,(-9 + 22 \,\nu) + 15 \,\kappa_{-} \,(-9 + 40 \,\nu)\big) \,\bSigma^2\Big] + \,(nv) \,\Big[\,(-336 \,\delta + 12 \,\kappa_{-} - 168 \,\delta \,\kappa_{+}) \,(vS)^2 
+ \,\bS^2 \,(-708 \,\delta \,\bv^2 - 48 \,\kappa_{-} \,\bv^2 - 354 \,\delta \,\kappa_{+} \,\bv^2) + \,(nS)^2 \,(4020 \,\delta \,\bv^2 + 2010 \,\delta \,\kappa_{+} \,\bv^2)
+ \,(nS) \,(n\Sigma) \,(4020 \,\delta^2 \,\bv^2 - 2010 \,\delta \,\kappa_{-} \,\bv^2 + 2010 \,\delta^2 \,\kappa_{+} \,\bv^2) + \,(n\Sigma)^2 \,\big(-1005 \,\delta^2 \,\kappa_{-} \,\bv^2
- 4020 \,\delta \,\nu \,\bv^2 - 1005 \,\delta \,\kappa_{+} \,(-1 + 2 \,\nu) \,\bv^2\big) + \,(S\Sigma) \,\big(-708 \,\delta^2 \,\bv^2 + 306 \,\delta \,\kappa_{-} \,\bv^2 + 6 \,\kappa_{+} \,(-51 + 236 \,\nu) \,\bv^2\big) 
+ \,\big(-336 \,\delta^2 + 180 \,\delta \,\kappa_{-} + 12 \,\kappa_{+} \,(-15 + 56 \,\nu)\big) \,(vS) \,(v\Sigma)
+ \,\big(336 \,\delta \,\nu + 6 \,\delta \,\kappa_{+} \,(-15 + 28 \,\nu) - 6 \,\kappa_{-} \,(-15 + 58 \,\nu)\big) \,(v\Sigma)^2 
+ \,\big(708 \,\delta \,\nu \,\bv^2 + 3 \,\delta \,\kappa_{+} \,(-51 + 118 \,\nu) \,\bv^2 - 3 \,\kappa_{-} \,(-51 + 220 \,\nu) \,\bv^2\big) \,\bSigma^2\Big]\Biggr) 
+ \,\bn^i \Biggl(\,(vS)^2 \,(-636 \,\delta \,\bv^2 + 30 \,\kappa_{-} \,\bv^2 - 318 \,\delta \,\kappa_{+} \,\bv^2) + \,\bS^2 \,(-714 \,\delta \,(\bv^2)^2 - 6 \,\kappa_{-} \,(\bv^2)^2 
- 357 \,\delta \,\kappa_{+} \,(\bv^2)^2) + \,(nS)^2 \,(3630 \,\delta \,(\bv^2)^2 + 1815 \,\delta \,\kappa_{+} \,(\bv^2)^2) + \,(nS) \,(n\Sigma) \,(3630 \,\delta^2 \,(\bv^2)^2 - 1815 \,\delta \,\kappa_{-} \,(\bv^2)^2 
+ 1815 \,\delta^2 \,\kappa_{+} \,(\bv^2)^2) + \,(n\Sigma)^2 \,\big(- \tfrac{1815}{2} \,\delta^2 \,\kappa_{-} \,(\bv^2)^2 - 3630 \,\delta \,\nu \,(\bv^2)^2
-  \tfrac{1815}{2} \,\delta \,\kappa_{+} \,(-1 + 2 \,\nu) \,(\bv^2)^2\big) + \,(S\Sigma) \,\big(-714 \,\delta^2 \,(\bv^2)^2 + 351 \,\delta \,\kappa_{-} \,(\bv^2)^2 + 3 \,\kappa_{+} \,(-117 + 476 \,\nu) \,(\bv^2)^2\big)
+ \,(vS) \,\big(-636 \,\delta^2 \,\bv^2 + 348 \,\delta \,\kappa_{-} \,\bv^2 + 12 \,\kappa_{+} \,(-29 + 106 \,\nu) \,\bv^2\big) \,(v\Sigma) + \,\big(636 \,\delta \,\nu \,\bv^2
+ 6 \,\delta \,\kappa_{+} \,(-29 + 53 \,\nu) \,\bv^2 - 6 \,\kappa_{-} \,(-29 + 111 \,\nu) \,\bv^2\big) \,(v\Sigma)^2 + \,(nv)^3 \,\bigg\{\,\big(\,(-16800 \,\delta - 8400 \,\delta \,\kappa_{+}) \,(nS) 
+ \,(-8400 \,\delta^2 + 4200 \,\delta \,\kappa_{-} - 4200 \,\delta^2 \,\kappa_{+}) \,(n\Sigma)\big) \,(vS) + \,\Big[\,(-8400 \,\delta^2 + 4200 \,\delta \,\kappa_{-} - 4200 \,\delta^2 \,\kappa_{+}) \,(nS) 
+ \,\big(4200 \,\delta^2 \,\kappa_{-} + 16800 \,\delta \,\nu + 4200 \,\delta \,\kappa_{+} \,(-1 + 2 \,\nu)\big) \,(n\Sigma)\Big] \,(v\Sigma)\bigg\}
+ \,(nv) \,\bigg\{\,(vS) \,\big(\,(nS) \,(10680 \,\delta \,\bv^2 + 5340 \,\delta \,\kappa_{+} \,\bv^2) + \,(n\Sigma) \,(5340 \,\delta^2 \,\bv^2 - 2670 \,\delta \,\kappa_{-} \,\bv^2
+ 2670 \,\delta^2 \,\kappa_{+} \,\bv^2)\big) + \,\Big[\,(nS) \,(5340 \,\delta^2 \,\bv^2 - 2670 \,\delta \,\kappa_{-} \,\bv^2 + 2670 \,\delta^2 \,\kappa_{+} \,\bv^2)
+ \,(n\Sigma) \,\big(-2670 \,\delta^2 \,\kappa_{-} \,\bv^2 - 10680 \,\delta \,\nu \,\bv^2 - 2670 \,\delta \,\kappa_{+} \,(-1 + 2 \,\nu) \,\bv^2\big)\Big] \,(v\Sigma)\bigg\}
+ \,\big(714 \,\delta \,\nu \,(\bv^2)^2 + \tfrac{3}{2} \,\delta \,\kappa_{+} \,(-117 + 238 \,\nu) \,(\bv^2)^2 -  \tfrac{3}{2} \,\kappa_{-} \,(-117 + 472 \,\nu) \,(\bv^2)^2\big) \,\bSigma^2
+ \,(nv)^4 \,\Big[\,(28350 \,\delta + 14175 \,\delta \,\kappa_{+}) \,(nS)^2 + \,(28350 \,\delta^2 - 14175 \,\delta \,\kappa_{-} + 14175 \,\delta^2 \,\kappa_{+}) \,(nS) \,(n\Sigma)
+ \,\big(- \tfrac{14175}{2} \,\delta^2 \,\kappa_{-} - 28350 \,\delta \,\nu -  \tfrac{14175}{2} \,\delta \,\kappa_{+} \,(-1 + 2 \,\nu)\big) \,(n\Sigma)^2 + \,(-3150 \,\delta - 1575 \,\delta \,\kappa_{+}) \,\bS^2 
+ \,(-3150 \,\delta^2 + 1575 \,\delta \,\kappa_{-} - 1575 \,\delta^2 \,\kappa_{+}) \,(S\Sigma) + \,\big(\tfrac{1575}{2} \,\delta^2 \,\kappa_{-} + 3150 \,\delta \,\nu
+ \frac{1575}{2} \,\delta \,\kappa_{+} \,(-1 + 2 \,\nu)\big) \,\bSigma^2\Big] + \,(nv)^2 \,\Big[\,(2100 \,\delta - 150 \,\kappa_{-} + 1050 \,\delta \,\kappa_{+}) \,(vS)^2
+ \,(nS)^2 \,(-27300 \,\delta \,\bv^2 - 13650 \,\delta \,\kappa_{+} \,\bv^2) + \,\bS^2 \,(3840 \,\delta \,\bv^2 + 30 \,\kappa_{-} \,\bv^2 + 1920 \,\delta \,\kappa_{+} \,\bv^2) 
+ \,(nS) \,(n\Sigma) \,(-27300 \,\delta^2 \,\bv^2 + 13650 \,\delta \,\kappa_{-} \,\bv^2 - 13650 \,\delta^2 \,\kappa_{+} \,\bv^2) + \,(n\Sigma)^2 \,\big(6825 \,\delta^2 \,\kappa_{-} \,\bv^2
+ 27300 \,\delta \,\nu \,\bv^2 + 6825 \,\delta \,\kappa_{+} \,(-1 + 2 \,\nu) \,\bv^2\big) + \,(S\Sigma) \,\big(3840 \,\delta^2 \,\bv^2 - 1890 \,\delta \,\kappa_{-} \,\bv^2 
- 30 \,\kappa_{+} \,(-63 + 256 \,\nu) \,\bv^2\big) + \,\big(2100 \,\delta^2 - 1200 \,\delta \,\kappa_{-} - 600 \,\kappa_{+} \,(-2 + 7 \,\nu)\big) \,(vS) \,(v\Sigma) 
+ \,\big(-2100 \,\delta \,\nu - 150 \,\delta \,\kappa_{+} \,(-4 + 7 \,\nu) + 150 \,\kappa_{-} \,(-4 + 15 \,\nu)\big) \,(v\Sigma)^2 + \,\big(-3840 \,\delta \,\nu \,\bv^2 
- 15 \,\delta \,\kappa_{+} \,(-63 + 128 \,\nu) \,\bv^2 + 15 \,\kappa_{-} \,(-63 + 254 \,\nu) \,\bv^2\big) \,\bSigma^2\Big]\Biggr)\,,
\end{autobreak}
\\[2ex]
\begin{autobreak}
k^{1\,i}_6
=\,\bS^i \,\bigg\{\,(nv)^2 \,\Big[\,(1212 \,\delta + 108 \,\kappa_{-} + 606 \,\delta \,\kappa_{+}) \,(nS) 
+ \,\big(-249 \,\delta \,\kappa_{-} - 6 \,(-121 + 404 \,\nu) - 3 \,\kappa_{+} \,(-83 + 404 \,\nu)\big) \,(n\Sigma)\Big]
+ \,(nS) \,(-1092 \,\delta \,\bv^2 - 80 \,\kappa_{-} \,\bv^2 - 546 \,\delta \,\kappa_{+} \,\bv^2) + \,(n\Sigma) \,\big(233 \,\delta \,\kappa_{-} \,\bv^2 + 26 \,(-25 + 84 \,\nu) \,\bv^2 
+ \,\kappa_{+} \,(-233 + 1092 \,\nu) \,\bv^2\big) + \,(nv) \,\big(68 \,\kappa_{-} \,(vS) + \,(-16 + 34 \,\delta \,\kappa_{-} - 34 \,\kappa_{+}) \,(v\Sigma)\big)\bigg\} 
+ \,\bigg\{\,(nv)^2 \,\Big[\,\big(-249 \,\delta \,\kappa_{-} - 6 \,(-93 + 404 \,\nu) - 3 \,\kappa_{+} \,(-83 + 404 \,\nu)\big) \,(nS) + \,\big(-12 \,\delta \,(-6 + 101 \,\nu) 
- 3 \,\delta \,\kappa_{+} \,(-83 + 202 \,\nu) + 3 \,\kappa_{-} \,(-83 + 368 \,\nu)\big) \,(n\Sigma)\Big] + \,(n\Sigma) \,\big(4 \,\delta \,(-17 + 273 \,\nu) \,\bv^2 + \,\delta \,\kappa_{+} \,(-233 + 546 \,\nu) \,\bv^2 
-  \,\kappa_{-} \,(-233 + 1012 \,\nu) \,\bv^2\big) + \,(nS) \,\big(233 \,\delta \,\kappa_{-} \,\bv^2 + 6 \,(-85 + 364 \,\nu) \,\bv^2 + \,\kappa_{+} \,(-233 + 1092 \,\nu) \,\bv^2\big) 
+ \,(nv) \,\Big[\,(12 + 34 \,\delta \,\kappa_{-} - 34 \,\kappa_{+}) \,(vS) + \,\big(-4 \,\delta - 34 \,\delta \,\kappa_{+} - 34 \,\kappa_{-} \,(-1 + 2 \,\nu)\big) \,(v\Sigma)\Big]\bigg\} \,\bSigma^i
+ \,\bv^i \,\bigg\{\,\Big[\,(-480 \,\delta - 8 \,\kappa_{-} - 240 \,\delta \,\kappa_{+}) \,(nS) + \,\big(116 \,\delta \,\kappa_{-} + 8 \,(-37 + 120 \,\nu) + 4 \,\kappa_{+} \,(-29 + 120 \,\nu)\big) \,(n\Sigma)\Big] \,(vS)
+ \,\Big[\,\big(116 \,\delta \,\kappa_{-} + 64 \,(-2 + 15 \,\nu) + 4 \,\kappa_{+} \,(-29 + 120 \,\nu)\big) \,(nS) + \,\big(4 \,\delta \,\kappa_{+} \,(-29 + 60 \,\nu) 
+ 8 \,\delta \,(7 + 60 \,\nu) - 4 \,\kappa_{-} \,(-29 + 118 \,\nu)\big) \,(n\Sigma)\Big] \,(v\Sigma) + \,(nv) \,\Big[\,(-168 \,\delta - 78 \,\kappa_{-} - 84 \,\delta \,\kappa_{+}) \,(nS)^2
+ \,\big(6 \,\delta \,\kappa_{-} + 12 \,(-19 + 56 \,\nu) + 6 \,\kappa_{+} \,(-1 + 56 \,\nu)\big) \,(nS) \,(n\Sigma) + \,\big(12 \,\delta \,(-5 + 14 \,\nu) + 3 \,\delta \,\kappa_{+} \,(-1 + 28 \,\nu)
- 3 \,\kappa_{-} \,(-1 + 30 \,\nu)\big) \,(n\Sigma)^2 + \,(216 \,\delta + 6 \,\kappa_{-} + 108 \,\delta \,\kappa_{+}) \,\bS^2 
+ \,\big(-102 \,\delta \,\kappa_{-}- 6 \,\kappa_{+} \,(-17 + 72 \,\nu) - 8 \,(-25 + 108 \,\nu)\big) \,(S\Sigma) 
+ \,\big(-8 \,\delta \,(2 + 27 \,\nu) - 3 \,\delta \,\kappa_{+} \,(-17 + 36 \,\nu) + 3 \,\kappa_{-} \,(-17 + 70 \,\nu)\big) \,\bSigma^2\Big]\bigg\} 
+ \,\bn^i \Biggl(\,(-1176 \,\delta - 4 \,\kappa_{-} - 588 \,\delta \,\kappa_{+}) \,(vS)^2 + \,\bS^2 \,(-2460 \,\delta \,\bv^2 + 30 \,\kappa_{-} \,\bv^2 - 1230 \,\delta \,\kappa_{+} \,\bv^2) 
+ \,(nS)^2 \,(9648 \,\delta \,\bv^2 - 6 \,\kappa_{-} \,\bv^2 + 4824 \,\delta \,\kappa_{+} \,\bv^2) + \,(S\Sigma) \,\big(1260 \,\delta \,\kappa_{-} \,\bv^2 
+ 60 \,\kappa_{+} \,(-21 + 82 \,\nu) \,\bv^2 + 4 \,(-641 + 2460 \,\nu) \,\bv^2\big) + \,(nS) \,(n\Sigma) \,\big(-4830 \,\delta \,\kappa_{-} \,\bv^2 - 6 \,\kappa_{+} \,(-805 + 3216 \,\nu) \,\bv^2 
- 12 \,(-803 + 3216 \,\nu) \,\bv^2\big) + \,(n\Sigma)^2 \,\big(-12 \,\delta \,(1 + 804 \,\nu) \,\bv^2- 3 \,\delta \,\kappa_{+} \,(-805 + 1608 \,\nu) \,\bv^2 
+ 3 \,\kappa_{-} \,(-805 + 3218 \,\nu) \,\bv^2\big) + \,\big(584 \,\delta \,\kappa_{-} + 112 \,(-11 + 42 \,\nu) + 8 \,\kappa_{+} \,(-73 + 294 \,\nu)\big) \,(vS) \,(v\Sigma) 
+ \,\big(56 \,\delta \,(-1 + 21 \,\nu) + 4 \,\delta \,\kappa_{+} \,(-73 + 147 \,\nu) - 4 \,\kappa_{-} \,(-73 + 293 \,\nu)\big) \,(v\Sigma)^2 
+ \,(nv) \,\bigg\{\,\Big[\,(7080 \,\delta + 96 \,\kappa_{-} + 3540 \,\delta \,\kappa_{+}) \,(nS) + \,\big(-1722 \,\delta \,\kappa_{-} - 48 \,(-76 + 295 \,\nu) 
- 6 \,\kappa_{+} \,(-287 + 1180 \,\nu)\big) \,(n\Sigma)\Big] \,(vS)+ \,\Big[\,\big(-1722 \,\delta \,\kappa_{-} - 120 \,(-29 + 118 \,\nu) - 6 \,\kappa_{+} \,(-287 + 1180 \,\nu)\big) \,(nS) 
+ \,\big(-24 \,\delta \,(-2 + 295 \,\nu) - 6 \,\delta \,\kappa_{+} \,(-287 + 590 \,\nu) + 6 \,\kappa_{-} \,(-287 + 1164 \,\nu)\big) \,(n\Sigma)\Big] \,(v\Sigma)\bigg\}
+ \,\big(30 \,\delta \,\kappa_{+} \,(-21 + 41 \,\nu) \,\bv^2 - 30 \,\kappa_{-} \,(-21 + 83 \,\nu) \,\bv^2 + 4 \,\delta \,(-26 + 615 \,\nu) \,\bv^2\big) \,\bSigma^2
+ \,(nv)^2 \,\Big[\,(-16104 \,\delta + 12 \,\kappa_{-} - 8052 \,\delta \,\kappa_{+}) \,(nS)^2 + \,\big(8064 \,\delta \,\kappa_{-}+ 48 \,\kappa_{+} \,(-168 + 671 \,\nu)
+ 48 \,(-335 + 1342 \,\nu)\big) \,(nS) \,(n\Sigma)+ \,\big(12 \,\delta \,\kappa_{+} \,(-336 + 671 \,\nu) + 24 \,\delta \,(1 + 671 \,\nu) - 12 \,\kappa_{-} \,(-336 + 1343 \,\nu)\big) \,(n\Sigma)^2 
+ \,(2604 \,\delta - 72 \,\kappa_{-} + 1302 \,\delta \,\kappa_{+}) \,\bS^2 + \,\big(-1374 \,\delta \,\kappa_{-} - 6 \,\kappa_{+} \,(-229 + 868 \,\nu) - 12 \,(-227 + 868 \,\nu)\big) \,(S\Sigma)
+ \,\big(-12 \,\delta \,(-10 + 217 \,\nu)- 3 \,\delta \,\kappa_{+} \,(-229 + 434 \,\nu) + 3 \,\kappa_{-} \,(-229 + 892 \,\nu)\big) \,\bSigma^2\Big]\Biggr)\,,
\end{autobreak}
\\[2ex]
\begin{autobreak}
k^{2\,i}_6
=\,\Big[\,(-288 \,\delta + 32 \,\kappa_{-} - 144 \,\delta \,\kappa_{+}) \,(nS) + \,\big(88 \,\delta \,\kappa_{-} + 32 \,(-5 + 18 \,\nu) + 8 \,\kappa_{+} \,(-11 + 36 \,\nu)\big) \,(n\Sigma)\Big] \,\bS^i 
+ \,\Big[\,\big(88 \,\delta \,\kappa_{-} + 8 \,\kappa_{+} \,(-11 + 36 \,\nu) + 12 \,(-11 + 48 \,\nu)\big) \,(nS) + \,\big(8 \,\delta \,\kappa_{+} \,(-11 + 18 \,\nu) - 8 \,\kappa_{-} \,(-11 + 40 \,\nu) 
+ 4 \,\delta \,(-1 + 72 \,\nu)\big) \,(n\Sigma)\Big] \,\bSigma^i
+ \,\bn^i \,\Big[\,(1368 \,\delta + 4 \,\kappa_{-} + 684 \,\delta \,\kappa_{+}) \,(nS)^2 + \,\big(-680 \,\delta \,\kappa_{-} - 8 \,\kappa_{+} \,(-85 + 342 \,\nu) - 4 \,(-341 + 1368 \,\nu)\big) \,(nS) \,(n\Sigma)
+ \,\big(-4 \,\delta \,\kappa_{+} \,(-85 + 171 \,\nu) + 4 \,\kappa_{-} \,(-85 + 341 \,\nu) - 4 \,\delta \,(1 + 342 \,\nu)\big) \,(n\Sigma)^2 + \,(-360 \,\delta - 12 \,\kappa_{-} - 180 \,\delta \,\kappa_{+}) \,\bS^2
+ \,\big(168 \,\delta \,\kappa_{-} + 24 \,\kappa_{+} \,(-7 + 30 \,\nu) + 8 \,(-47 + 180 \,\nu)\big) \,(S\Sigma) + \,\big(12 \,\delta \,\kappa_{+} \,(-7 + 15 \,\nu)
- 12 \,\kappa_{-} \,(-7 + 29 \,\nu) + 8 \,\delta \,(-2 + 45 \,\nu)\big) \,\bSigma^2\Big]\,.
\end{autobreak}
\end{align}
\end{subequations}
\bibliographystyle{hunsrt}
\bibliography{Ref3PN}
\end{document}